\documentclass[aps,preprint]{revtex4}%
\usepackage{amsfonts}
\usepackage{amsmath}
\usepackage{amssymb}

\usepackage{subfigure}
\usepackage{graphicx}%
\begin{document}
\title{Joule-Thomson expansion of Reissner-Nordstr\"om-Anti-de Sitter black holes with cloud of strings and
quintessence}

\author{Rui Yin$^{a,b}$}
\email{yrphysics@126.com}
\author{Jing Liang$^{a,b}$}
\email{ljphysics@163.com}
\author{Benrong Mu$^{a,b}$}
\email{benrongmu@cdutcm.edu.cn}
\affiliation{$^{a}$ Physics Teaching and Research section, College of Medical Technology,
Chengdu University of Traditional Chinese Medicine, Chengdu, 611137, PR China}
\affiliation{$^{b}$Center for Theoretical Physics, College of Physics, Sichuan University,
Chengdu, 610064, PR China}

\begin{abstract}
The Joule-Thomson expansion is studied for Reissner-Nordstr\"om-Anti-de Sitter black holes with cloud of strings and quintessence, as well as its thermodynamics. The cosmological constant is treated as thermodynamic pressure, whose conjugate variable is considered as the volume. The characteristics of the Joule-Thomson expansion are studied in four main aspects with the case of $\omega=-1$ and $\omega=-\frac{2}{3}$, including the Joule-Thomson coefficient, the inversion curves, the isenthalpic curves and the ratio between $T_{i}^{min}$ and $T_{c}$. The sign of the Joule-Thomson coefficient is possible for determining the occurrence of heating or cooling. The scattering point of the Joule-Thomson coefficient corresponds to the zero point of the Hawking temperature. Unlike the van der Waals fluids, the inversion curve is the dividing line between heating and cooling regions, above which the slope of the isenthalpic curve is positive and cooling occurs, and the cooling-heating critical point is more sensitive to $Q$. Concerning the ratio $\frac{T_{i}^{min}}{T_{c}}$, we calculate it separately in the cases where only the cloud of strings, only quintessence and both are present.

\end{abstract}
\keywords{}\maketitle
\tableofcontents

{}

\bigskip{}



\section{Introduction}
A black hole is a structure that occupies an important place in the theory of gravity, which contributes to the understanding of an eventual formulation of quantum gravity as well as some of their thermodynamic properties issues. Then there were other people who obtained the solutions of the black hole in different gravitational theories, among which some exact solutions in the framework of Lovelock gravity were obtained \cite{Boulware:1985wk,Wheeler:1985nh,Cai:2003kt,Hennigar:2016xwd,Myers:1988ze}. When considering being surrounded by a cloud of strings, Letelier first performed a theoretical analysis and obtained the solutions of the Einstein equations corresponding to a black hole surrounded by a cloud of strings \cite{Letelier:1979ej}. String theory was extended to a major idea of considering a cloud of strings and studying its possible measurable effects on long range gravitational fields of a black hole. Ghosh et al. obtained a generalization for third-order Lovelock gravity \cite{Ghosh:2014pga}, and Herscovich et al. obtained the solution for Einstein-Gauss-Bonnet theory in the Letelier spacetime \cite{Herscovich:2010vr}. Just under such a background, others investigated the thermodynamic property aspects \cite{Lee:2015xlp} and the tensor quasinormal modes \cite{Graca:2016cbd}. It also had some interesting results that raised attention \cite{Letelier:1984dm,Li:2020zxi,Cai:2019nlo,intro-Toledo:2019szg,Ghaffarnejad:2018tpr,Ghaffarnejad:2018gbf,intro-Ghosh:2014dqa}.

Astronomical observations show that the universe may accelerate its expansion \cite{Perlmutter:1998np}, and the acceleration is due to the existence of a gravitationally repulsive energy component, i.e., a negative pressure. One of the possible factors leading to this negative pressure is the cosmological constant, and the other is the quintessence hypothetical form of the dark energy \cite{Kiselev:2002dx}. In the latter case, the pressure and energy density are proportional as $p_{q}=\rho_{q}\omega$, where the barotropic index $\omega$ takes the value interval $-1<\omega<-\frac{1}{3}$ \cite{Stern:1999ta,Bahcall:1999xn,Steinhardt:1999nw,Wang:1999fa}. When $\omega =-1$, the border state of the quintessence covers the cosmological constant regime. Then the solution of a black hole surrounded by quintessence was studied extensively. Kiselev obtained the analytical solutions with spherical
symmetry. Based on this, someone also obtained the generalization of the Kiselev solutions \cite{Godunov:2015nea,Toshmatov:2017btq}. There are some typical effects of quintessence on black holes that have been studied as well \cite{Chen:2008ra,Chen:2005qh,intro-Saleh:2011zz,intro-Li:2014ixn,intro-Singh:2020tkf,intro-Haldar:2020jmt,intro-Chen:2020rov,intro-Moinuddin:2019mzf,intro-Toledo:2019mlz,intro-Chabab:2017xdw,intro-Liu:2017baz,intro-Ghaderi:2016dpi,intro-Fernando:2014rsa,intro-Fernando:2014wma,intro-Guo:2019hxa,intro-Nandan:2016ksb,intro-Malakolkalami:2015cza,intro-WangChun-Yan:2012tcg,intro-Xi:2008ce,intro-Harada:2006dv,Ghaffarnejad:2018zlz}. Up to now, the effects of cloud of strings and quintessence on black holes have been comprehensively studied \cite{Yin:2021fsg,Toledo:2019amt,Chabab:2020ejk,intro-Sakti:2019iku,Ma:2019pya,Toledo:2018pfy}.

Black hole thermodynamics has been an area of great interest. The work of Bekenstein and Hawking \cite{Bekenstein:1973ur,Bardeen:1973gs,Bekenstein:1974ax,Hawking:1974rv,Hawking:1974sw} pioneered the research related to black hole thermodynamics. In black hole thermodynamics, a black hole is considered as a thermodynamic system with temperature and entropy. Another important development in black hole thermodynamics is phase transition. Among the research in this field, thermodynamics of AdS (Anti-de Sitter) black holes has been intensively studied in the extended phase space. The transition between the Schwarzschild AdS black hole and the thermal AdS space was discovered earlier by Hawking and Page \cite{Hawking:1982dh}. And then, a fact that the close relationship between charged AdS black holes and the liquid-gas system was discovered \cite{Chamblin:1999tk,Chamblin:1999hg}. With the discoveries of the reentrant phase transition \cite{Gunasekaran:2012dq,Altamirano:2013ane} and triple point \cite{Altamirano:2013uqa}, the behavior of AdS black holes was further shown to be similar to that of ordinary thermodynamic systems, where the cosmological constant and its conjugate quantity are treated as the thermodynamic pressure and volume, respectively. In this context, various interesting studies have been carried out. For instance, Johnson creatively introduced the concept of holographic heat engine \cite{Belhaj:2015hha}, which provides one way to extract mechanical work from black holes. It was explored that more researches on thermodynamic aspects of black holes, such as compressibility \cite{Dolan:2011jm,Dolan:2013dga}, critical phenomenon \cite{Wei:2012ui,Banerjee:2011cz,Niu:2011tb}, weak cosmic censorship conjecture \cite{Chen:2019pdj,Liang:2021voh,Mu:2020szg,Liang:2020uul,Liang:2020hjz,Bai:2020ieh,Mu:2019bim,Hong:2019yiz}, and behaviour of the quasi-normal modes \cite{Chabab:2017knz}.

Moreover, the Joule-Thomson (JT) extension has recently been cleverly applied to black holes in AdS spacetime, in which the analogy between the black holes and the van der Waals system is generalized. The Joule-Thomson expansion of black holes was investigated in Ref. \cite{Okcu:2016tgt} for the first time. In classical thermodynamics, Joule-Thomson expansion is described as gas at a high pressure passes through a porous plug to a section with a low pressure, during which the enthalpy remains constant. The study found that Joule-Thomson expansion is used as a isoenthalpic tool to show the thermal expansion where there are heating and cooling regimes. And the $T-P$ graph is divided into heating and cooling parts by the inversion curve. This research was soon generalized to the Kerr-AdS black holes \cite{Okcu:2017qgo}, quintessence charged AdS black holes \cite{Ghaffarnejad:2018exz}, holographic superfluids\cite{DAlmeida:2018ldi}, charged AdS black holes in $f(R)$ gravity \cite{Chabab:2018zix}, AdS black hole with a global monopole \cite{Rizwan:2018mpy} and AdS black holes in Lovelock gravity \cite{Mo:2018qkt}. There have recently been many studies on Joule-Thomson expansion for various black holes \cite{Liang:2021elg,Hegde:2020xlv,Mo:2018rgq,Lan:2018nnp,Wei:2017vqs,Kuang:2018goo,Yekta:2019wmt,Pu:2019bxf,Nam:2018sii,Zhao:2018kpz,Li:2019jcd,Hyun:2019gfz,Ghaffarnejad:2018tpr,Nam:2019zyk,Nam:2018ltb,Rostami:2019ivr,Haldar:2018cks,Guo:2019gkr,Lan:2019kak,Sadeghi:2020bon,Bi:2020vcg,Ranjbari:2019ktp,Guo:2019pzq,K.:2020rzl,Nam:2020gud,Meng:2020csd,Guo:2020qxy,Ghanaatian:2019xhi,Guo:2020zcr,Feng:2020swq,Debnath:2020zdv,Cao:2021dcq,Huang:2020xcs,Zhang:2021raw,Chen:2020igz,Jawad:2020mdc,Liang:2021xny,Debnath:2020inx,Mirza:2021kvi,Graca:2021izb}. All the papers above showed that the inversion curves are different from the inversion curves in van der Waals system. However, so far, Joule-Thomson expansion of RN-AdS black holes with cloud of strings and quintessence in extended phase space has never
been studied, which is the main subject of our research.

This paper is organized as follows. In section \ref{sec:B}, the metric of RN-AdS black holes with cloud of strings and quintessence is proposed, whose thermodynamic properties are discussed. Then comes section \ref{sec:C}, we review the well-known results of Joule-Thomson expansion for van der Waals fluids in classical thermodynamics in section \ref{sec:CA}. In section \ref{sec:CB}, we explore Joule-Thomson expansion for RN-AdS black holes with cloud of strings and quintessence, where the Joule-Thomson coefficient, the inversion curves and the isenthalpic curves are studied in detail. Furthermore, we also calculate the ratio between minimum inversion temperature $T_{i}^{min}$ and the critical temperature $T_{c}$ in different cases. Section \ref{sec:D} denotes to conclusion.

\section{Quintessence surrounding Reissner-Nordstr\"om-Anti-de Sitter black holes with a
cloud of strings}
\label{sec:B}
In this section, it is obtained that the metric corresponding to the spacetime generated by a charged static black hole with cosmological constant and surrounded by a cloud of strings and quintessence.
The solution corresponding to a black hole with quintessence and cloud of strings is derived in Ref. \cite{Toledo:2019amt}, where it is assumed that the cloud of strings and quintessence do not interact. On this basis, the energy-momentum tensor of the two sources is considered as a linear superposition, whereupon we obtain
\begin{equation}\label{eqn:Q1}
 T_{t}^{t}=T_{r}^{r}=\rho_{q}+\frac{a}{r^{2}},
\end{equation}
\begin{equation}\label{eqn:Q2}
  T_{\theta}^{\theta}=T_{\phi}^{\phi}=-\frac{1}{2}\rho_{q}(3\omega+1).
\end{equation}
Here the pressure and density of quintessence are related as $p_{q}=\rho_{q}\omega$, where $\omega$ is the quintessential state parameter. By means of the static spherically symmetric line element, we can obtain the line element associated with a charged black hole surrounded by a cloud of strings and quintessence as follows \cite{Ma:2019pya,Chabab:2020ejk}
\begin{equation}\label{eqn:Q3}
\begin{aligned}
 & ds^{2}=(1-a-\frac{2M}{r}+\frac{Q^{2}}{r^{2}}-\frac{\alpha}{r^{3\omega+1}}-\frac{\varLambda r^{2}}{3})dt^{2}\\
 &-(1-a-\frac{2M}{r}+\frac{Q^{2}}{r^{2}}-\frac{\alpha}{r^{3\omega+1}}-\frac{\varLambda r^{2}}{3})^{-1}dr^{2}-r^{2}d\varOmega^{2},\\
 \end{aligned}
\end{equation}
where the solution of the main equation is
\begin{equation}\label{eqn:Q4}
  f(r)=1-a-\frac{2M}{r}+\frac{Q^{2}}{r^{2}}-\frac{\alpha}{r^{3\omega+1}}-\frac{\varLambda r^{2}}{3}.
\end{equation}
In the equation above, $M$ and $Q$ are the mass and electric charge of the black hole, respectively. Where $\varLambda$ is the cosmological constant, $a$ is an integration constant caused by the cloud of strings, and $\alpha$ is a normalization constant related to the quintessence as
\begin{equation}\label{eqn:Q5}
\rho_{q}=-\frac{\alpha}{2}\frac{3\omega}{r^{3(\omega+1)}},
\end{equation}
in order to get the scenario of accelerated expansion, the barotropic index will have $-1<\omega<-\frac{1}{3}$. In general, when $-\frac{1}{3}<\omega<0$, the free quintessence generates the horizon of the black hole; when $-1<\omega<-\frac{2}{3}$, the free quintessence generates the AdS radius $l$. In what follows, we will fix $\omega= -1$ for the cosmological constant regime of the quintessence and fix $\omega=-\frac{2}{3}$ for the quintessence regime of the dark energy.

The black hole has three positive horizons. The first two are the black hole event horizon $r_{+}$ and the internal (Cauchy) horizon $r_{-}$, which correspond to equation $f(r)=0$ under a non-extreme black hole. The last one is a quintessential cosmological horizon $r_{q}$. When the black hole is extremal, $f(r)=0$ only has a single root $r_{+}$.

Recently, the cosmological constant is treated as a variable related to pressure in the thermodynamics of black holes, whose relationship is expressed as \cite{Dolan:2011xt,Kubiznak:2012wp,Cvetic:2010jb,Caceres:2015vsa,Hendi:2012um,Pedraza:2018eey}
\begin{equation}\label{eqn:Q6}
  P=-\frac{\varLambda}{8\pi}=\frac{3}{8\pi l^{2}}.
\end{equation}
Then, the mass of the black hole can be represented by
\begin{equation}\label{eqn:Q7}
  M=\frac{1}{2}(r_{+}-r_{+}a+\frac{8}{3}\pi Pr_{+}^{3}+\frac{Q^{2}}{r_{+}}-\alpha r_{+}^{-3\omega}),
\end{equation}
and the Hawking temperature of the black hole is
\begin{equation}\label{eqn:Q8}
  T=\frac{f'(r_{+})}{4\pi}=\frac{\frac{2M}{r_{+}^{2}}+\frac{16\pi Pr_{+}}{3}-\frac{2Q^{2}}{r_{+}^{3}}+\alpha(3\omega+1)r_{+}^{-3\omega-2}}{4\pi}.
\end{equation}
The first law of thermodynamics in the extended phase space takes on the form as
\begin{equation}\label{eqn:Q9}
  dM=TdS+VdP+\varphi dQ+\gamma d\alpha+\eta da,
\end{equation}
where
\begin{equation}\label{eqn:Q10}
  \gamma=-\frac{1}{2r_{+}^{3\omega}},\eta=-\frac{r_{+}}{2}.
\end{equation}
At the event horizon, one can also derive the volume as well as the entropy of this kind of black hole
\begin{equation}\label{eqn:Q11}
  S=\pi r_{+}^{2},
\end{equation}
\begin{equation}\label{eqn:Q12}
 V=(\frac{\partial M}{\partial P})_{S,Q}=\frac{4\pi r_{+}^{3}}{3}.
\end{equation}

\section{Joule-Thomson expansion}
\label{sec:C}
In this section, the Joule-Thomson expansion of the RN-AdS black hole with cloud of strings and
quintessence is studied and compared with the well known expansion of the van der Waals fluids. Joule-Thomson effect is an irreversible adiabatic expansion of gas from a high pressure section to a low pressure section through a porous plug. In this process, the nonideal gas changes in the final state temperature due to a continuous throttling process, while the usual thermodynamic coordinates cannot be used to describe the gas through the dissipative nonequilibrium state. But the sum of internal energy and the pressure volume product remains the same in the final state, based on which the enthalpy can be obtained
\begin{equation}\label{eqn:J1}
H=PV+U.
\end{equation}
The enthalpy remains constant during the expansion process. Hence, the Joule-Thomson expansion of a black hole is an isenthalpy process in the extended phase space. The Joule-Thomson coefficient serves as an important physical quantity on investigating the Joule-Thomson expansion, whose sign can be used to determine whether heating or cooling occurs. The Joule-Thomson coefficient can describe temperature changes with respect to pressure, which is given by
\begin{equation}\label{eqn:J2}
 \mu=(\frac{\partial T}{\partial P})_{H}.
\end{equation}
The change in pressure during expansion is negative because the pressure is always decreasing, while the change in temperature is uncertain. If the change in temperature is positive, the coefficient $\mu$ is negative, which means that the gas becomes warmer; conversely, if the change in temperature is negative, the coefficient $\mu$ is positive and the gas becomes colder.

The following relationship for constant particle number $N$ can be obtained from the first law of thermodynamics
\begin{equation}\label{eqn:J33}
dU=TdS-PdV,
\end{equation}
relating the differential form of the enthalpy, one can obtain
\begin{equation}\label{eqn:J3}
 dH=VdP+TdS,
\end{equation}
when $dH=0$, Eq. $\left(\ref{eqn:J3}\right)$ becomes
\begin{equation}\label{eqn:J4}
  T(\frac{\partial S}{\partial P})_{H}+V=0.
\end{equation}
The differential form of the entropy for the state function can be written
\begin{equation}\label{eqn:J5}
  dS=(\frac{\partial S}{\partial P})_{T}dP+(\frac{\partial S}{\partial T})_{P}dT,
\end{equation}
based on which another expression can be obtained
\begin{equation}\label{eqn:J6}
 (\frac{\partial S}{\partial P})_{H}=(\frac{\partial S}{\partial P})_{T}+(\frac{\partial S}{\partial T})_{P}(\frac{\partial T}{\partial P})_{H}.
\end{equation}
Substituting Eq. $\left(\ref{eqn:J6}\right)$ into Eq. $\left(\ref{eqn:J4}\right)$, the following relationship can be obtained
\begin{equation}\label{eqn:J7}
  T[(\frac{\partial S}{\partial P})_{T}+(\frac{\partial S}{\partial T})_{P}(\frac{\partial T}{\partial P})_{H}]+V=0.
\end{equation}
With the assistance of $(\frac{\partial S}{\partial P})_{T}=-(\frac{\partial V}{\partial T})_{P}$ and $C_{P}=T(\frac{\partial S}{\partial T})_{P}$, Eq. $\left(  \ref{eqn:J7}\right)$ can be rewritten as
\begin{equation}\label{eqn:J8}
 -T(\frac{\partial V}{\partial T})_{P}+C_{P}(\frac{\partial T}{\partial P})_{H}+V=0,
\end{equation}
which can be rearranged to give the Joule-Thomson coefficient \cite{Hawking:1982dh} as follows
\begin{equation}\label{eqn:J9}
  \mu=(\frac{\partial T}{\partial P})_{H}=\frac{1}{C_{P}}[T(\frac{\partial V}{\partial T})_{P}-V],
\end{equation}
setting $\mu = 0$, we can obtain the inversion temperature, i.e.
\begin{equation}\label{eqn:J10}
  T_{i}=V(\frac{\partial T}{\partial V})_{P}.
\end{equation}
\subsection{ Van der Waals fluids}
\label{sec:CA}
The van der Waals gas is the simplest model used to explain the behavior of the real gases, which is obtained by taking into account the size of the gas molecules and the attraction between them. The equation of state for van der Waals fluids is given by
\begin{equation}\label{eqn:V1}
  (P+\frac{a}{V_{m}^{2}})(V_{m}-b)=k_{B}T,
\end{equation}
where $k_{B}$ denotes Boltzmann constant, the constants $a$ and $b$ parametrize the strength of the intermolecular interaction and the volume that is excluded owing to the finite size of the molecule, respectively. The constants $a$ and $b$ are determined from experimental data. In the case where $a$ and $b$ tend to 0, the equation of state can be simplified to the ideal gas equation. The gas-liquid phase transition behavior of the actual fluids is better described \cite{Johnston:2014dc,Landau:1980dc}, it is given by
\begin{equation}\label{eqn:V2}
  P=\frac{k_{B}T}{V-b}-\frac{a}{V^{2}}.
\end{equation}
The internal energy of van der Waals gas is given by
\begin{equation}\label{eqn:V3}
  U(T,V)=\frac{3k_{B}T}{2}-\frac{a}{V},
\end{equation}
using the Legendre transformation and Eq. $\left(\ref{eqn:J1}\right)$, the following equation can be obtained
\begin{equation}\label{eqn:V4}
  H(T,V)=\frac{3k_{B}T}{2}+\frac{k_{B}TV}{V-b}-\frac{2a}{V}.
\end{equation}
Using Eq. $\left(\ref{eqn:J10}\right) $, the inversion temperature of the van der Waals system is obtained as
\begin{equation}\label{eqn:V5}
  T_{i}=\frac{1}{k_{B}}(P_{i}V-\frac{a}{V^{2}}(V-2b)),
\end{equation}
and from the state Eq. $\left(\ref{eqn:V2}\right) $ we have
\begin{equation}\label{eqn:V6}
  T_{i}=\frac{1}{k_{B}}(P_{i}+\frac{a}{V^{2}})(V-b).
\end{equation}
Subtracting Eq. $\left(  \ref{eqn:V5}\right) $ from Eq. $\left(\ref{eqn:V6}\right) $ yields
\begin{equation}\label{eqn:V7}
  bP_{i}V^{2}-2aV+3ab=0,
\end{equation}
then, it can obtain two roots by solving this equation for $V(P_{i})$,
\begin{equation}\label{eqn:V8}
  V=\frac{a\pm\sqrt{a^{2}-3ab^{2}P_{i}}}{bP_{i}},
\end{equation}
 substituting these roots into Eq. $\left( \ref{eqn:V5}\right) $, one can obtain
 \begin{equation}\label{eqn:V9}
   T_{i}=\frac{2(5a-3b^{2}P_{i}\pm4\sqrt{a^{2}-3ab^{2}P_{i}})}{9bk_{B}}.
 \end{equation}

 \begin{figure}
  \centering
  \includegraphics[width=0.45\textwidth]{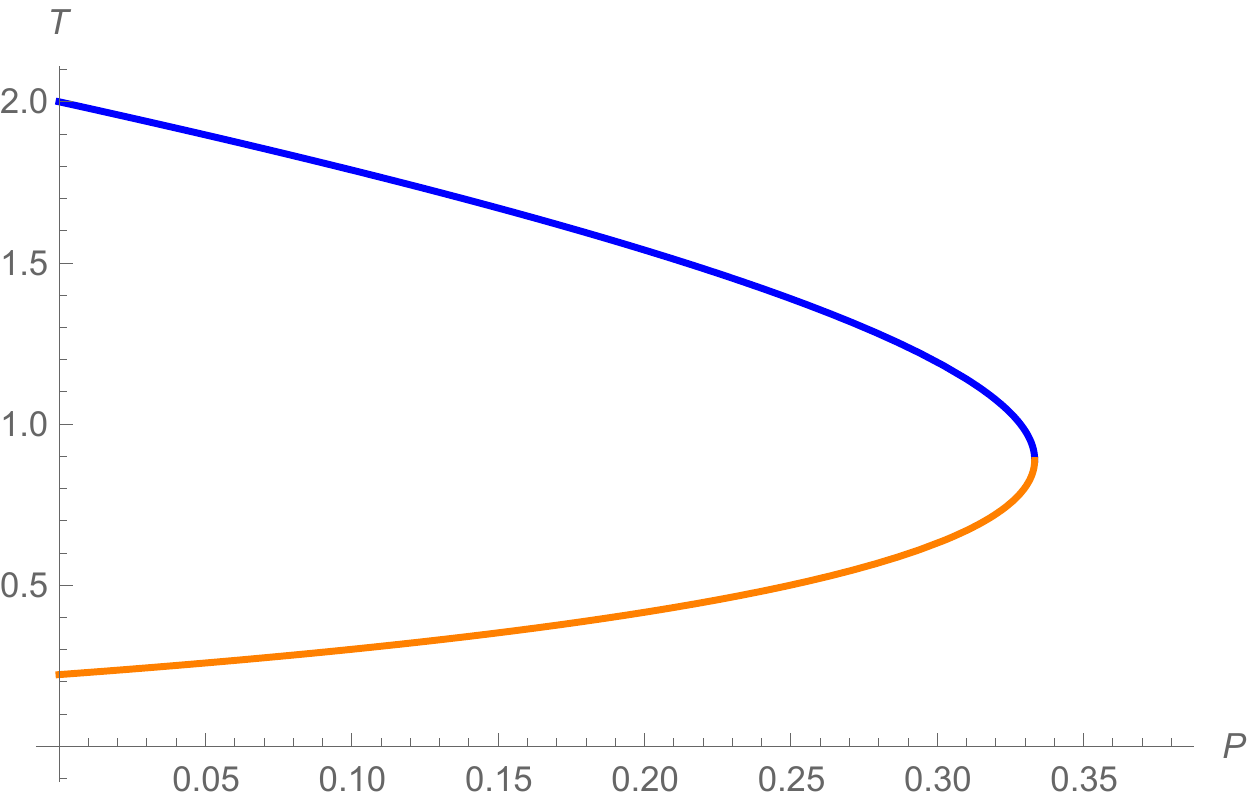}\\
  \caption{Lower (solid orange line) and upper (dashed blue line) inversion
curves. We fix $a=b=k_{B}= 1$.}\label{fig.V1}
\end{figure}

\begin{figure}
  \centering
  \includegraphics[width=0.45\textwidth]{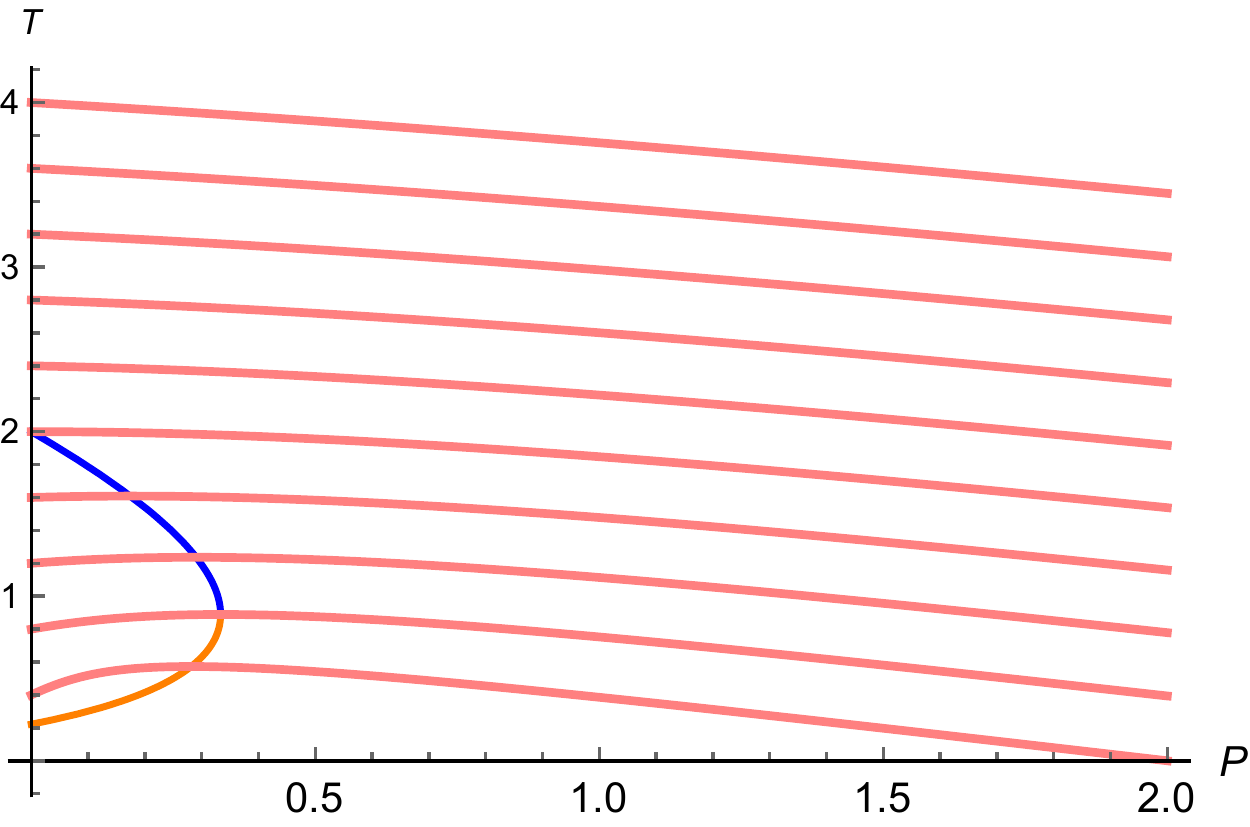}\\
  \caption{ The inversion curves and
the isenthalpic curves. The enthalpies of isenthalpic curves increase
from bottom to top and correspond to $H = 1, 2, 3, 4, 5, 6, 7, 8, 9, 10$.
 fix $a=b=k_{B}= 1$.}\label{fig.V2}
\end{figure}

Using above the inversion curves for the van der Waals system are plotted in Fig. \ref{fig.V1}. In Fig. \ref{fig.V2}, the isenthalpic and inversion curves are shown together. It can be clearly seen that the slope of the isenthalpic curves changes when it passes through the inversion curves, i.e., from positive to negative. It can be concluded that the Joule-Thomson coefficient is positive within the inversion curves, and so cooling occurs in this region.

At the point $P_{i} = 0$, the minimum and maximum inversion temperatures can be obtained as
\begin{equation}\label{eqn:V99}
 T_{i}^{min}=\frac{2a}{9bk_{B}},T_{i}^{max}=\frac{2a}{bk_{B}},
\end{equation}
the critical temperature of the van der Waals fluids is given
\begin{equation}\label{eqn:V10}
T_{c}=\frac{8a}{27bk_{B}},
\end{equation}
then, the ratio between the inversion temperature and the critical temperature is
\begin{equation}\label{eqn:V11}
  \frac{T_{i}^{min}}{T_{c}}=\frac{3}{4},\frac{T_{i}^{max}}{T_{c}}=\frac{27}{4}.
\end{equation}
\begin{figure}
\begin{center}
  \includegraphics[width=0.45\textwidth]{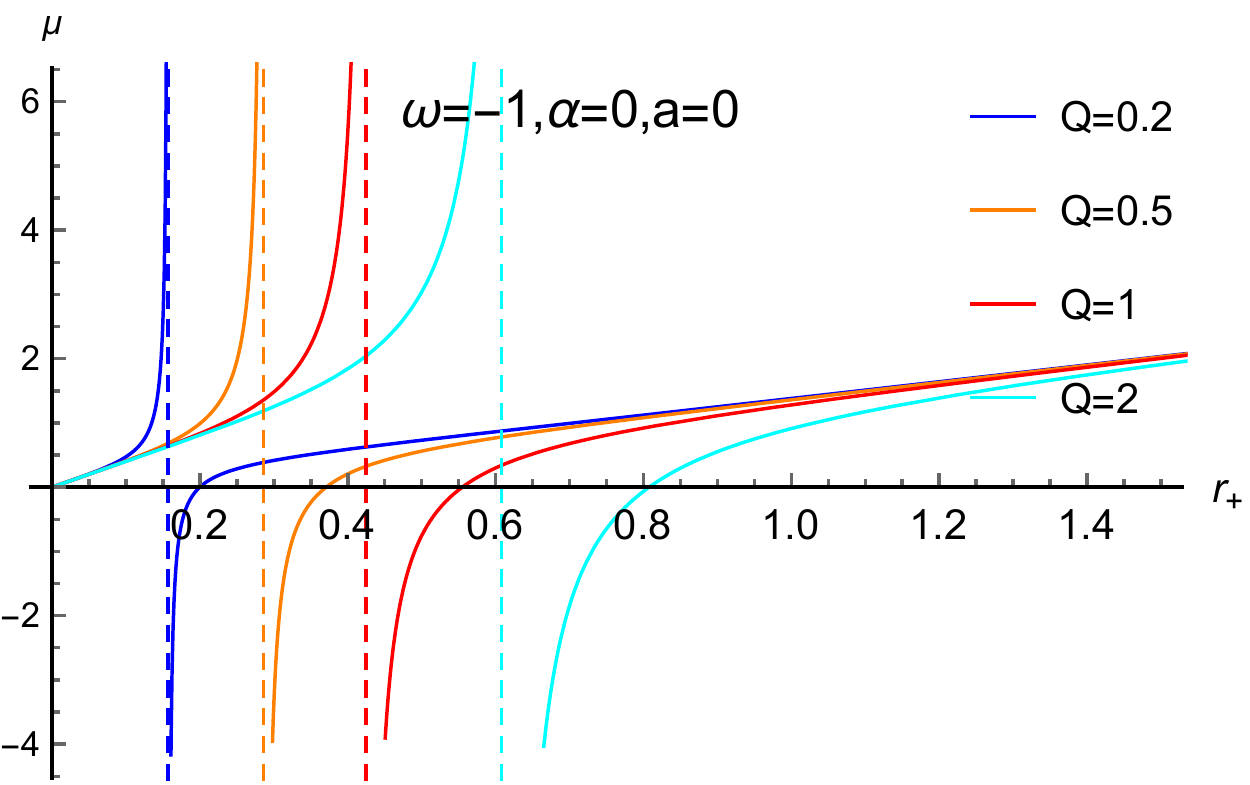}
 \label{fig:j1}
 \includegraphics[width=0.4\textwidth]{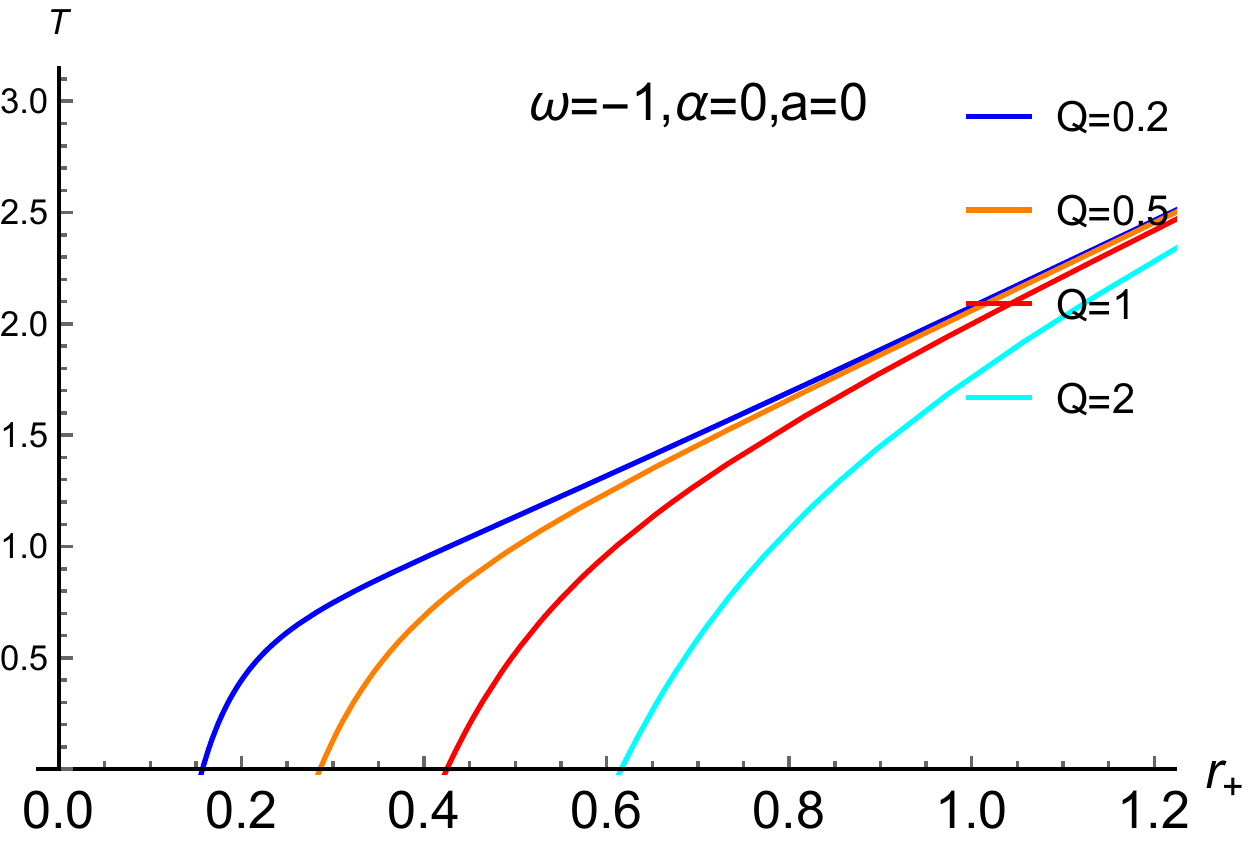}
 \label{fig:h1}
   \includegraphics[width=0.45\textwidth]{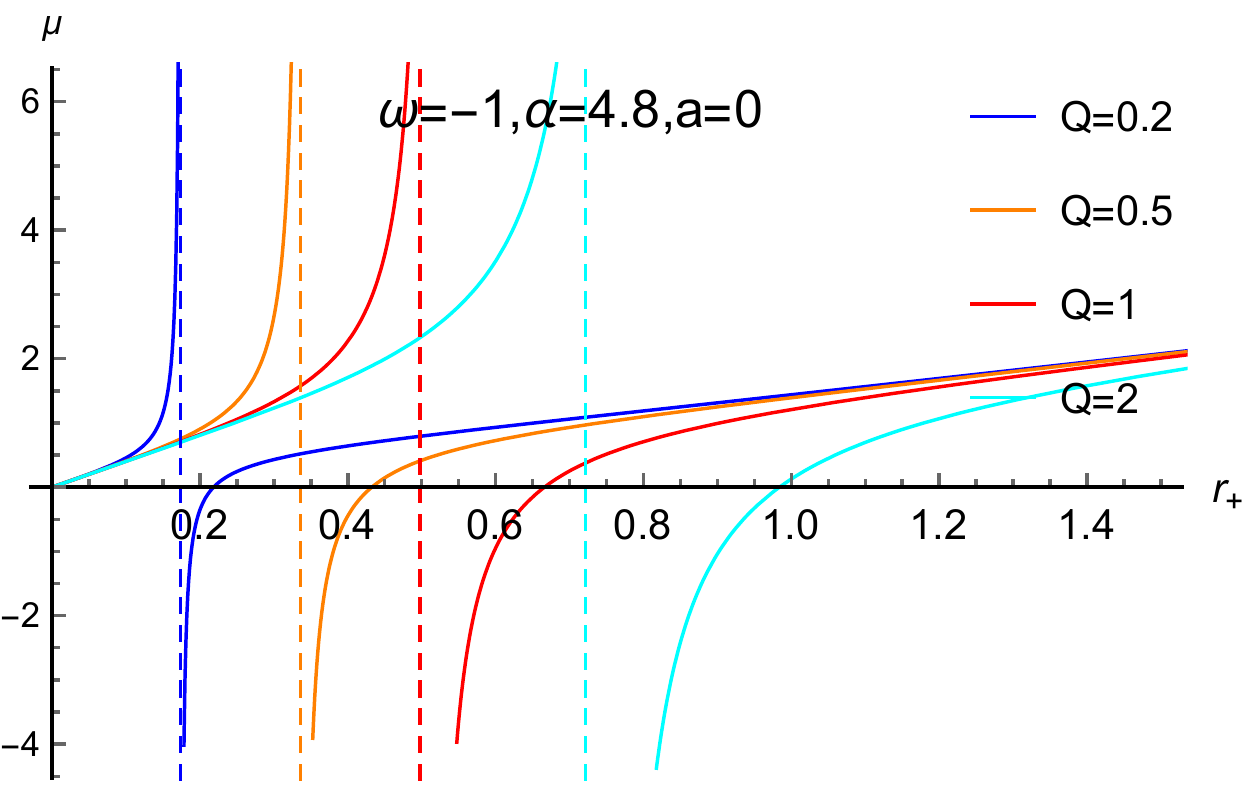}
 \label{fig:j2}
 \includegraphics[width=0.4\textwidth]{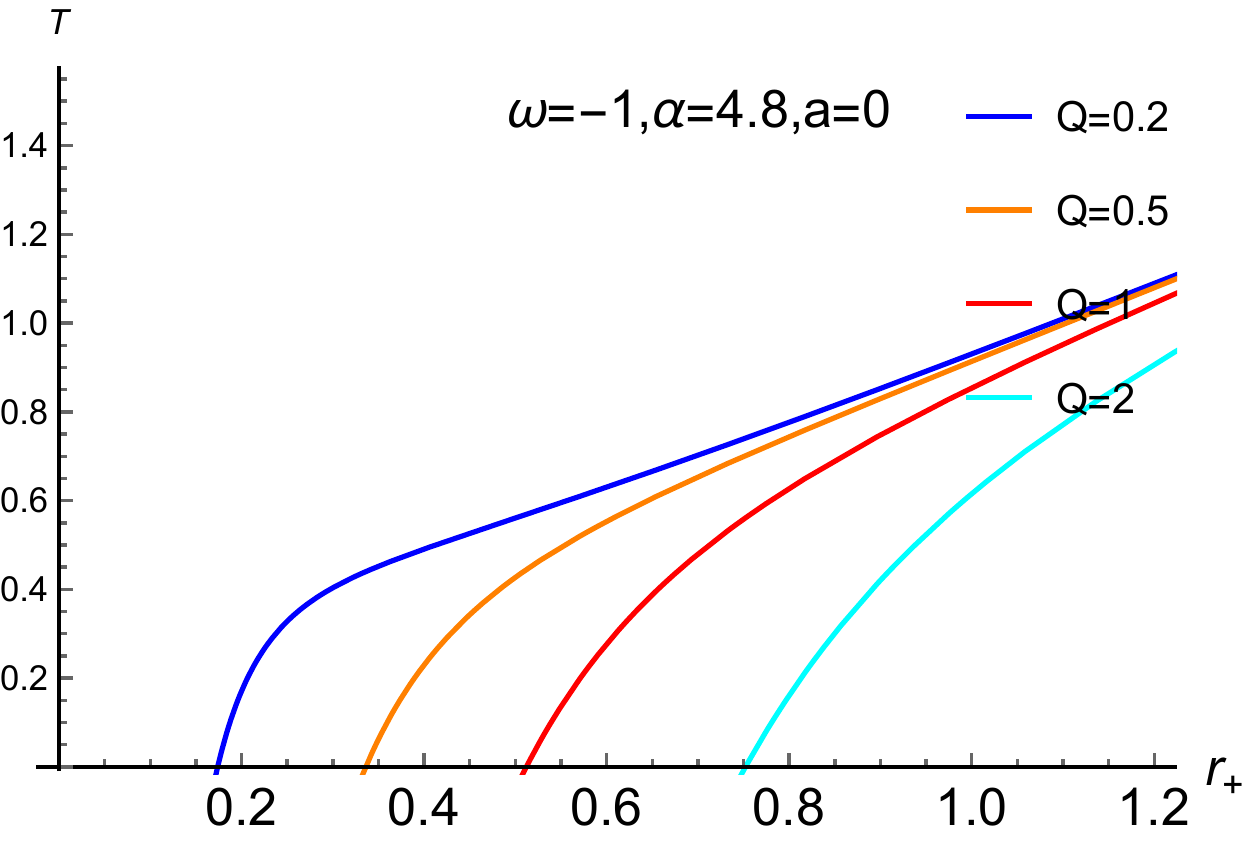}
 \label{fig:h2}
  \includegraphics[width=0.45\textwidth]{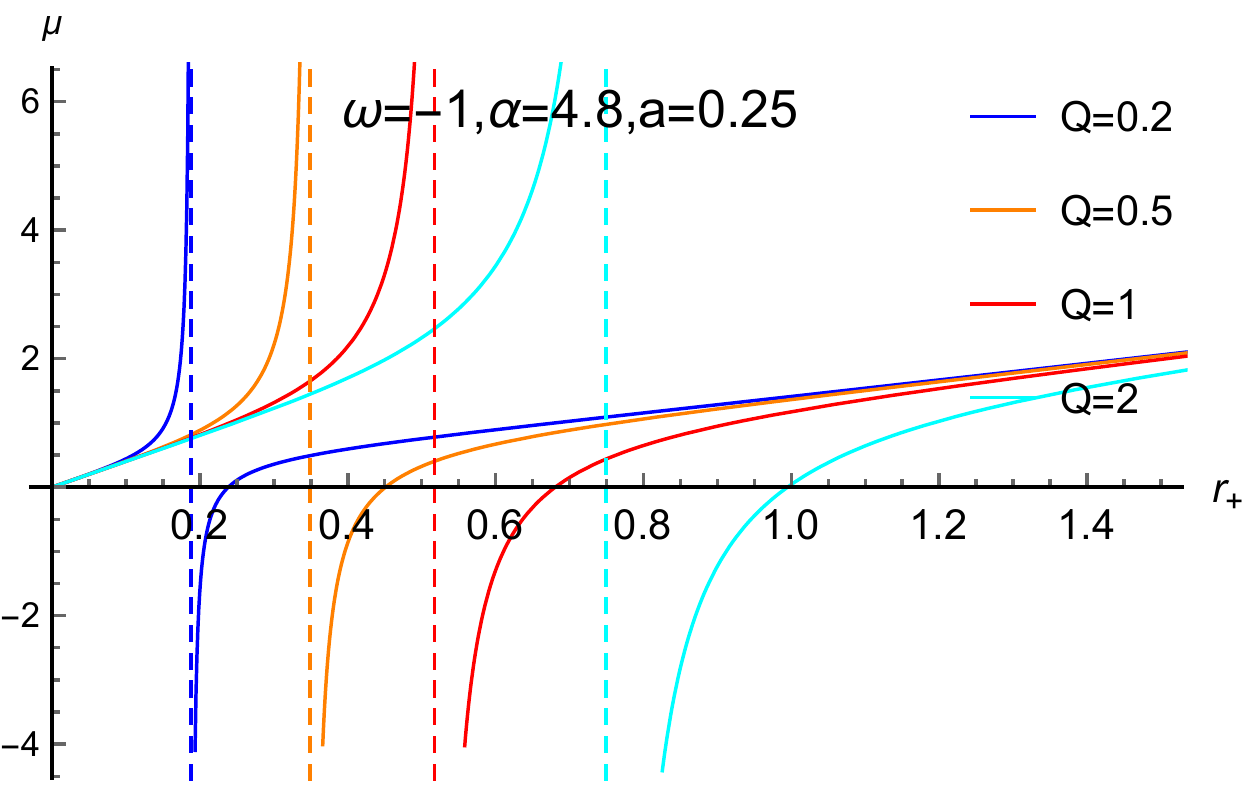}
 \label{fig:j3}
\includegraphics[width=0.4\textwidth]{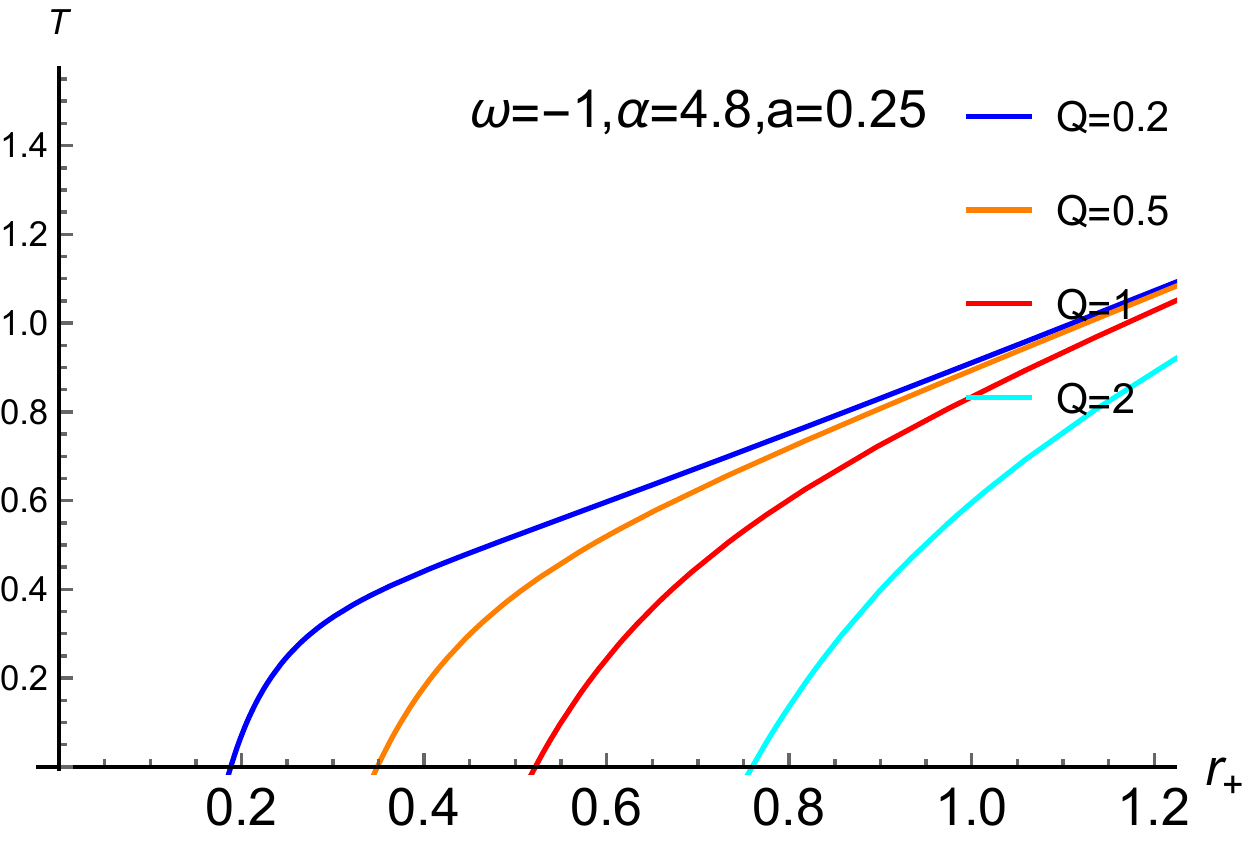}
 \label{fig:h3}
  \includegraphics[width=0.45\textwidth]{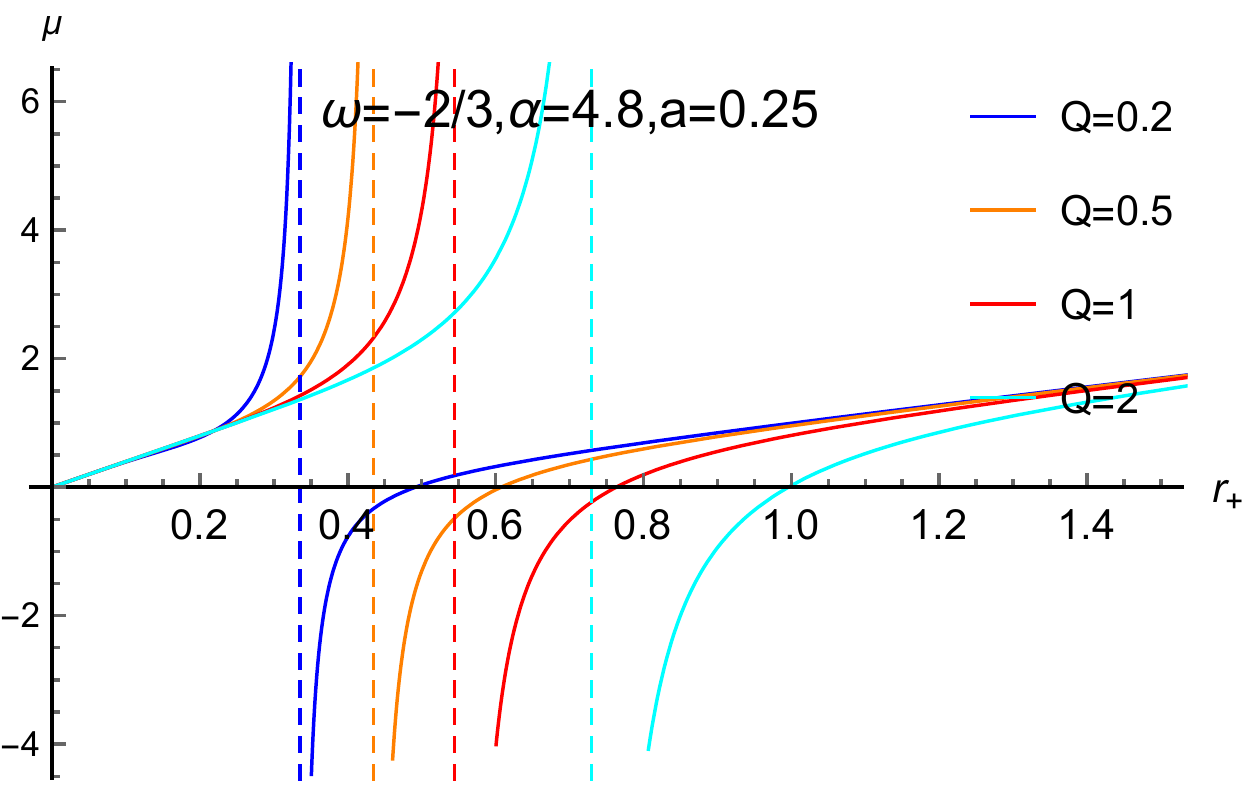}
 \label{fig:j4}
   \includegraphics[width=0.4\textwidth]{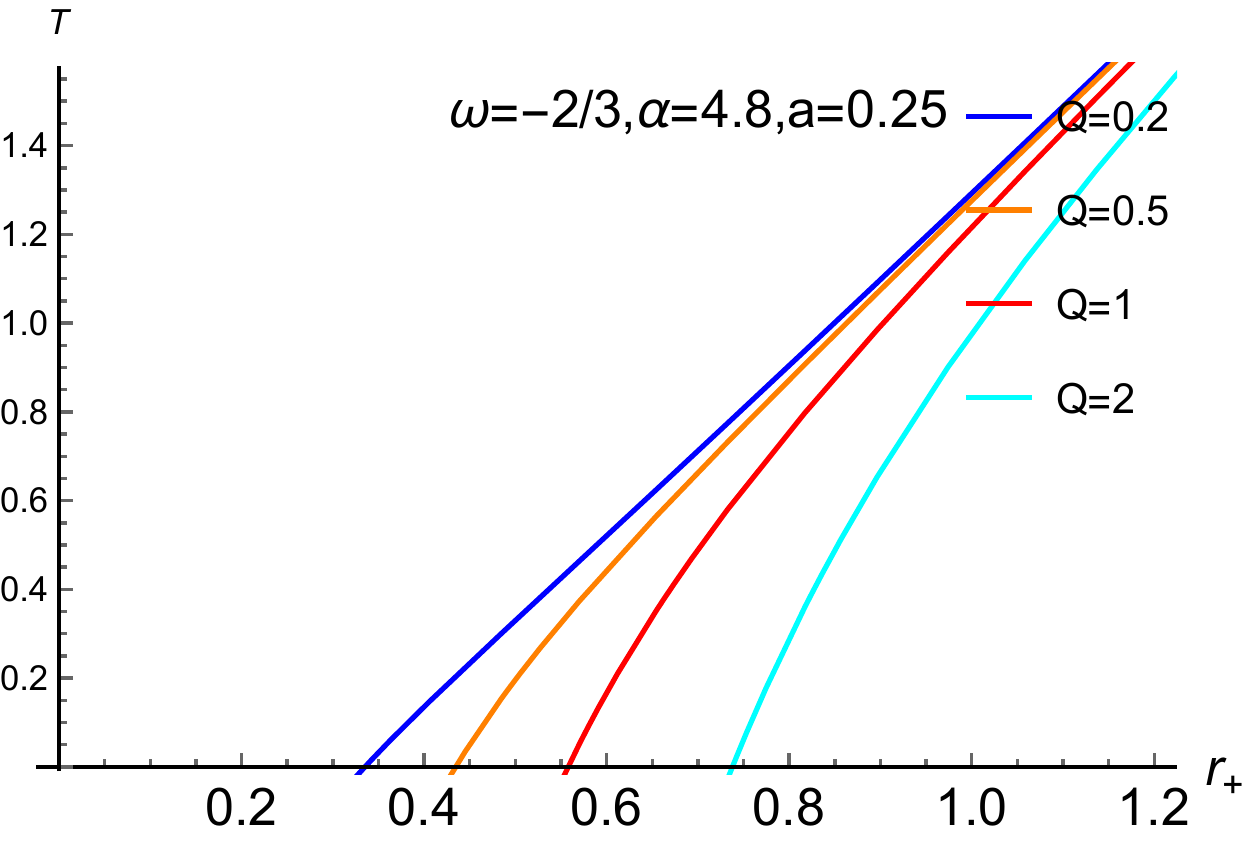}
 \label{fig:h4}
 \end{center}
 \caption{Joule-Thomson coefficient $\mu$ and Hawking temperature and $T$ versus the event horizon $r_{+}$, here $P=1$, from the left to the right, the curves correspond to $Q = 0.2, 0.5, 1, 2$.}%
\label{fig:J}
\end{figure}

\begin{figure}
\begin{center}
\subfigure[{}]{
\includegraphics[width=0.3\textwidth]{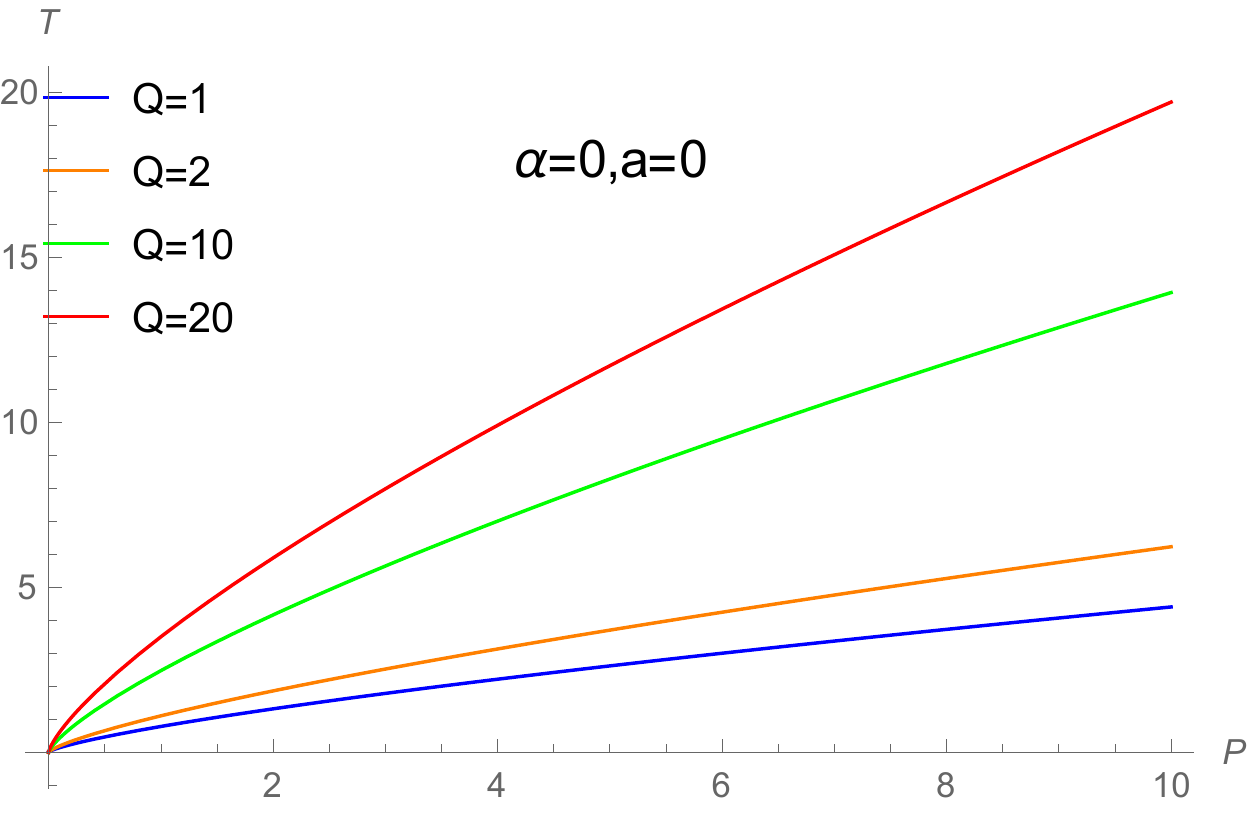}\label{fig:t11}}
\subfigure[{}]{
\includegraphics[width=0.3\textwidth]{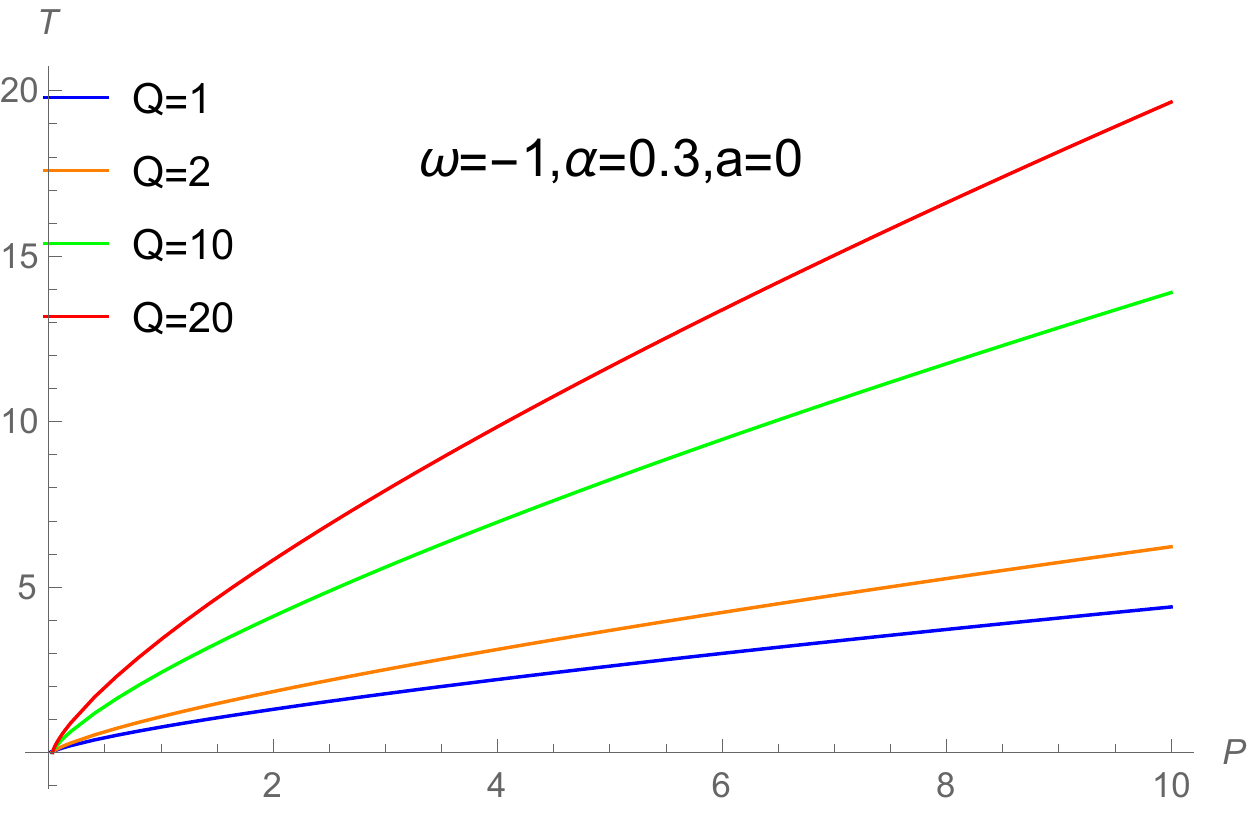}\label{fig:t15}}
\subfigure[{}]{
\includegraphics[width=0.3\textwidth]{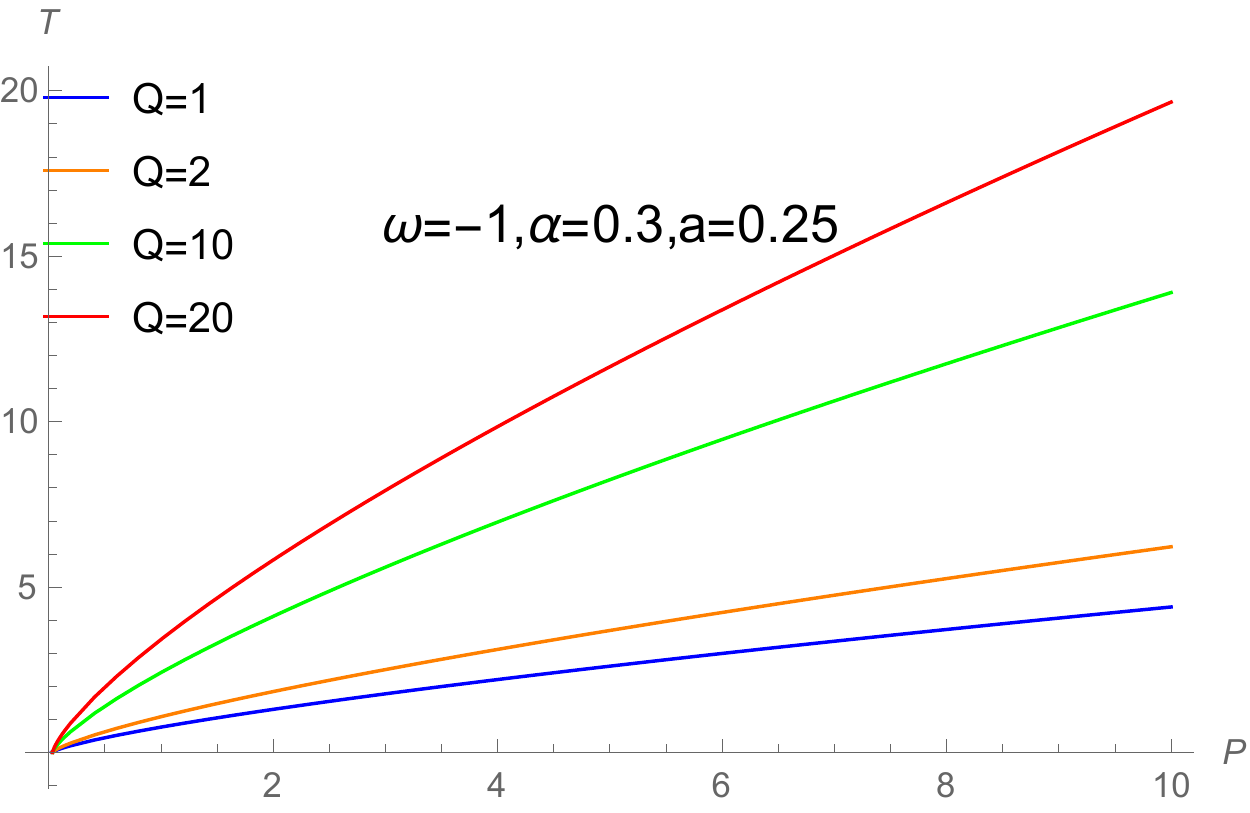}\label{fig:t12}}
\subfigure[{}]{
\includegraphics[width=0.3\textwidth]{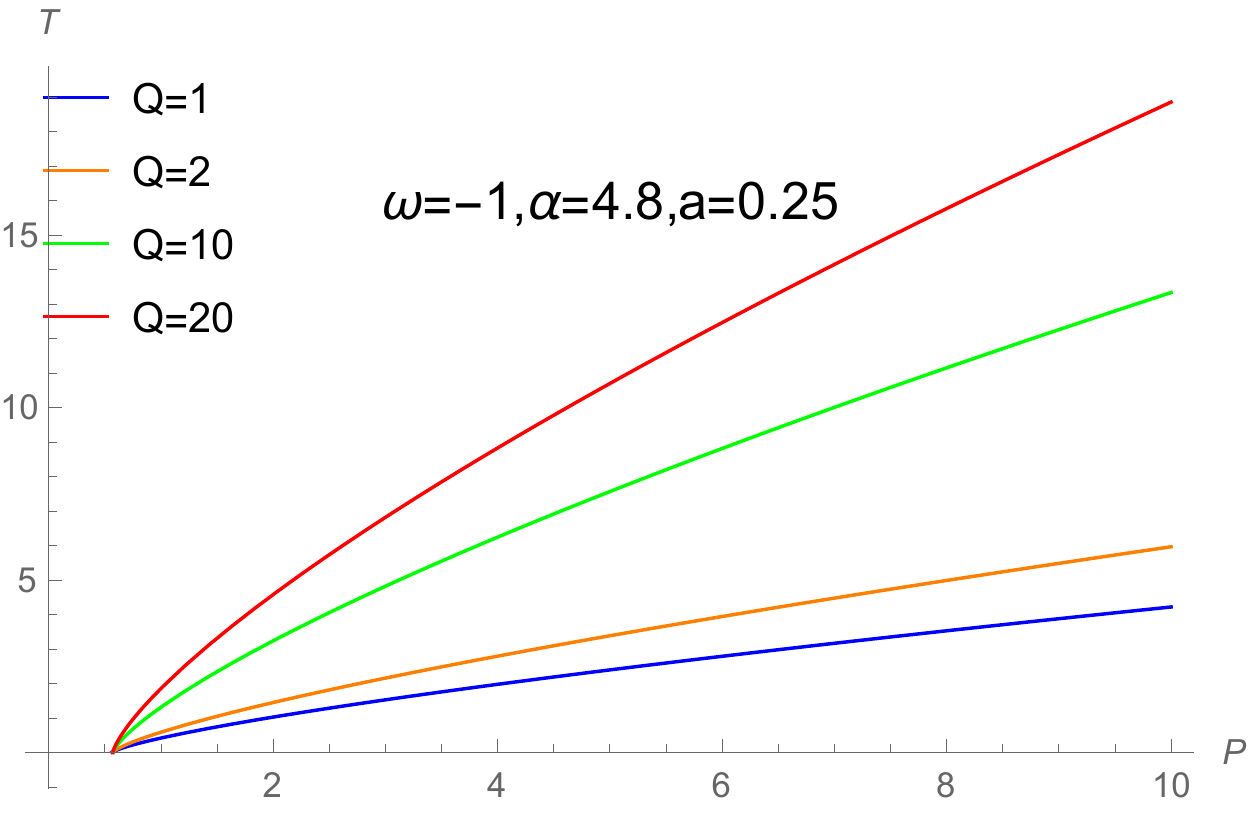}\label{fig:t13}}
\subfigure[{}]{
\includegraphics[width=0.3\textwidth]{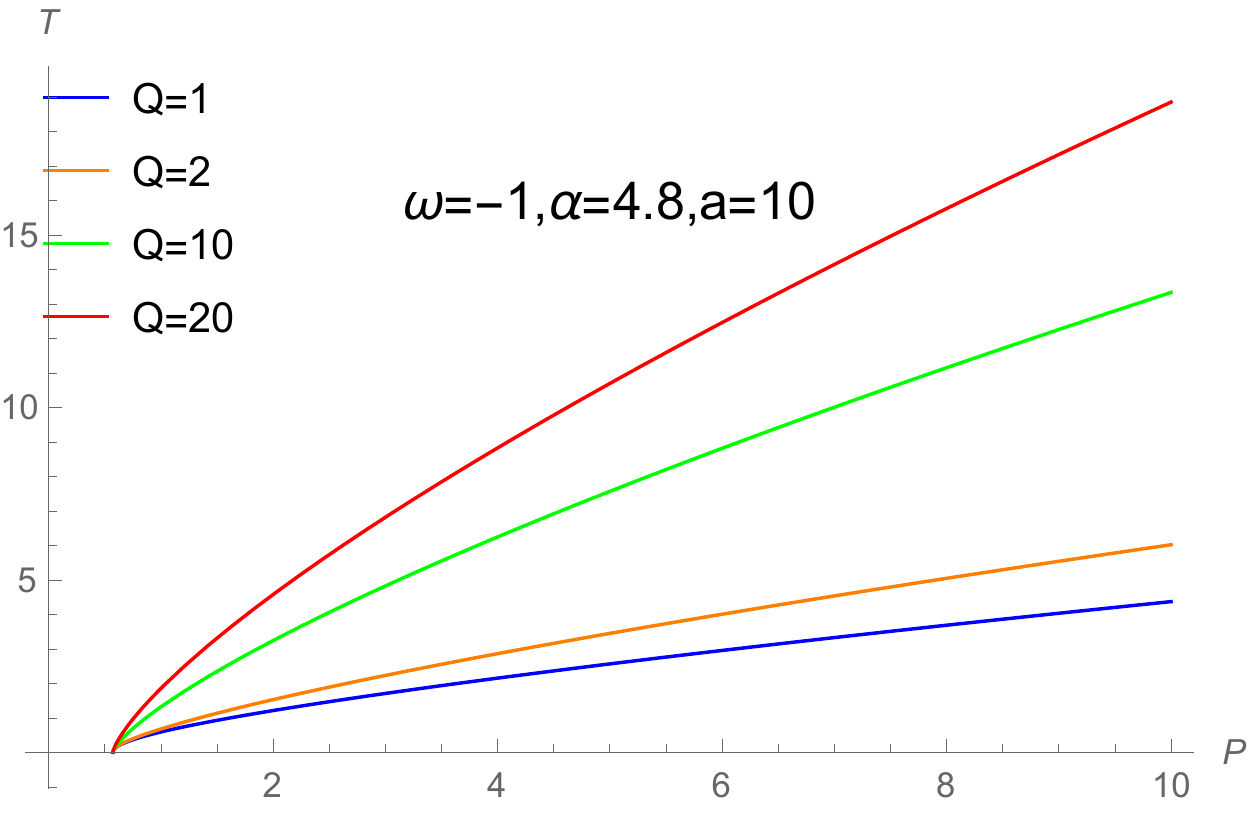}\label{fig:t14}}
\subfigure[{}]{
\includegraphics[width=0.3\textwidth]{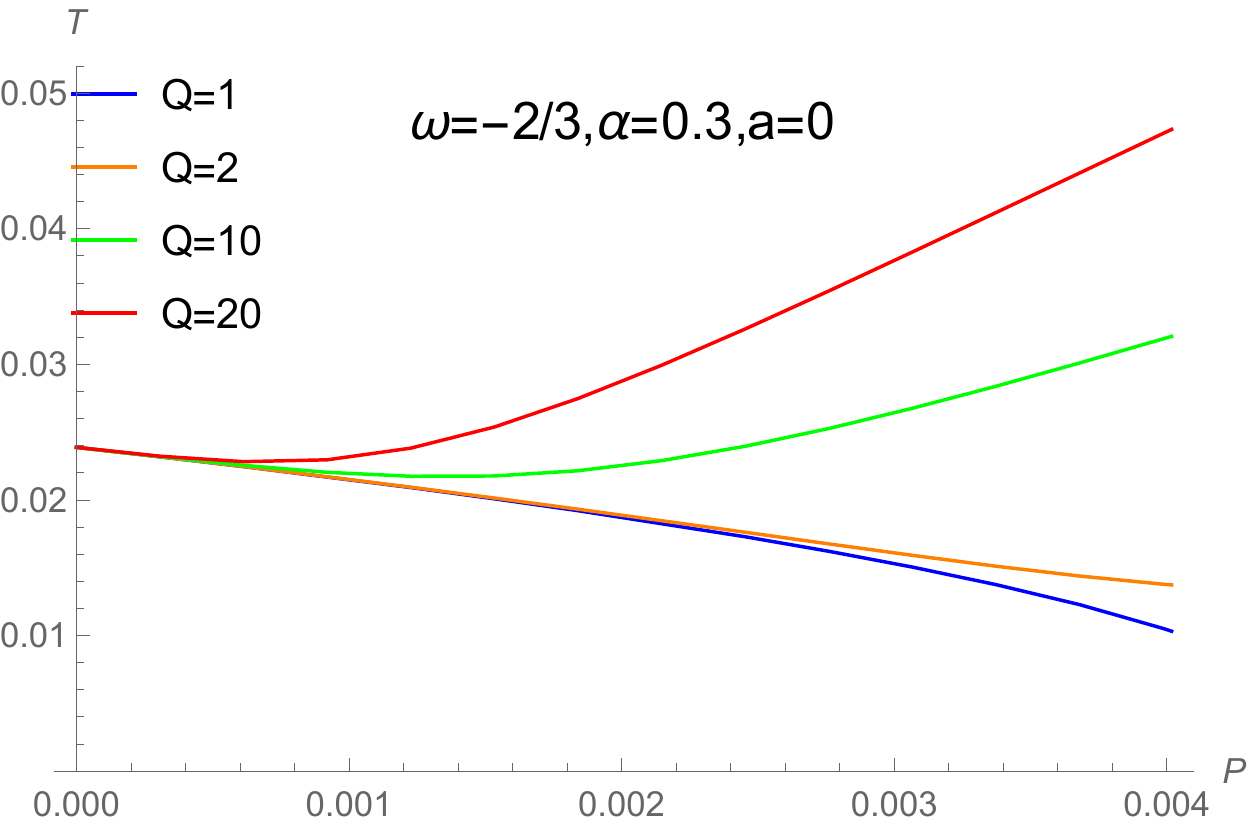}\label{fig:t21}}
\subfigure[{}]{
\includegraphics[width=0.3\textwidth]{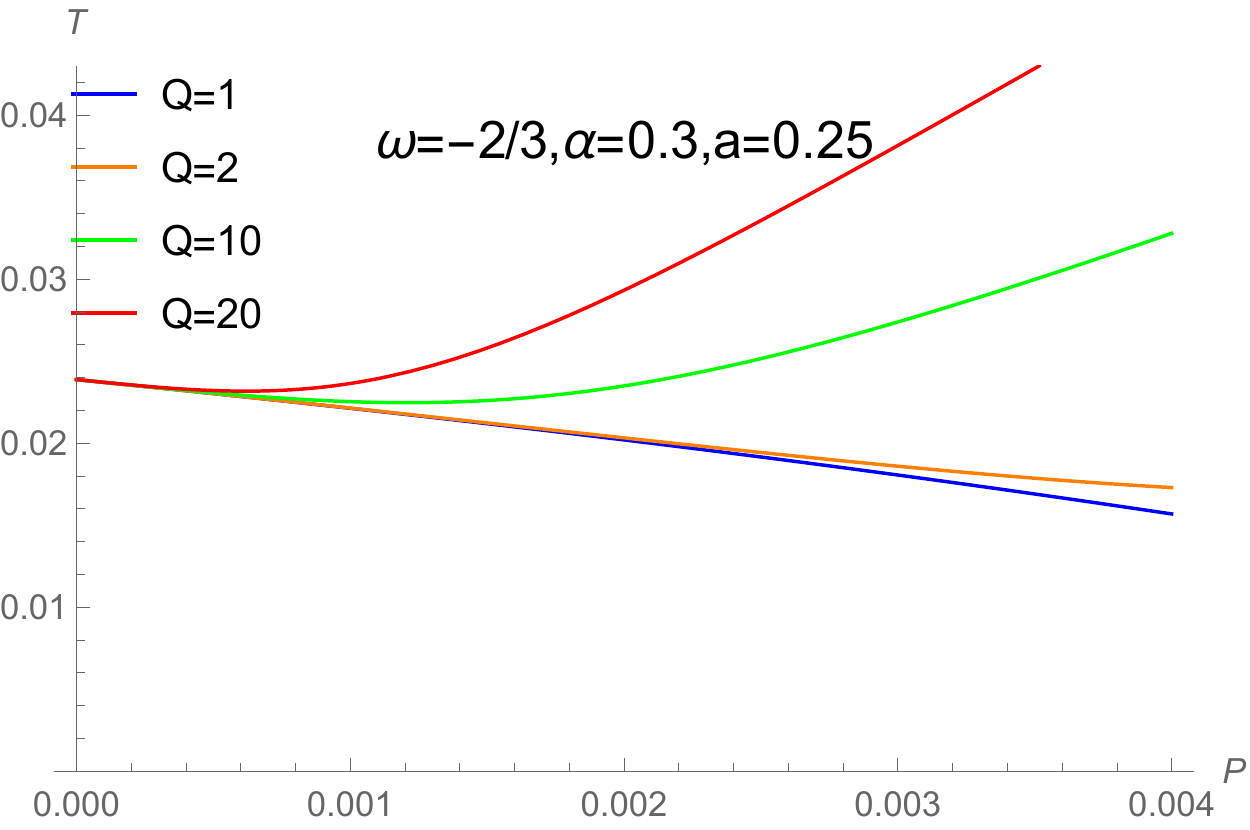}\label{fig:t22}}
\subfigure[{}]{
\includegraphics[width=0.3\textwidth]{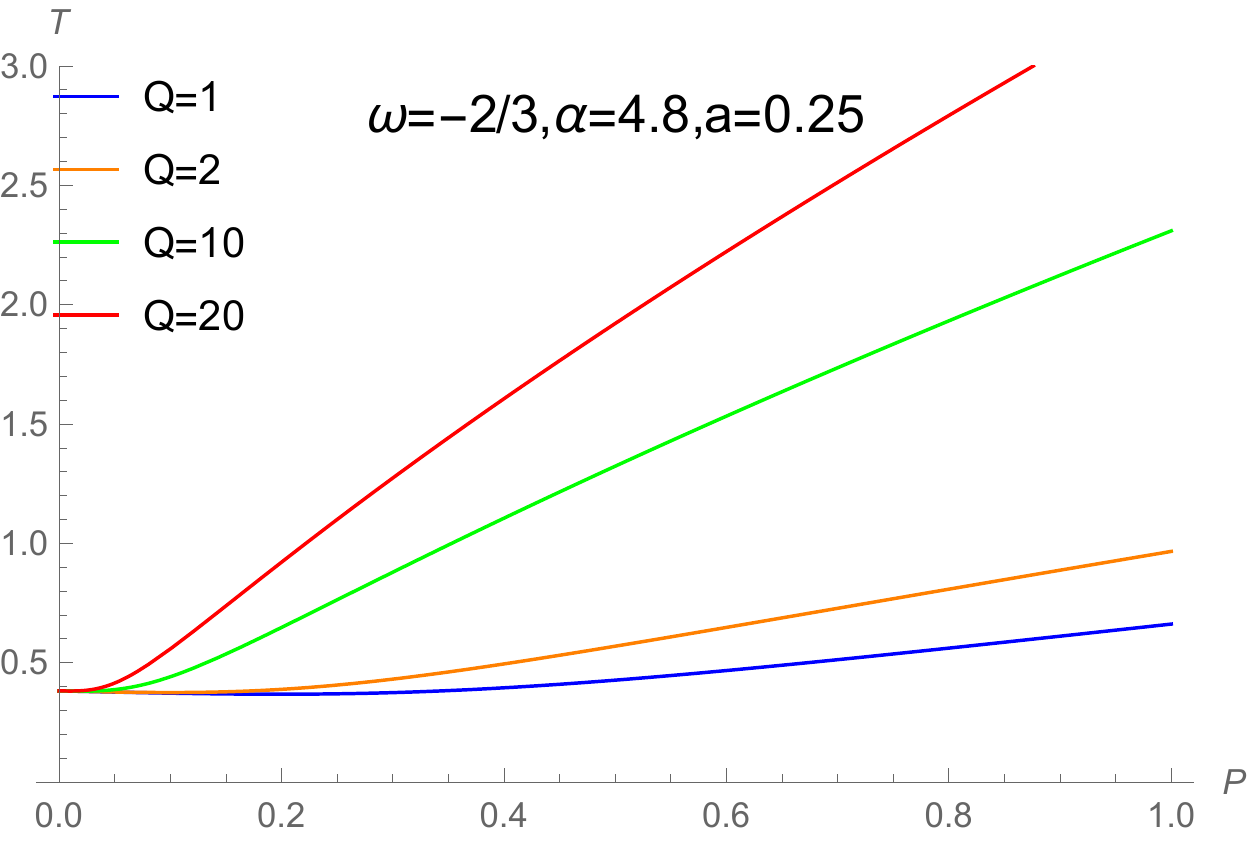}\label{fig:t23}}
\subfigure[{}]{
\includegraphics[width=0.3\textwidth]{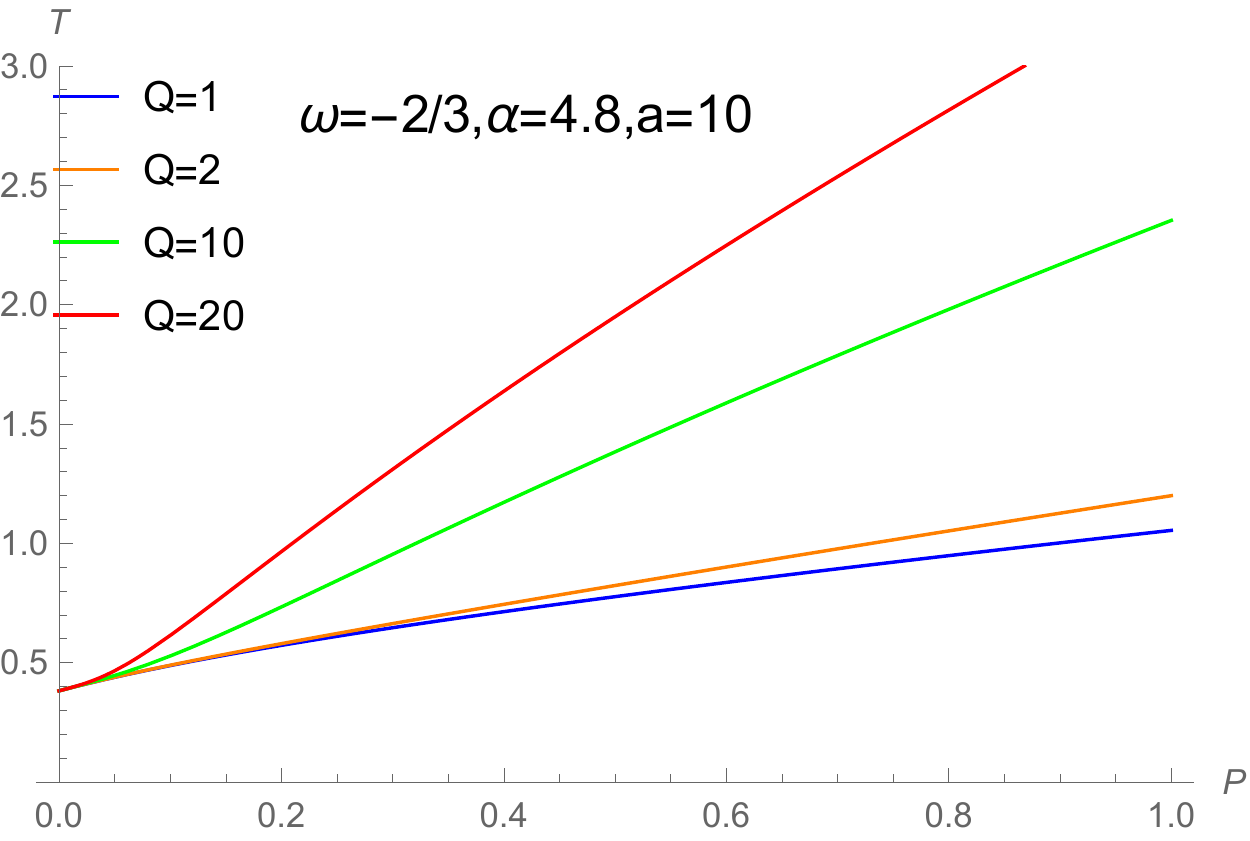}\label{fig:t24}}
\end{center}
\caption{Diagrams of the inversion curves of the RN-AdS black hole with cloud of
strings and quintessence for $Q=1,2,10,20$.}%
\label{fig:T1}
\end{figure}

\subsection{ Reissner-Nordstr\"om-Anti-de Sitter black holes with cloud of
strings and quintessence}
\label{sec:CB}
In this section, we will investigate Joule-Thomson expansion for RN-AdS black holes with cloud of strings and quintessence. As showed in Ref. \cite{Hawking:1982dh}, a large-small black hole phase transition is predicted, i.e., there is a critical temperature where a large stable black hole reaches to thermal gas in the AdS background. The Joule-Thomson expansion of AdS gas at constant enthalpy conditions is related to the instability of small AdS black holes. In general, the large AdS black holes are thermodynamically stable, while small black holes are unstable, and this is where the thermodynamic properties of AdS black holes differ from those of the asymptotically flat or de Sitter spacetime. Expansion is a process of temperature changes with respect to pressure, during which the enthalpy remains constant. Since the black hole mass is determined by the enthalpy in AdS space, the Joule-Thomson coefficient is defined as

\begin{equation}\label{eqn:R1}
  \mu=(\frac{\partial T}{\partial P})_{M}=\frac{1}{C_{P}}[T(\frac{\partial V}{\partial T})_{P}-V].
\end{equation}
Where the heat capacity at constant pressure is
\begin{equation}\label{eqn:R2}
\begin{aligned}
 &C_{P}=T(\frac{\partial S}{\partial T})_{P}\\
 &=\frac{2\pi\left(\pi\left(r_{+}^{2}\right){}^{\frac{3\omega}{2}+1}\left(r_{+}^{2}\left(-a+8\pi Pr_{+}^{2}+1\right)-Q^{2}\right)+3\pi\alpha\left(r_{+}^{2}\right){}^{3/2}\omega\right)}{\left(r_{+}^{2}\right){}^{\frac{3\omega}{2}}\left(\pi r_{+}^{2}\left(a+8\pi Pr_{+}^{2}-1\right)+3\pi Q^{2}\right)-3\pi\alpha\sqrt{r_{+}^{2}}\omega(3\omega+2)},\\
\end{aligned}
\end{equation}
and one can derive
\begin{equation}\label{eqn:R3}
\begin{aligned}
 &\mu=\\
 &\{\frac{2r_{+}^{3}\left(-4a\left(r_{+}^{3}\right){}^{\omega+1}+16\pi P\left(r_{+}^{3}\right){}^{\omega+\frac{5}{3}}-6Q^{2}\left(r_{+}^{3}\right){}^{\omega+\frac{1}{3}}\right)}{3\left(a\left(r_{+}^{3}\right){}^{\omega+1}+8\pi P\left(r_{+}^{3}\right){}^{\omega+\frac{5}{3}}+3Q^{2}\left(r_{+}^{3}\right){}^{\omega+\frac{1}{3}}-9\alpha\left(r_{+}^{3}\right){}^{2/3}\omega^{2}-6\alpha\left(r_{+}^{3}\right){}^{2/3}\omega-\left(r_{+}^{3}\right){}^{\omega+1}\right)}\\&+\frac{2r_{+}^{3}\left(9\alpha\left(r_{+}^{3}\right){}^{2/3}\omega^{2}+15\alpha\left(r_{+}^{3}\right){}^{2/3}\omega+4\left(r_{+}^{3}\right){}^{\omega+1}\right)}{3\left(a\left(r_{+}^{3}\right){}^{\omega+1}+8\pi P\left(r_{+}^{3}\right){}^{\omega+\frac{5}{3}}+3Q^{2}\left(r_{+}^{3}\right){}^{\omega+\frac{1}{3}}-9\alpha\left(r_{+}^{3}\right){}^{2/3}\omega^{2}-6\alpha\left(r_{+}^{3}\right){}^{2/3}\omega-\left(r_{+}^{3}\right){}^{\omega+1}\right)}\}\\
 &\times\{\frac{\sqrt[3]{r_{+}^{3}}\left(a(4\pi)^{\omega}\left(r_{+}^{3}\right){}^{\omega+\frac{1}{3}}-9\alpha\omega^{2}(4\pi)^{\omega}-3\alpha2^{2\omega+1}\pi^{\omega}\omega+P2^{2\omega+3}\pi^{\omega+1}\left(r_{+}^{3}\right){}^{\omega+1}-(4\pi)^{\omega}\left(r_{+}^{3}\right){}^{\omega+\frac{1}{3}}\right)}{\left((4\pi)^{\omega}\left(-a\left(r_{+}^{3}\right){}^{\omega+\frac{4}{3}}-Q^{2}\left(r_{+}^{3}\right){}^{\omega+\frac{2}{3}}+3\alpha r_{+}^{3}\omega+\left(r_{+}^{3}\right){}^{\omega+\frac{4}{3}}\right)+Pr_{+}^{6}2^{2\omega+3}\pi^{\omega+1}\left(r_{+}^{3}\right){}^{\omega}\right)}\\&+\frac{3Q^{2}(4\pi)^{\omega}\left(r_{+}^{3}\right){}^{\omega}}{\left((4\pi)^{\omega}\left(-a\left(r_{+}^{3}\right){}^{\omega+\frac{4}{3}}-Q^{2}\left(r_{+}^{3}\right){}^{\omega+\frac{2}{3}}+3\alpha r_{+}^{3}\omega+\left(r_{+}^{3}\right){}^{\omega+\frac{4}{3}}\right)+Pr_{+}^{6}2^{2\omega+3}\pi^{\omega+1}\left(r_{+}^{3}\right){}^{\omega}\right)}\}.\\
\end{aligned}
\end{equation}
\begin{figure}
\begin{center}
\subfigure[{$Q=1$, $M=1.5,2,2.5,3$.}]{
\includegraphics[width=0.35\textwidth]{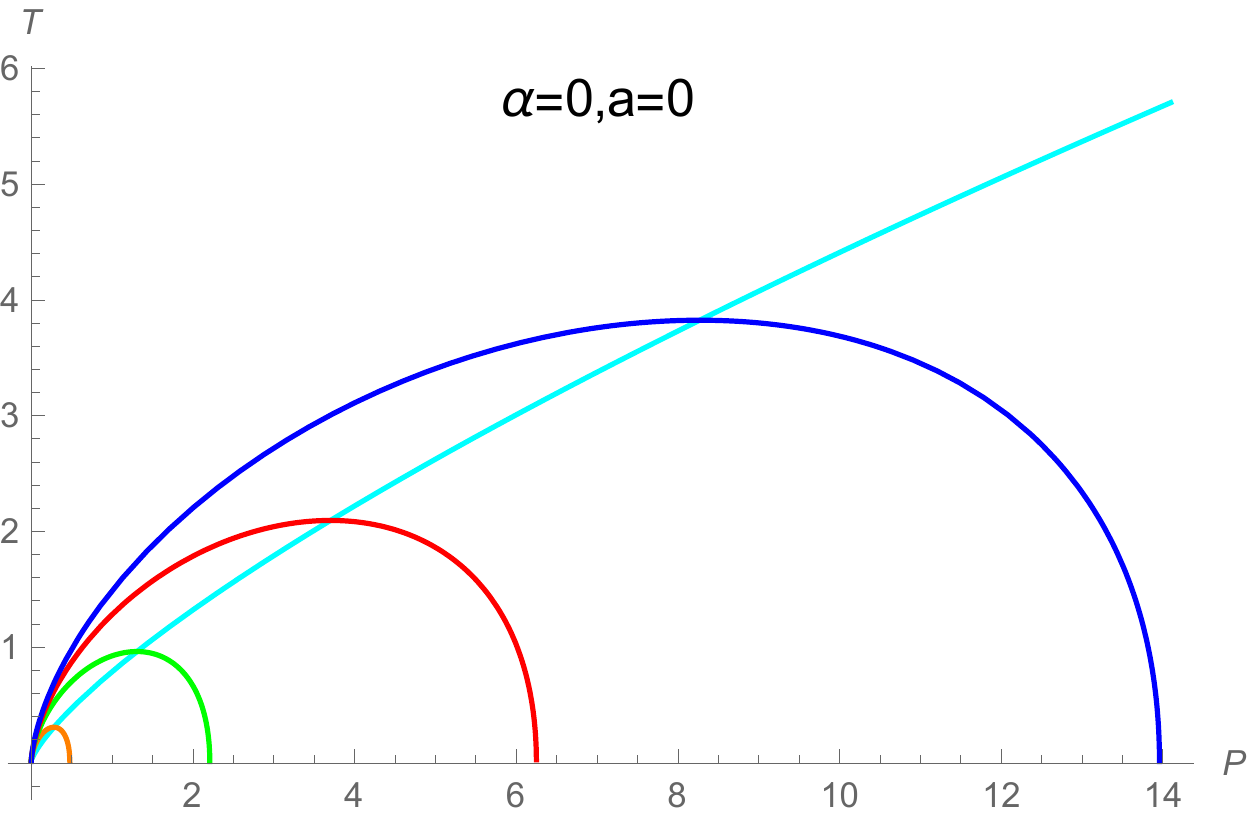}\label{fig:TP01}}
\subfigure[{$Q=2$, $M=2.5,3,3.5,4$.}]{
\includegraphics[width=0.35\textwidth]{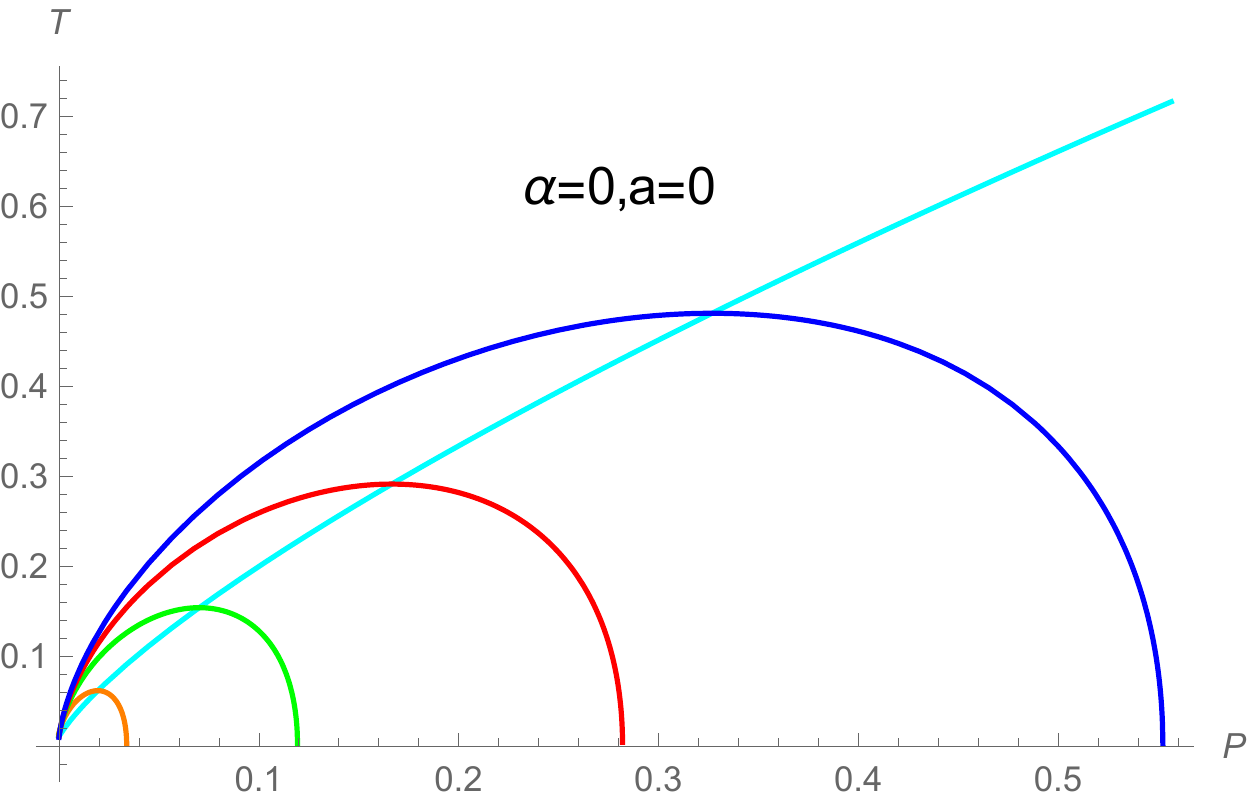}\label{fig:TP02}}
\subfigure[{$Q=10$, $M=10.5,11,11.5,12$.}]{
\includegraphics[width=0.35\textwidth]{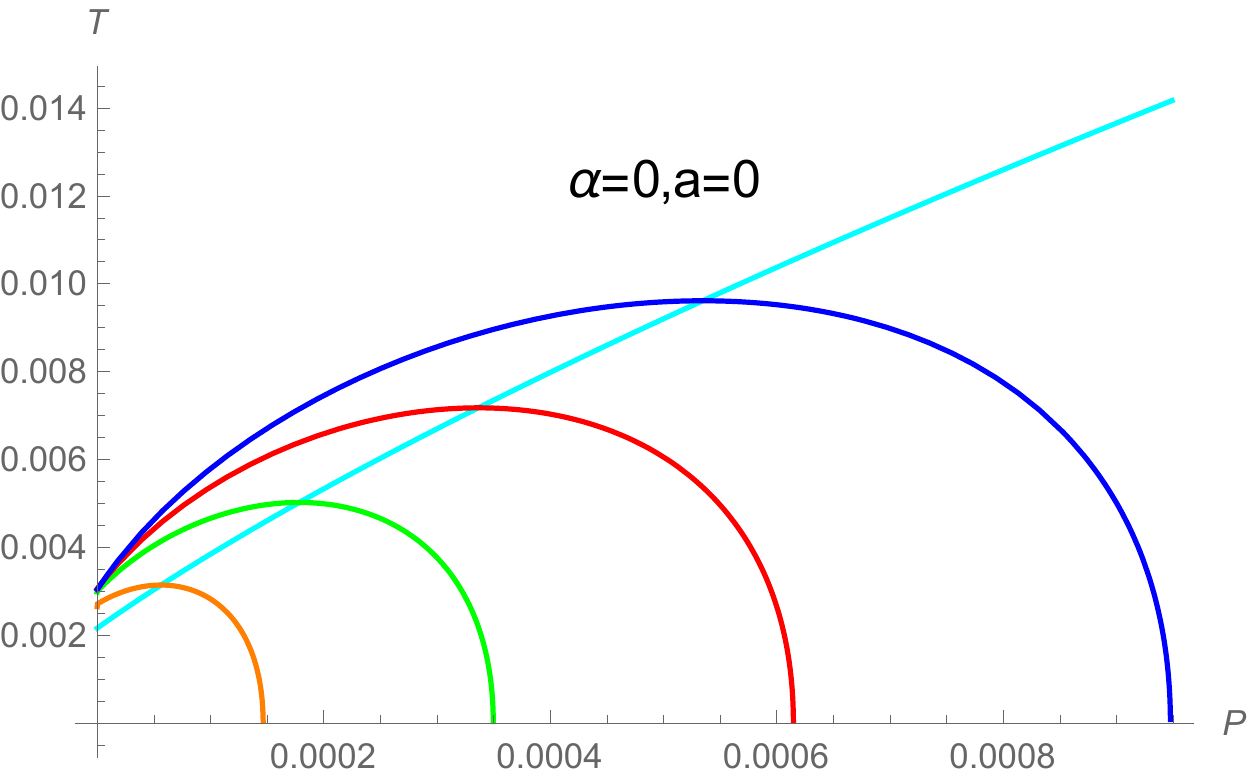}\label{fig:TP03}}
\subfigure[{$Q=20$, $M=20.5,21,21.5,22$.}]{
\includegraphics[width=0.35\textwidth]{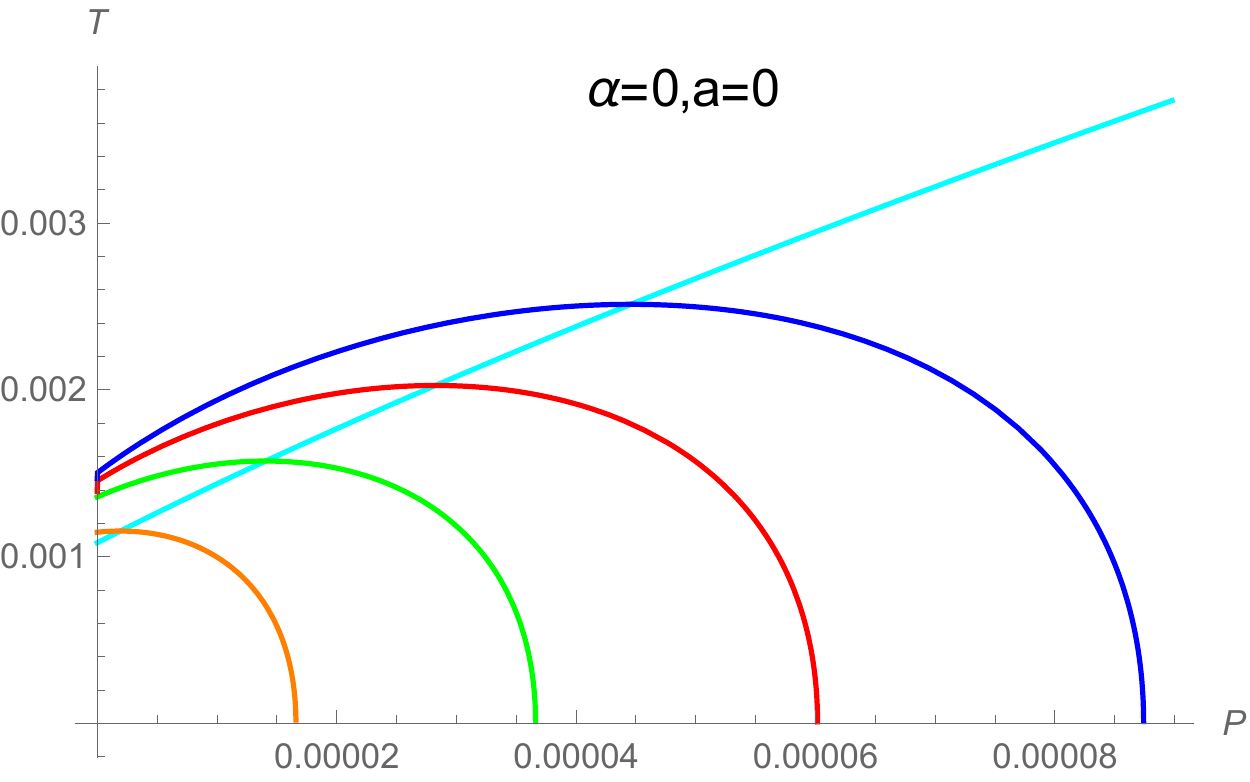}\label{fig:TP04}}
\end{center}
\caption{Diagrams of the inversion curves and the isoenthalpic curves of the RN-AdS black hole with cloud of strings and
quintessence. From bottom to top, these curves correspond to the increase in $M$.}%
\label{fig:J00}
\end{figure}

\begin{figure}
\begin{center}
\subfigure[{$Q=1$, $M=1.5,2,2.5,3$.}]{
\includegraphics[width=0.35\textwidth]{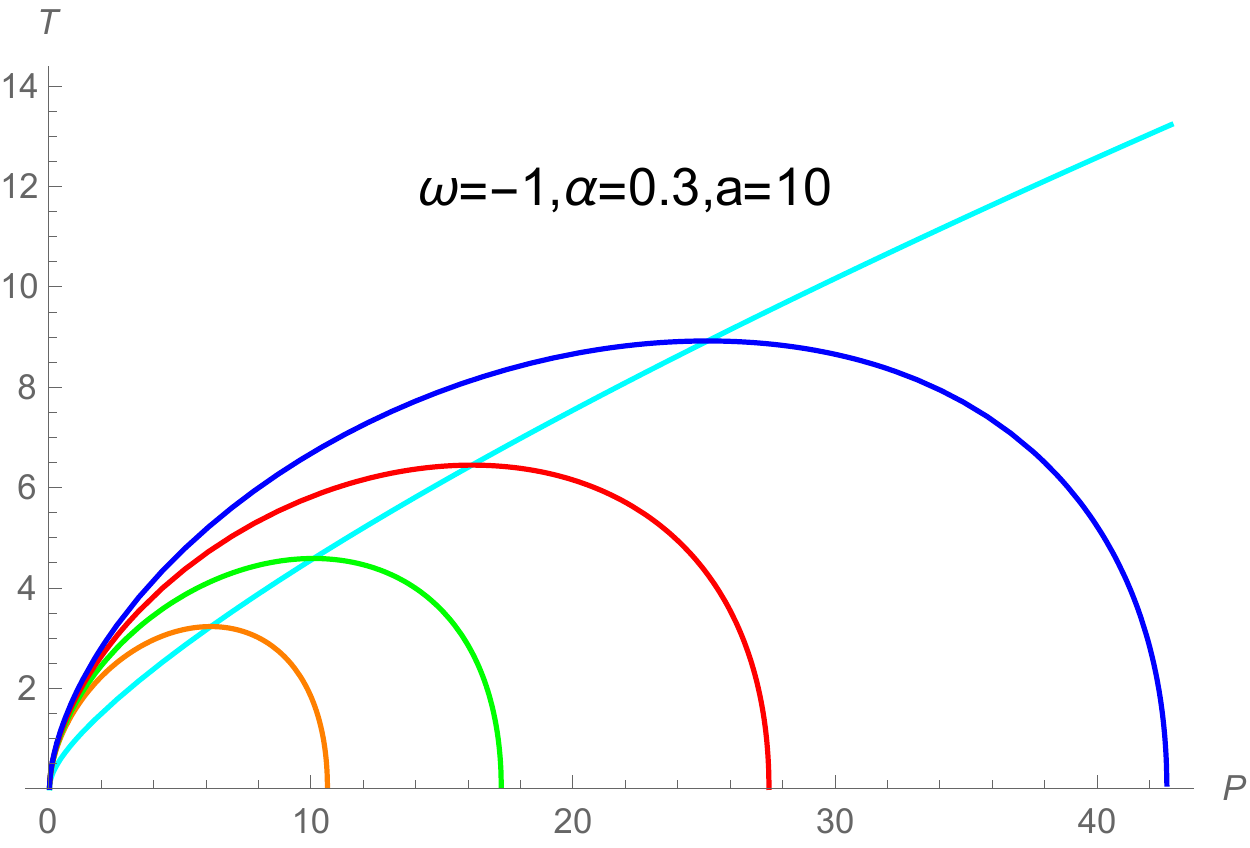}\label{fig:TP11}}
\subfigure[{$Q=2$, $M=2.5,3,3.5,4$.}]{
\includegraphics[width=0.35\textwidth]{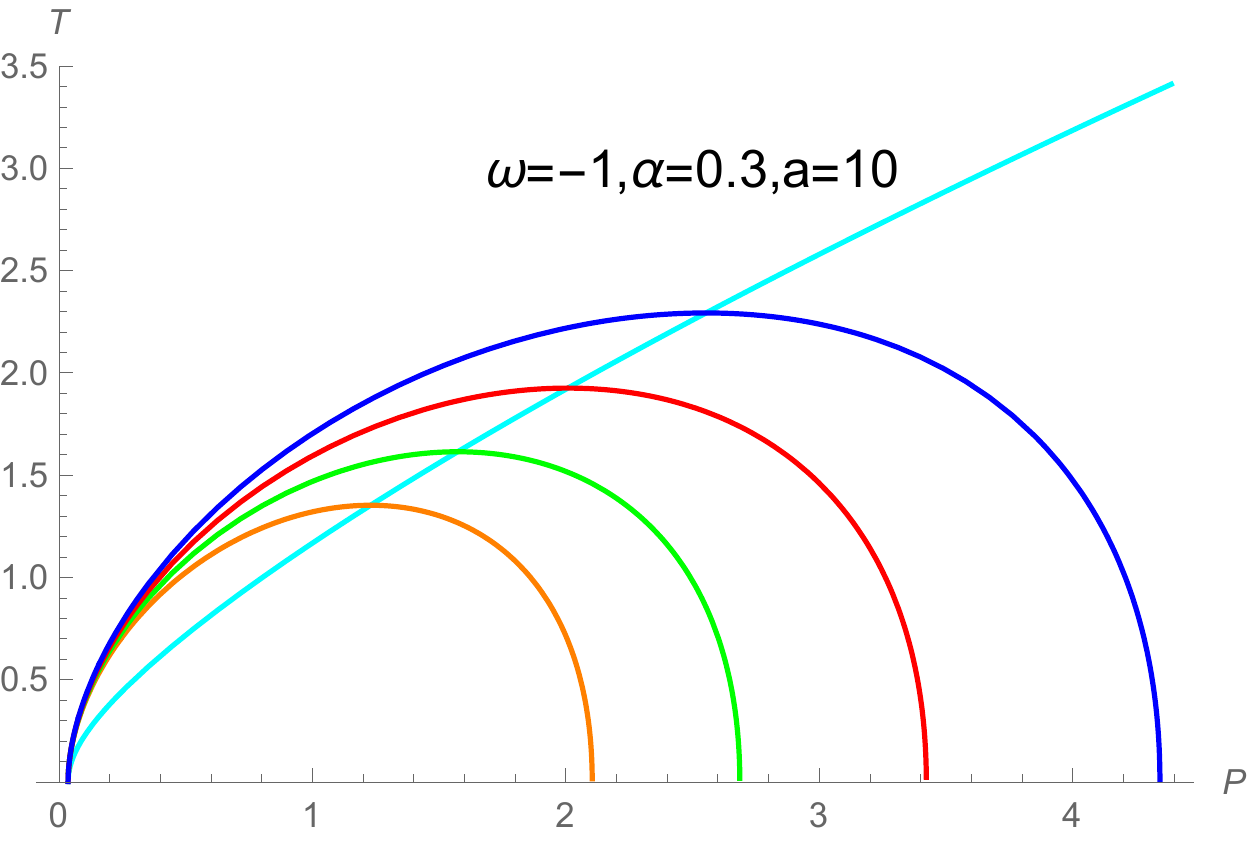}\label{fig:TP12}}
\subfigure[{$Q=1$, $M=1.5,2,2.5,3$.}]{
\includegraphics[width=0.35\textwidth]{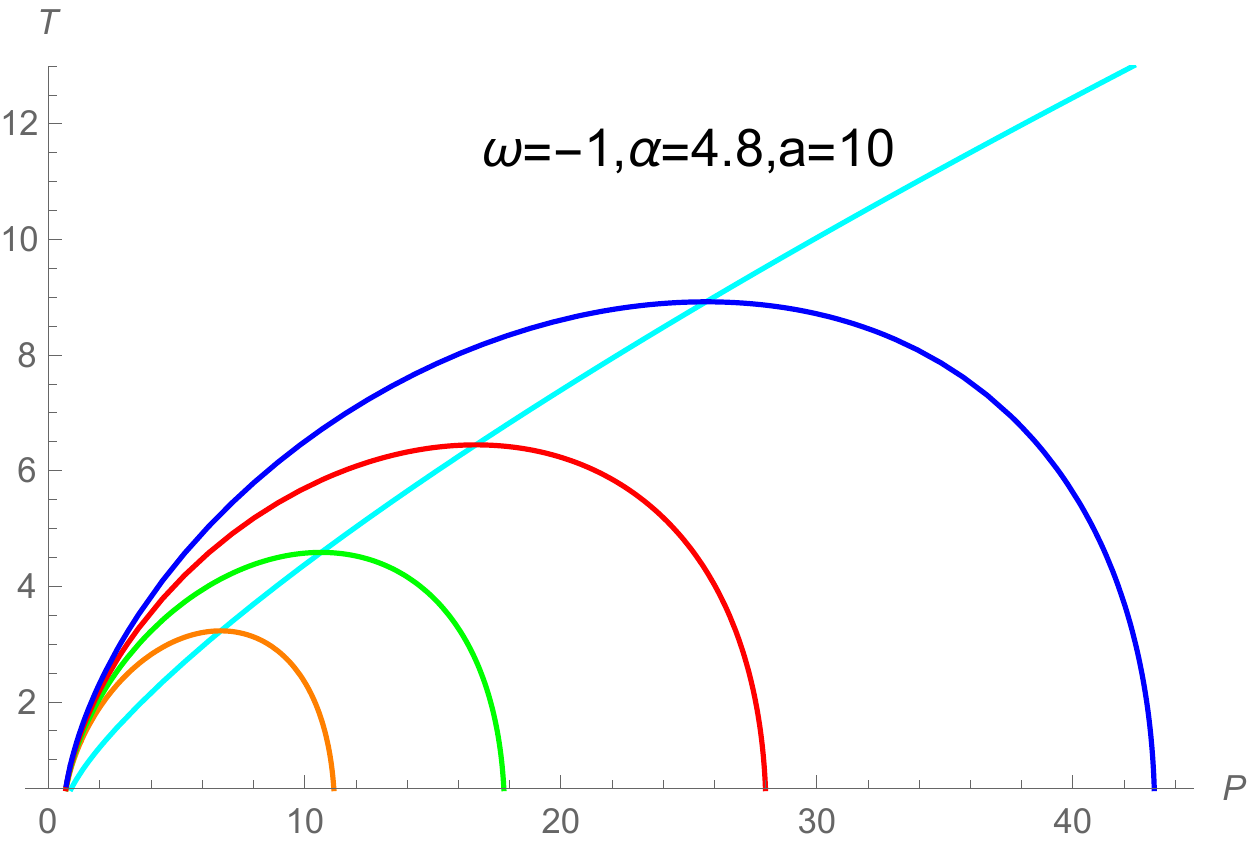}\label{fig:TP13}}
\subfigure[{$Q=2$, $M=2.5,3,3.5,4$.}]{
\includegraphics[width=0.35\textwidth]{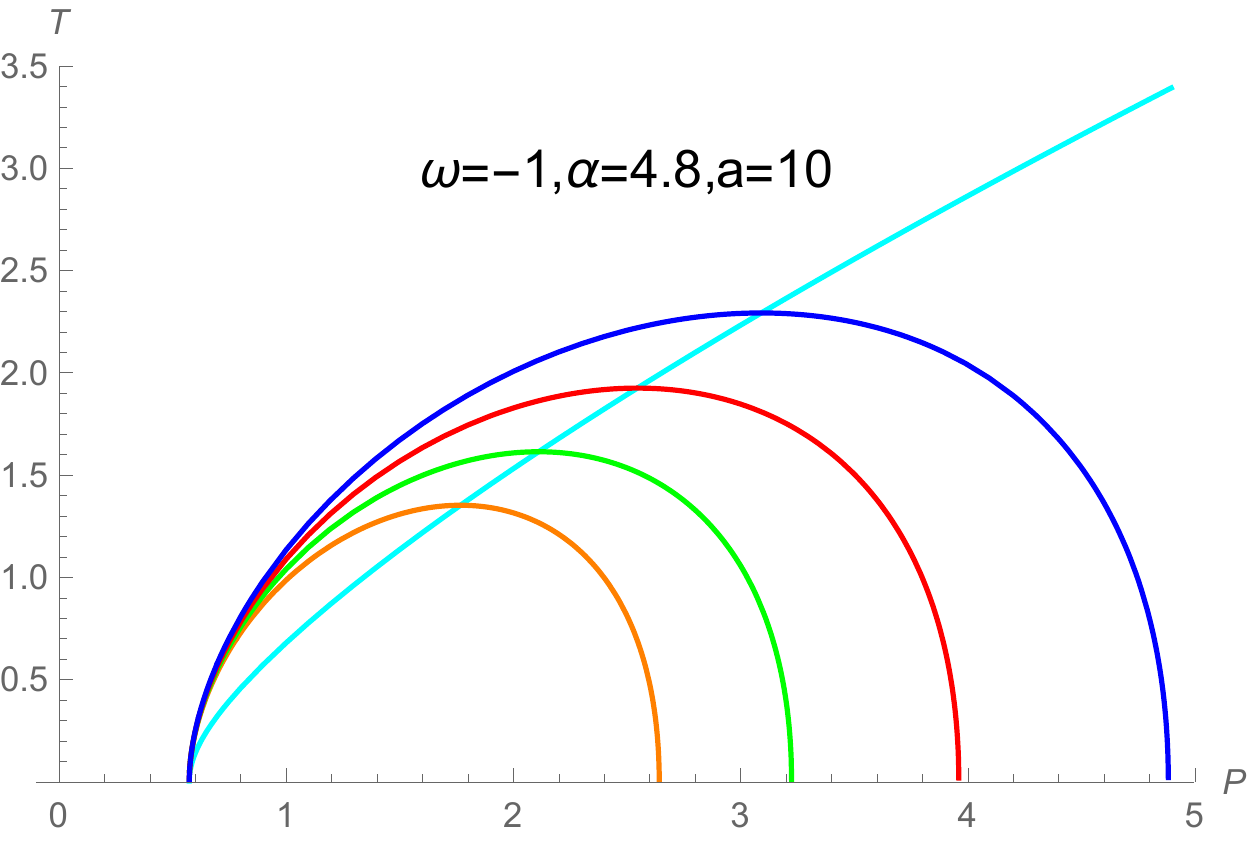}\label{fig:TP14}}
\subfigure[{$Q=1$, $M=1.5,2,2.5,3$.}]{
\includegraphics[width=0.35\textwidth]{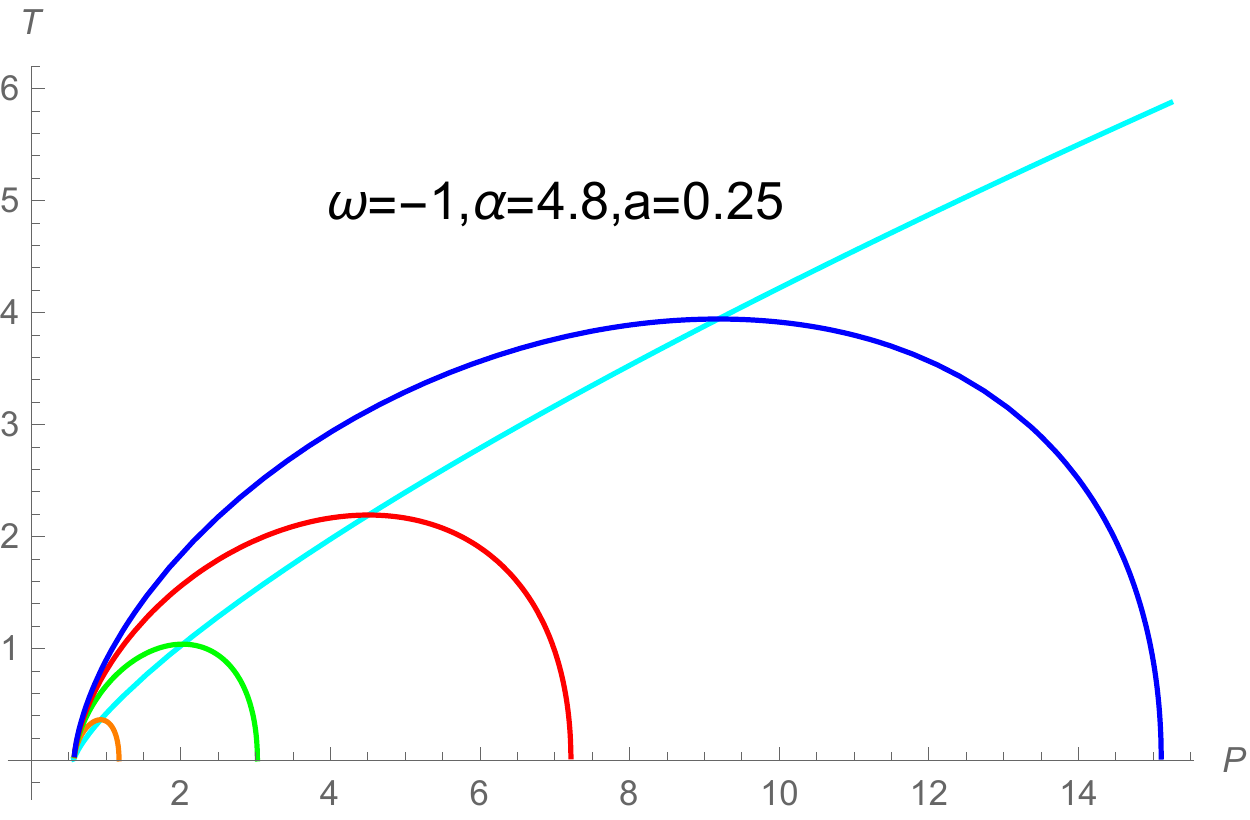}\label{fig:TP15}}
\subfigure[{$Q=2$, $M=2.5,3,3.5,4$.}]{
\includegraphics[width=0.35\textwidth]{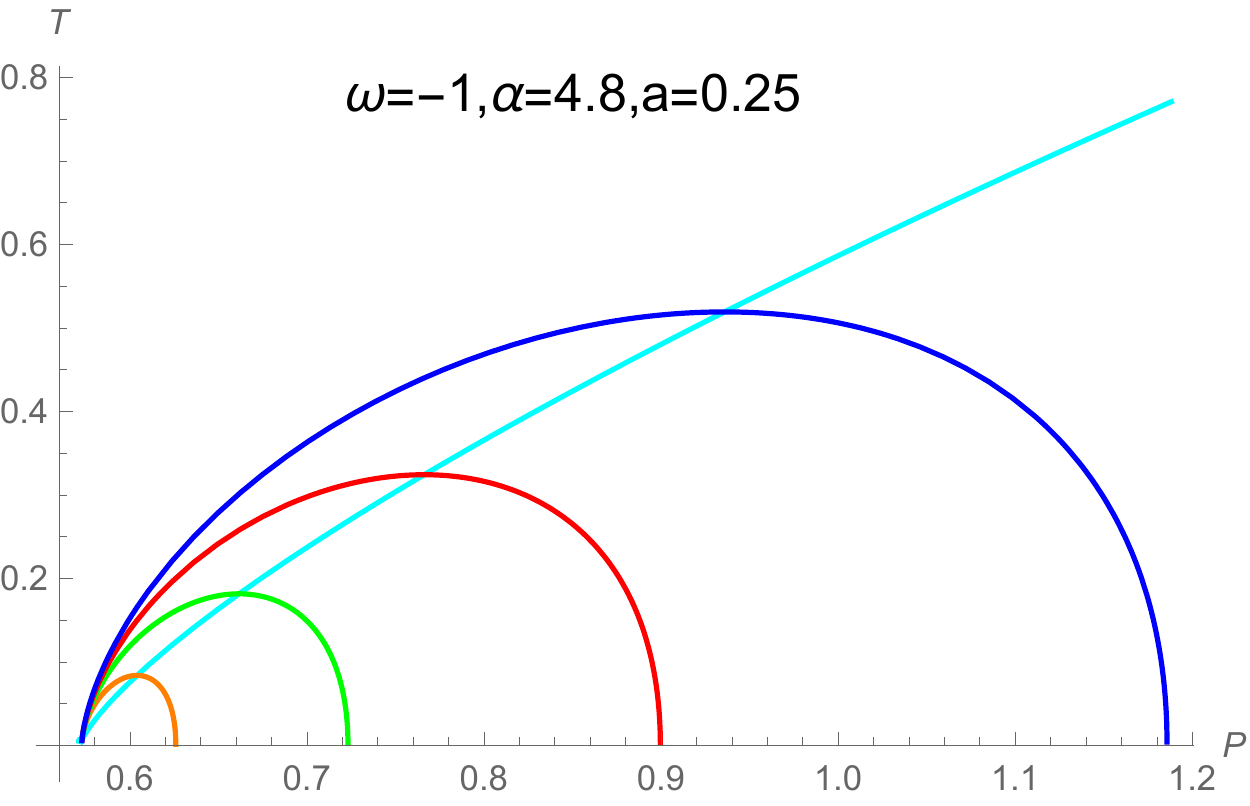}\label{fig:TP16}}
\end{center}
\caption{Diagrams of the inversion curves and the isoenthalpic curves of the RN-AdS black hole with cloud of
strings and quintessence for $\omega=-1$. From bottom to top, these curves correspond to the increase in $M$.}%
\label{fig:J11}
\end{figure}

\begin{figure}
\begin{center}
\subfigure[{$Q=10$, $M=10.5,11,11.5,12$.}]{
\includegraphics[width=0.35\textwidth]{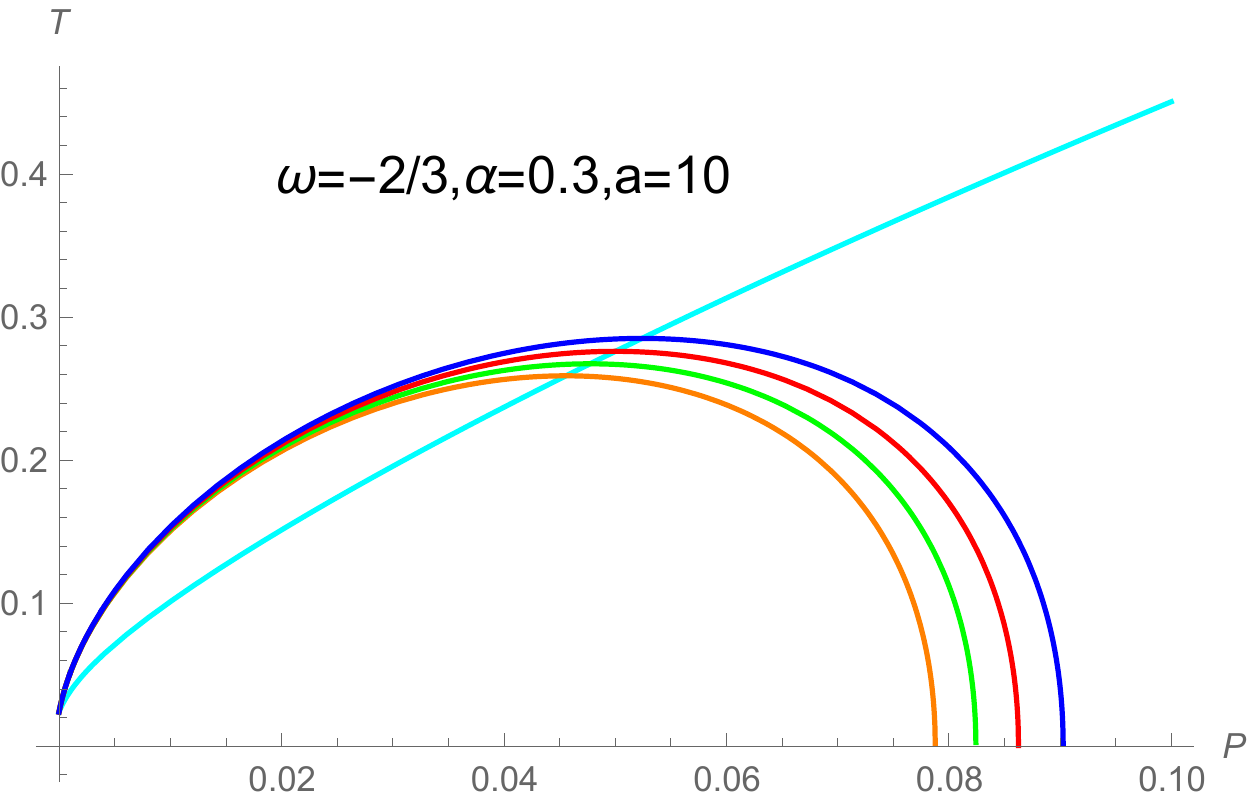}\label{fig:TP21}}
\subfigure[{$Q=20$, $M=20.5,21,21.5,22$.}]{
\includegraphics[width=0.35\textwidth]{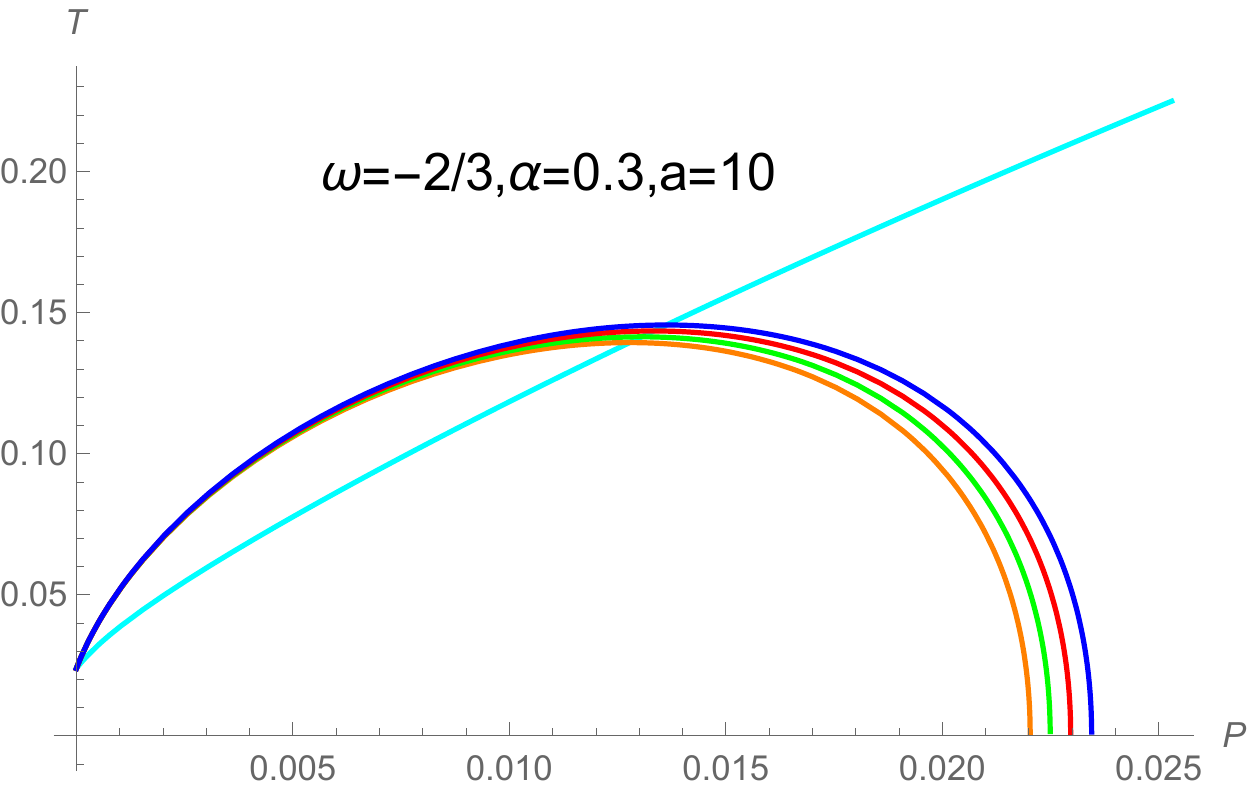}\label{fig:TP22}}
\subfigure[{$Q=10$, $M=10.5,11,11.5,12$.}]{
\includegraphics[width=0.35\textwidth]{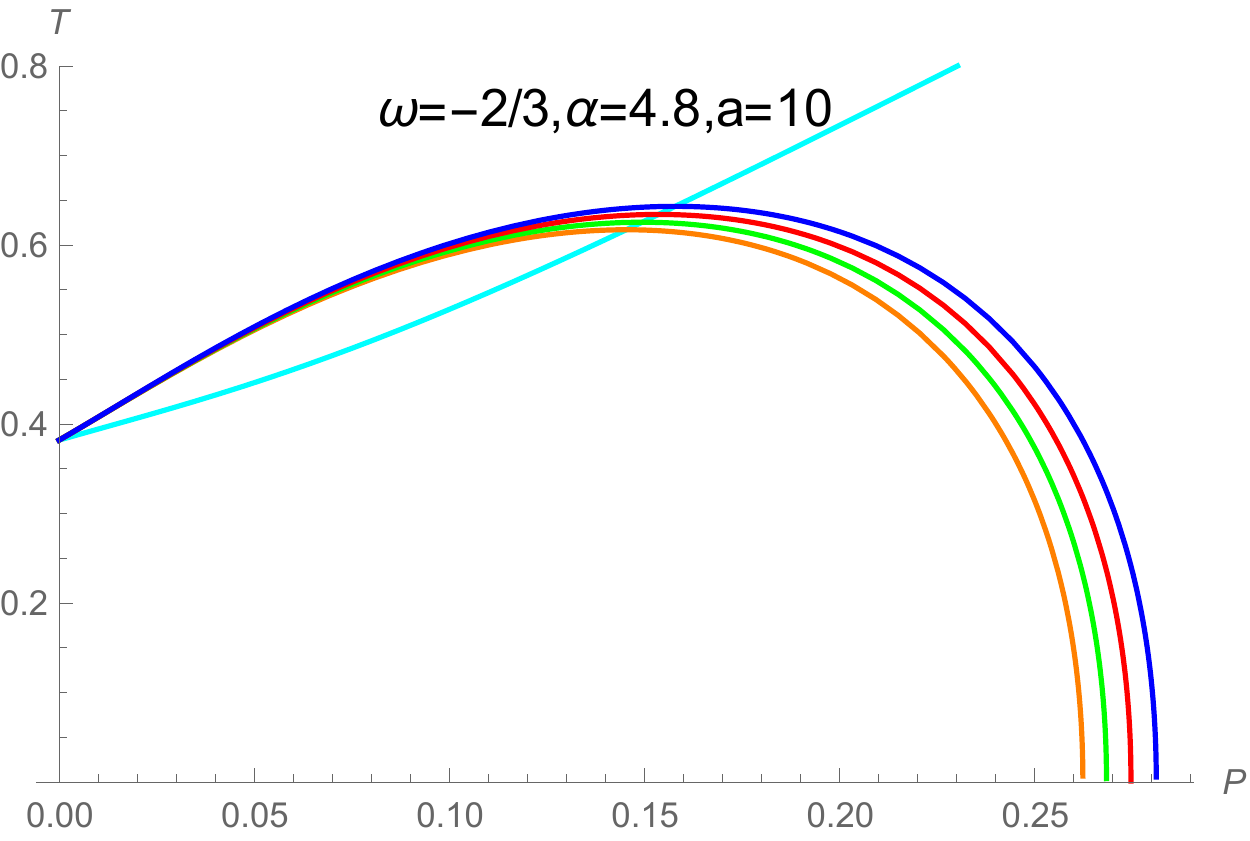}\label{fig:TP23}}
\subfigure[{$Q=20$, $M=20.5,21,21.5,22$.}]{
\includegraphics[width=0.35\textwidth]{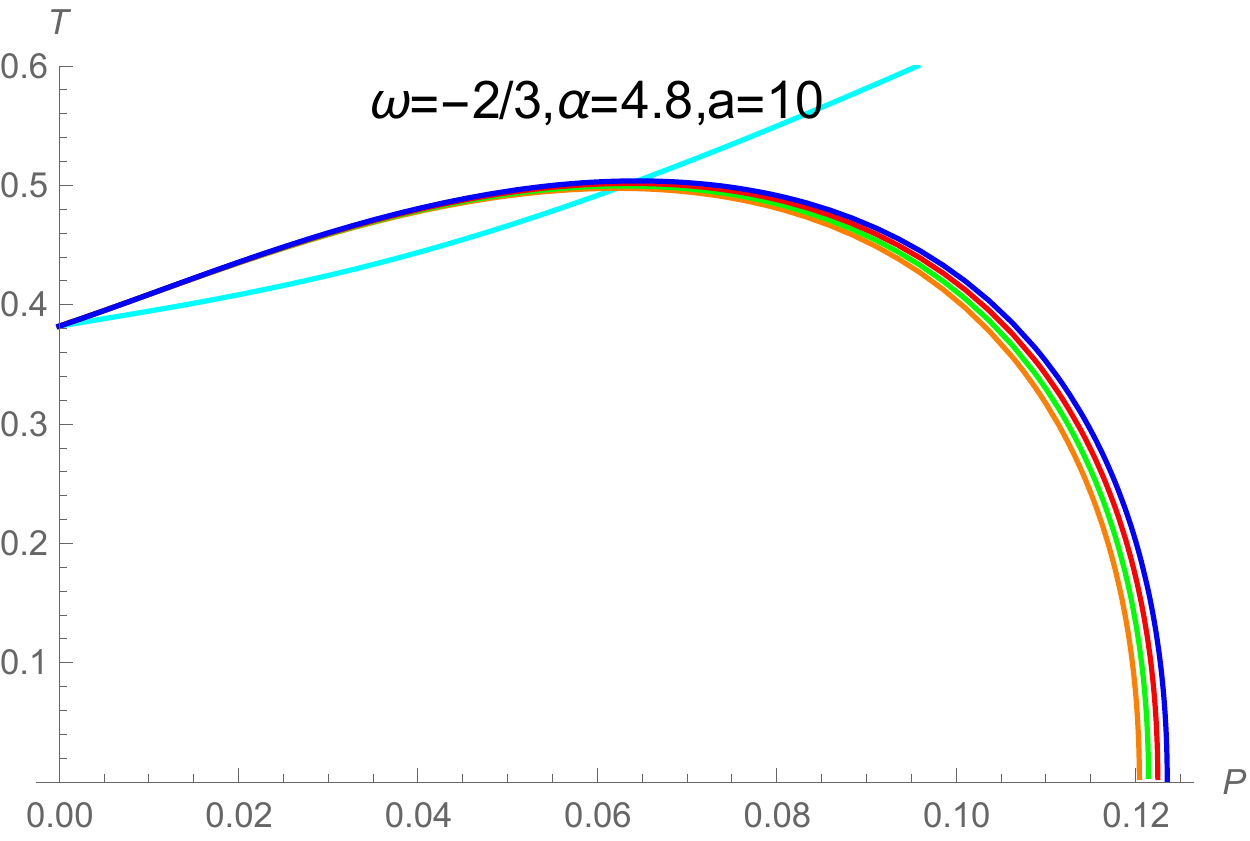}\label{fig:TP24}}
\subfigure[{$Q=10$, $M=10.5,11,11.5,12$.}]{
\includegraphics[width=0.35\textwidth]{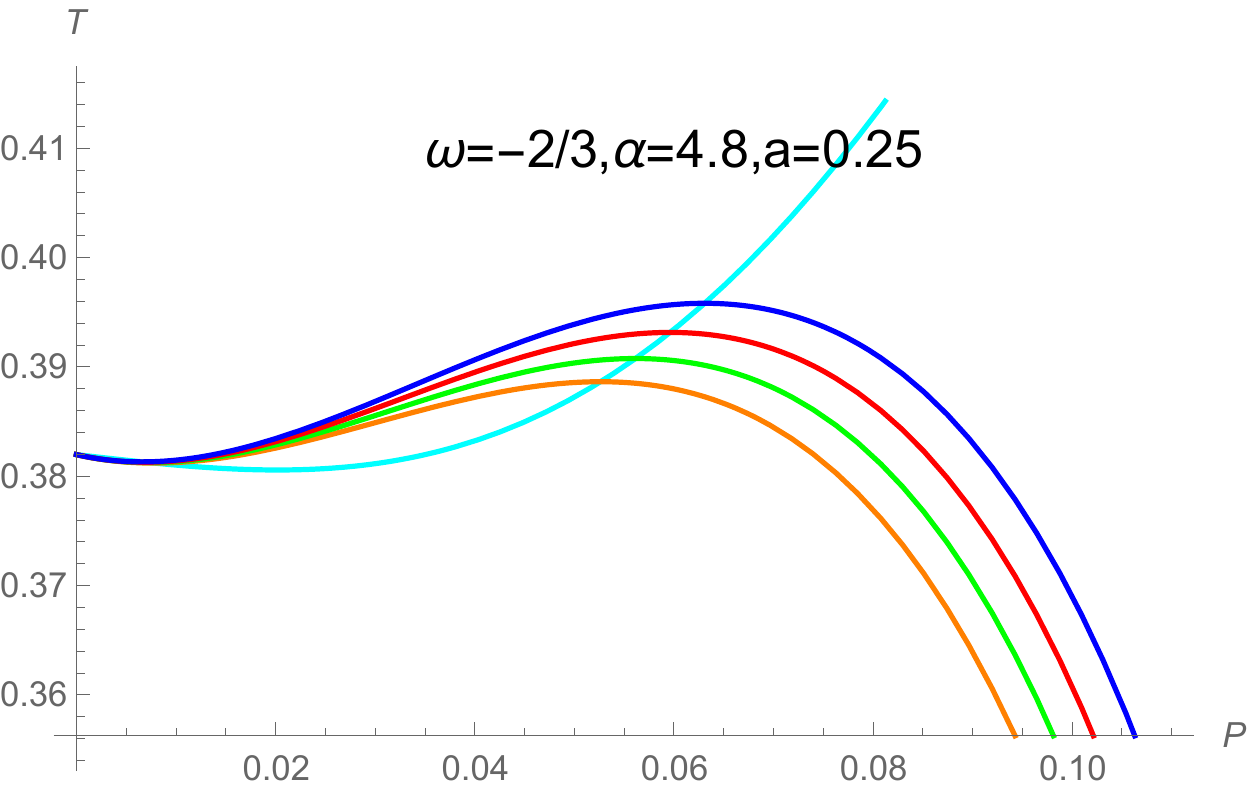}\label{fig:TP25}}
\subfigure[{$Q=20$, $M=20.5,21,21.5,22$.}]{
\includegraphics[width=0.35\textwidth]{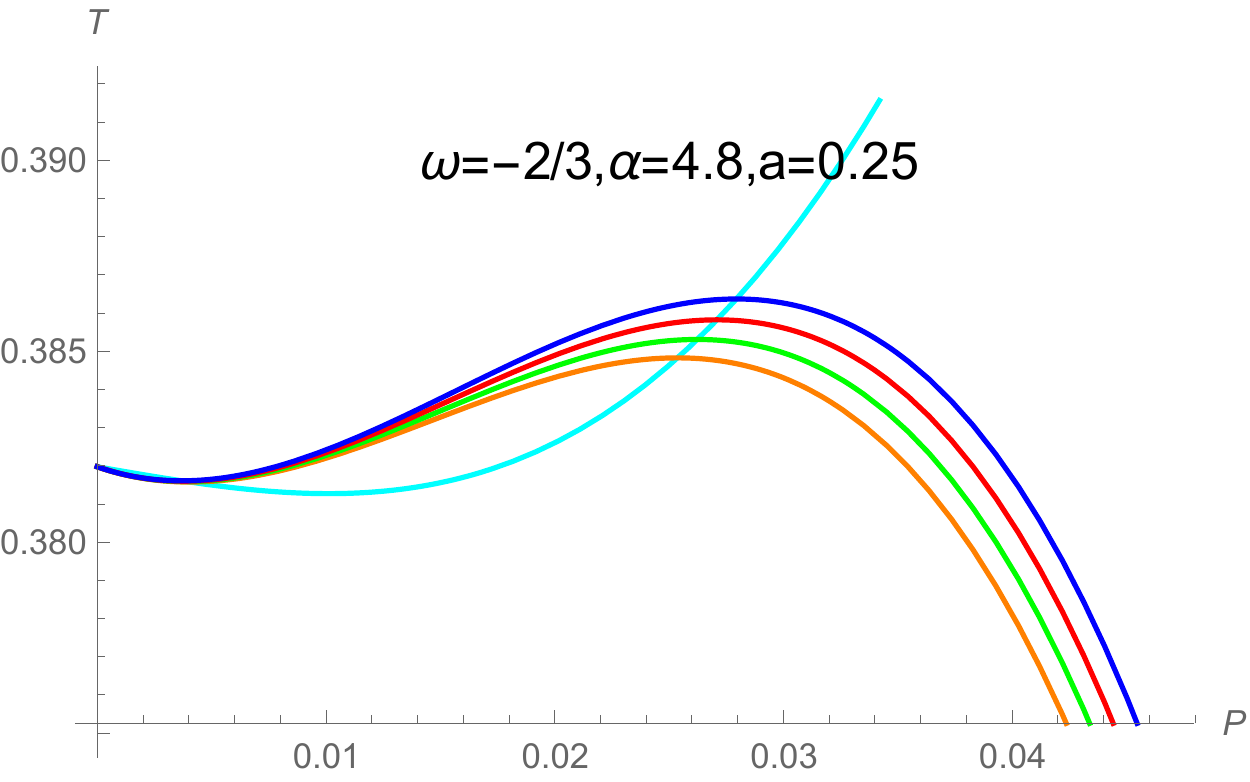}\label{fig:TP26}}
\end{center}
\caption{Diagrams of the inversion curves and the isoenthalpic curves of the RN-AdS black hole with cloud of
strings and quintessence for $\omega=-2/3$. From bottom to top, these curves correspond to the increase in $M$.}%
\label{fig:J12}
\end{figure}
The relationship between the Joule-Thomson coefficient $\mu$ and the event horizon $r_{+}$ is shown in Fig. \ref{fig:J}. It is clear that the Joule-Thomson coefficient curve has a dispersion point, which divides the curve into positive and negative regions. In order to further reveal the relationship between the scattering point of the Joule-Thomson coefficient and the zero point of the Hawking temperature, the Hawking temperature is also plotted for $P=1$ in Fig. \ref{fig:J}. Through comparative analysis, we find that the scattering point of the Joule-Thomson coefficient corresponds to the zero point of the Hawking temperature, which reveals the relevant information of the extremal black hole.

The equation of state can be expressed as
\begin{equation}\label{eqn:R22}
T=(\frac{\partial M}{\partial S})_{P,Q}=\frac{r_{+}\left(-ar_{+}+8\pi Pr_{+}^{3}+3\alpha\omega r_{+}^{-3\omega}+r_{+}\right)-Q^{2}}{4\pi r_{+}^{3}},
\end{equation}
from which we can obtain the equation $P=P(V,T)$ for the black hole as
\begin{equation}\label{eqn:PFT1}
  P=\frac{a}{8\pi r^{2}}+\frac{Q^{2}}{8\pi r^{4}}-\frac{3\alpha\omega r^{-3\omega-3}}{8\pi}-\frac{1}{8\pi r^{2}}+\frac{T}{2r},
\end{equation}
then we can then obtain the critical point using the following conditions \cite{Kubiznak:2012wp}
\begin{equation}\label{eqn:ljd1}
  \frac{\partial P}{\partial r_{+}}=0,\frac{\partial^{2}P}{\partial r_{+}^{2}}=0.
\end{equation}
 The first condition leads to the following critical values of the thermodynamics quantities
 \begin{equation}\label{eqn:TC1}
   T_{c}=\frac{r^{-3(\omega+1)}\left(-2(a-1)r^{3\omega+2}-4Q^{2}r^{3\omega}+9\alpha r\omega(\omega+1)\right)}{4\pi},
 \end{equation}
 the critical pressure is given by
 \begin{equation}\label{eqn:PC1}
   P_{c}=\frac{r^{-3\omega-4}\left(-(a-1)r^{3\omega+2}-3Q^{2}r^{3\omega}+3\alpha r\omega(3\omega+2)\right)}{8\pi}.
 \end{equation}
Combining the second condition we can obtain
\begin{equation}\label{eqn:RC1}
  r_{c}=\frac{\sqrt{6}Q}{\sqrt{1-a}},
\end{equation}
 when $\omega=-1$, the critical physical quantities obtained as
\begin{equation}\label{eqn:TTCC1}
 T_{c}=\frac{(1-a)^{3/2}}{3\sqrt{6}\pi Q},P_{c}=\frac{a^{2}-2a+36\alpha Q^{2}+1}{96\pi Q^{2}},
\end{equation}
when $\omega=-2/3$, the critical physical quantities obtained as
\begin{equation}\label{eqn:TTCC2}
  T_{c}=\frac{2a^{2}-4a+2}{6\pi\sqrt{6-6a}Q}-\frac{\alpha}{2\pi},P_{c}=\frac{(a-1)^{2}}{96\pi Q^{2}}.
\end{equation}
Applying Eq. $\left(  \ref{eqn:J10}\right) $ one can calculate the inversion temperature
\begin{equation}\label{eqn:R33}
  T_{i}=\frac{r_{+}^{-3(\omega+1)}\left((a-1)r_{+}^{3\omega+2}+8\pi P_{i}r_{+}^{3\omega+4}+3Q^{2}r_{+}^{3\omega}-3\alpha r_{+}\omega(3\omega+2)\right)}{12\pi}.
\end{equation}
From Eq. $\left(  \ref{eqn:R22}\right) $, one can get
\begin{equation}\label{eqn:R4}
  T_{i}=\frac{r_{+}\left(-ar_{+}+8\pi P_{i}r_{+}^{3}+3\alpha\omega r_{+}^{-3\omega}+r_{+}\right)-Q^{2}}{4\pi r_{+}^{3}},
\end{equation}
Subtracting Eq. $\left(  \ref{eqn:R3}\right) $ from Eq. $\left(  \ref{eqn:R4}\right) $ we can obtain
\begin{equation}\label{eqn:R5}
 \frac{r_{+}\left(4r_{+}\left(a-4\pi r_{+}^{2}P_{i}-1\right)-3\alpha\omega(3\omega+5)r_{+}^{-3\omega}\right)+6Q^{2}}{12\pi r_{+}^{3}}=0.
\end{equation}

By eliminating the variable $r_{+}$ between Eq. $\left(  \ref{eqn:R4}\right) $ and Eq. $\left(  \ref{eqn:R5}\right) $, the equation for T versus P can be obtained thus plotting the inverse temperature profile. Usually, in order to obtain the analytical solution it is necessary to first set numerical values on the barotropic index $\omega$, so we obtain two cases about $\omega$
\begin{equation}\label{eqn:R6}
\begin{aligned}
&-3\alpha a^{4}+8\pi a^{4}P_{i}+12\alpha a^{3}-32\pi a^{3}P_{i}-32\pi^{2}a^{3}T_{i}^{2}-18\alpha a^{2}-384\pi\alpha a^{2}Q^{2}P_{i}\\&+512\pi^{2}a^{2}Q^{2}P_{i}^{2}+48\pi a^{2}P_{i}+96\pi^{2}a^{2}T_{i}^{2}+72\alpha^{2}a^{2}Q^{2}+12\alpha a+768\pi\alpha aQ^{2}P_{i}\\
&-1024\pi^{2}aQ^{2}P_{i}^{2}-32\pi aP_{i}-96\pi^{2}aT_{i}^{2}-144\alpha^{2}aQ^{2}-3\alpha+3456\pi\alpha^{2}Q^{4}P_{i}\\
&-9216\pi^{2}\alpha Q^{4}P_{i}^{2}+8192\pi^{3}Q^{4}P_{i}^{3}-384\pi\alpha Q^{2}P_{i}+512\pi^{2}Q^{2}P_{i}^{2}+8\pi P_{i}\\
&-6912\pi^{4}Q^{2}T_{i}^{4}+32\pi^{2}T_{i}^{2}-432\alpha^{3}Q^{4}+72\alpha^{2}Q^{2}=0,\\
\end{aligned}
\end{equation}
where $\omega=-1$.
\begin{equation}\label{eqn:R7}
\begin{aligned}
&-8\pi a^{4}P_{i}-\alpha^{2}a^{3}+32\pi a^{3}P_{i}-4\pi\alpha a^{3}T_{i}+32\pi^{2}a^{3}T_{i}^{2}+3\alpha^{2}a^{2}-512\pi^{2}a^{2}Q^{2}P_{i}^{2}\\
&-48\pi a^{2}P_{i}+12\pi\alpha a^{2}T_{i}-96\pi^{2}a^{2}T_{i}^{2}+\alpha^{2}-3\alpha^{2}a-288\pi\alpha^{2}aQ^{2}P_{i}-1152\pi^{2}\alpha aQ^{2}P_{i}T_{i}\\
&+1024\pi^{2}aQ^{2}P_{i}^{2}+32\pi aP_{i}-12\pi\alpha aT_{i}+96\pi^{2}aT_{i}^{2}-8192\pi^{3}Q^{4}P_{i}^{3}+288\pi\alpha^{2}Q^{2}P_{i}\\
&+1152\pi^{2}\alpha Q^{2}P_{i}T_{i}-512\pi^{2}Q^{2}P_{i}^{2}-8\pi P_{i}-216\pi\alpha^{3}Q^{2}T_{i}+3456\pi^{3}\alpha Q^{2}T_{i}^{3}\\&+6912\pi^{4}Q^{2}T_{i}^{4}+4\pi\alpha T_{i}-32\pi^{2}T_{i}^{2}-27\alpha^{4}Q^{2}=0,\\
\end{aligned}
\end{equation}
where $\omega=-2/3$.

The inversion curves for different values of $a$, $Q$, $\alpha$ and $\omega$ are shown in Fig. \ref{fig:T1}. It is obvious that the inversion curves are
not closed. The inversion temperature increases monotonically with increasing inversion pressure, and the slope of the inversion curve also changes. The parameters $a$, $Q$, $\alpha$ and $\omega$ also have different effects on the inversion curve respectively, with $\alpha$ showing a more significant effect. In contrast to van der Waals fluids, this curve does not terminate at a point. It shows that during Joule-Thomson expansion, the Joule-Thomson coefficient is positive above the inversion curve, i.e., the black hole is always cooling above the inversion curve. This is the same as the case of Kerr-AdS black hole \cite{Okcu:2017qgo}, Born-Infeld AdS black hole \cite{Bi:2020vcg} and other AdS black holes described previously.

In order to plot an isenthalpic curve in the $T-P$ plane, a relational formula for $T(P)$ has to be obtained by eliminating $r_{+}$ between Eq. $\left(  \ref{eqn:Q7}\right) $ and Eq. $\left(  \ref{eqn:R4}\right) $. Setting numerical values on the barotropic index $\omega$, we can obtain the analytical solution of $T(P)$, which leads to
\begin{equation}\label{eqn:TM1}
\begin{aligned}
&-72\pi a^{4}Q^{2}P_{i}+27a^{4}\alpha Q^{2}-72\pi a^{3}M^{2}P_{i}+288\pi a^{3}Q^{2}P_{i}+108\pi^{2}a^{3}Q^{2}T_{i}^{2}+27a^{3}\alpha M^{2}\\
&-108a^{3}\alpha Q^{2}+216\pi a^{2}M^{2}P_{i}+108\pi^{2}a^{2}M^{2}T_{i}^{2}-1152\pi a^{2}\alpha Q^{4}P_{i}+1536\pi^{2}a^{2}Q^{4}P_{i}^{2}\\
&-432\pi a^{2}Q^{2}P_{i}-324\pi^{2}a^{2}Q^{2}T_{i}^{2}-81a^{2}\alpha M^{2}+216a^{2}\alpha^{2}Q^{4}+162a^{2}\alpha Q^{2}\\
&+6912\pi^{2}aM^{2}Q^{2}P_{i}^{2}-216\pi aM^{2}P_{i}-216\pi^{2}aM^{2}T_{i}^{2}-864\pi^{3}aMQ^{2}T_{i}^{3}+2304\pi a\alpha Q^{4}P_{i}\\
&-1152\pi^{3}aQ^{4}P_{i}T_{i}^{2}-3072\pi^{2}aQ^{4}P_{i}^{2}+288\pi aQ^{2}P_{i}+432\pi^{2}a\alpha Q^{4}T_{i}^{2}+324\pi^{2}aQ^{2}T_{i}^{2}\\
&+81a\alpha M^{2}+972a\alpha^{2}M^{2}Q^{2}-432a\alpha^{2}Q^{4}-108a\alpha Q^{2}-3888\pi\alpha M^{4}P_{i}+5184\pi^{2}M^{4}P_{i}^{2}\\
&+5184\pi\alpha M^{2}Q^{2}P_{i}-1728\pi^{3}M^{2}Q^{2}P_{i}T_{i}^{2}-6912\pi^{2}M^{2}Q^{2}P_{i}^{2}+72\pi M^{2}P_{i}\\
&+108\pi^{2}M^{2}T_{i}^{2}+864\pi^{3}MQ^{2}T_{i}^{3}-3456\pi\alpha^{2}Q^{6}P_{i}+9216\pi^{2}\alpha Q^{6}P_{i}^{2}-8192\pi^{3}Q^{6}P_{i}^{3}\\
&-1152\pi\alpha Q^{4}P_{i}+1152\pi^{3}Q^{4}P_{i}T_{i}^{2}+1536\pi^{2}Q^{4}P_{i}^{2}-72\pi Q^{2}P_{i}-432\pi^{2}\alpha Q^{4}T_{i}^{2}\\
&-108\pi^{2}Q^{2}T_{i}^{2}+729\alpha^{2}M^{4}-27\alpha M^{2}-972\alpha^{2}M^{2}Q^{2}-5184\pi a\alpha M^{2}Q^{2}P_{i}-432\pi^{4}Q^{4}T_{i}^{4}\\
&+432\alpha^{3}Q^{6}+216\alpha^{2}Q^{4}+27\alpha Q^{2}-864\pi^{3}M^{3}T_{i}^{3}+648\pi^{2}\alpha M^{2}Q^{2}T_{i}^{2}=0,\\
\end{aligned}
\end{equation}
where $\omega=-1$.
\begin{equation}\label{eqn:TM2}
\begin{aligned}
&-1152\pi a^{4}Q^{2}P_{i}-1152\pi a^{3}M^{2}P_{i}+4608\pi a^{3}Q^{2}P_{i}+1728\pi^{2}a^{3}Q^{2}T_{i}^{2}-108a^{3}\alpha^{2}Q^{2}\\
&+3456\pi a^{2}M^{2}P_{i}+1728\pi^{2}a^{2}M^{2}T_{i}^{2}+11520\pi a^{2}\alpha MQ^{2}P_{i}+24576\pi^{2}a^{2}Q^{4}P_{i}^{2}\\
&-6912\pi a^{2}Q^{2}P_{i}-5184\pi^{2}a^{2}Q^{2}T_{i}^{2}-108a^{2}\alpha^{2}M^{2}+324a^{2}\alpha^{2}Q^{2}+10368\pi a\alpha M^{3}P_{i}\\
&+110592\pi^{2}aM^{2}Q^{2}P_{i}^{2}-3456\pi aM^{2}P_{i}-3456\pi^{2}aM^{2}T_{i}^{2}-23040\pi a\alpha MQ^{2}P_{i}\\
&-12096\pi^{2}a\alpha MQ^{2}T_{i}^{2}-13824\pi^{3}aMQ^{2}T_{i}^{3}+10368\pi a\alpha^{2}Q^{4}P_{i}-18432\pi^{3}aQ^{4}P_{i}T_{i}^{2}\\
&-49152\pi^{2}aQ^{4}P_{i}^{2}+4608\pi aQ^{2}P_{i}+5184\pi^{2}aQ^{2}T_{i}^{2}+216a\alpha^{2}M^{2}+972a\alpha^{3}MQ^{2}\\&-324a\alpha^{2}Q^{2}
+82944\pi^{2}M^{4}P_{i}^{2}-10368\pi\alpha M^{3}P_{i}-10368\pi^{2}\alpha M^{3}T_{i}^{2}-13824\pi^{3}M^{3}T_{i}^{3}\\
&+1728\pi\alpha^{2}M^{2}Q^{2}P_{i}
-27648\pi^{3}M^{2}Q^{2}P_{i}T_{i}^{2}-110592\pi^{2}M^{2}Q^{2}P_{i}^{2}+1152\pi M^{2}P_{i}\\
&+1728\pi^{2}M^{2}T_{i}^{2}+73728\pi^{2}\alpha MQ^{4}P_{i}^{2}
+11520\pi\alpha MQ^{2}P_{i}+12096\pi^{2}\alpha MQ^{2}T_{i}^{2}\\
&-131072\pi^{3}Q^{6}P_{i}^{3}-10368\pi\alpha^{2}Q^{4}P_{i}
+18432\pi^{3}Q^{4}P_{i}T_{i}^{2}+24576\pi^{2}Q^{4}P_{i}^{2}-1152\pi Q^{2}P_{i}\\
&-7776\pi^{2}\alpha^{2}Q^{4}T_{i}^{2}-13824\pi^{3}\alpha Q^{4}T_{i}^{3}
-6912\pi^{4}Q^{4}T_{i}^{4}-1728\pi^{2}Q^{2}T_{i}^{2}+864\alpha^{3}M^{3}\\
&-108\alpha^{2}M^{2}-972\alpha^{3}MQ^{2}+729\alpha^{4}Q^{4}+108\alpha^{2}Q^{2}+13824\pi^{3}MQ^{2}T_{i}^{3}=0,\\
\end{aligned}
\end{equation}
where $\omega=-2/3$. Then we can plot the isenthalpic curves in the $T-P$ plane by fixing the mass of the black hole. The setting $M>Q$ is to avoid particular hypersurfaces in $T-P$ plane which exhibits with naked singularity \cite{Okcu:2016tgt,Ghaffarnejad:2013cma}. The isenthalpic curves and the inversion curves are shown in Fig. \ref{fig:J00}, Fig. \ref{fig:J11} and Fig. \ref{fig:J12}, which show that the Joule-Thomson coefficient happens for all states where $M>Q$. The inversion curve is the dividing line between heating and cooling, above which the slope of the isenthalpic curve is positive and cooling occurs, below which the slope of the isenthalpic curve is negative and heating occurs. On the other hand, cooling (heating) does not happen on the inversion curve. As exhibited in Fig. \ref{fig:J00} in the case of $a=\alpha=0$, the quintessence and cloud of strings corrections are removed, thus our work reached the results of Ref. \cite{Okcu:2016tgt}. The comparison shows that the values of $a$, $Q$, $\alpha$ and $\omega$ all affect the cooling-heating critical point, but the cooling-heating critical point is more sensitive to $Q$, which is the same as the result in Ref. \cite{Ghaffarnejad:2018exz}. By comparing the effect of $a$ and $\alpha$ on the cooling-heating critical point of the black hole, the effect of $\alpha$, i.e., a normalization constant of the quintessence, is greater.

Next we studied the ratio between $T_{i}^{min}$ and $T_{c}$. Note that $T_{i}^{min}$ can be obtained by demanding $P_{i}=0$. Thus, one can obtain
\begin{equation}\label{eqn:TIM1}
\begin{aligned}
&-3\alpha a^{4}+12\alpha a^{3}-32\pi^{2}a^{3}\left(T_{i}^{\min}\right){}^{2}-18\alpha a^{2}+96\pi^{2}a^{2}\left(T_{i}^{\min}\right){}^{2}+72\alpha^{2}a^{2}Q^{2}\\
&+12\alpha a-96\pi^{2}a\left(T_{i}^{\min}\right){}^{2}-144\alpha^{2}aQ^{2}-3\alpha-6912\pi^{4}Q^{2}\left(T_{i}^{\min}\right){}^{4}\\
&+32\pi^{2}\left(T_{i}^{\min}\right){}^{2}-432\alpha^{3}Q^{4}+72\alpha^{2}Q^{2}=0,\\
\end{aligned}
\end{equation}
where $\omega=-1$.
\begin{equation}\label{eqn:TIM2}
\begin{aligned}
&-a^{3}\alpha^{2}-4\pi a^{3}\alpha T_{i}^{\min}+32\pi^{2}a^{3}\left(T_{i}^{\min}\right){}^{2}+3a^{2}\alpha^{2}+12\pi a^{2}\alpha T_{i}^{\min}\\
&-96\pi^{2}a^{2}\left(T_{i}^{\min}\right){}^{2}+\alpha^{2}-3a\alpha^{2}-12\pi a\alpha T_{i}^{\min}+96\pi^{2}a\left(T_{i}^{\min}\right){}^{2}\\
&-216\pi\alpha^{3}Q^{2}T_{i}^{\min}+3456\pi^{3}\alpha Q^{2}\left(T_{i}^{\min}\right){}^{3}+6912\pi^{4}Q^{2}\left(T_{i}^{\min}\right){}^{4}\\
&+4\pi\alpha T_{i}^{\min}-32\pi^{2}\left(T_{i}^{\min}\right){}^{2}-27\alpha^{4}Q^{2}=0,\\
\end{aligned}
\end{equation}
where $\omega=-2/3$. For visualization, we set different values of $a$ and $\alpha$ to calculate the ratio between $T_{i}^{min}$ and $T_{c}$. When $a=0$, $\alpha=0$, the following equation can be obtained
\begin{equation}\label{eqn:RAIO1}
 T_{i}^{min}=\frac{1}{6\sqrt{6}\pi Q},T_{c}=\frac{1}{3\sqrt{6}\pi Q},\frac{T_{i}^{min}}{T_{c}}=\frac{1}{2}.
\end{equation}
The above recover the results in Ref. \cite{Okcu:2016tgt}. If we set $a=0$,$\alpha\neq0$, one can obtain
\begin{equation}\label{eqn:RAIO2}
  -3\alpha\left(1-12\alpha Q^{2}\right)^{2}-6912\pi^{4}Q^{2}(T_{i}^{min})^{4}+32\pi^{2}(T_{i}^{min})^{2}=0,T_{c}=\frac{1}{3\sqrt{6}\pi Q},
\end{equation}
where $\omega=-1$.
\begin{equation}\label{eqn:RAIO22}
\begin{aligned}
&\left(-\alpha+27\alpha^{3}Q^{2}+1728\pi^{3}Q^{2}(T_{i}^{min})^{3}+1296\pi^{2}\alpha Q^{2}(T_{i}^{min})^{2}+4\pi T_{i}^{min}\left(81\alpha^{2}Q^{2}-2\right)\right)\\&(4\pi T_{i}^{min}-\alpha)=0,\\
&T_{c}=\frac{1}{3\sqrt{6}\pi Q}-\frac{\alpha}{2\pi},\\
\end{aligned}
\end{equation}
where $\omega=-2/3$.
\begin{figure}
\begin{center}
\subfigure[{}]{
\includegraphics[width=0.45\textwidth]{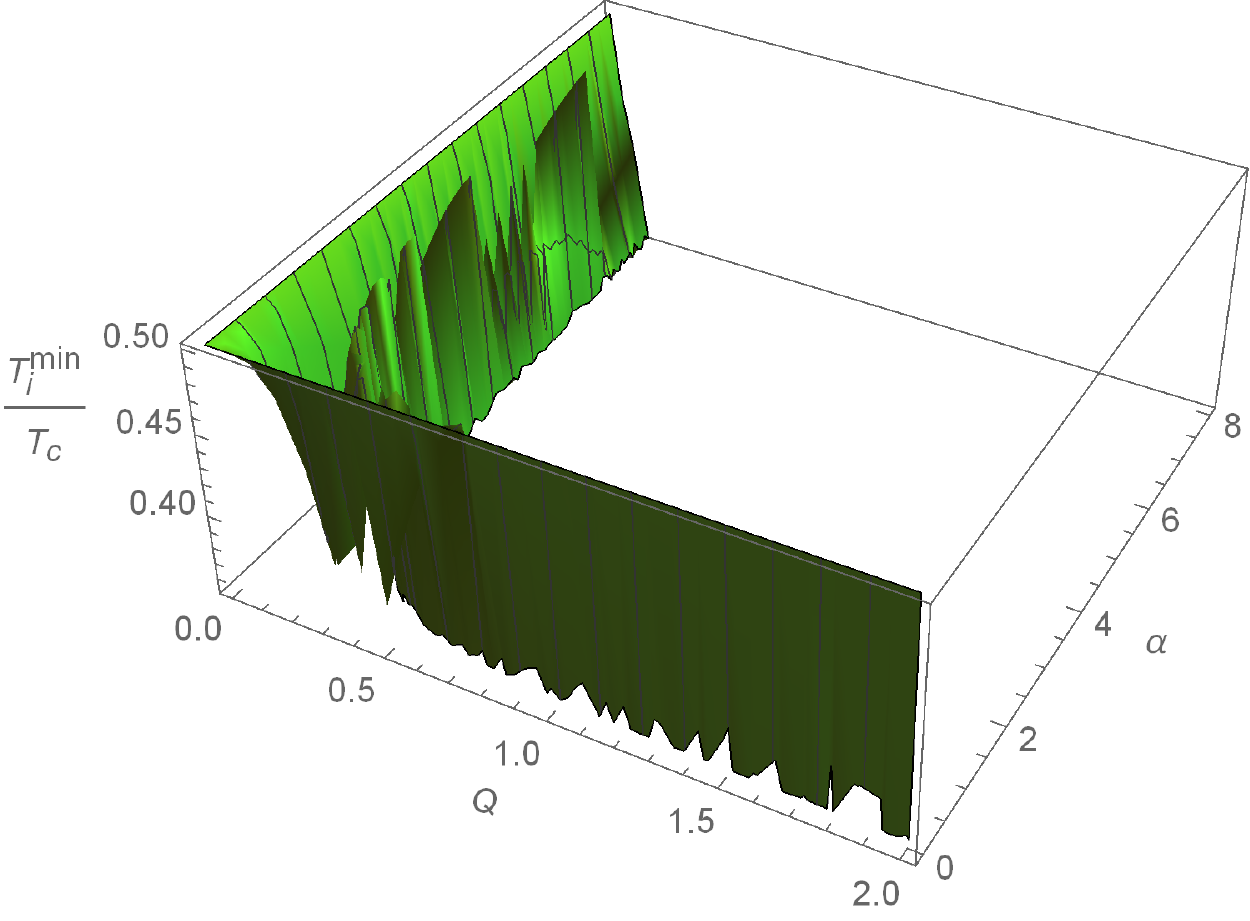}\label{TMC11}}
\subfigure[{}]{
\includegraphics[width=0.45\textwidth]{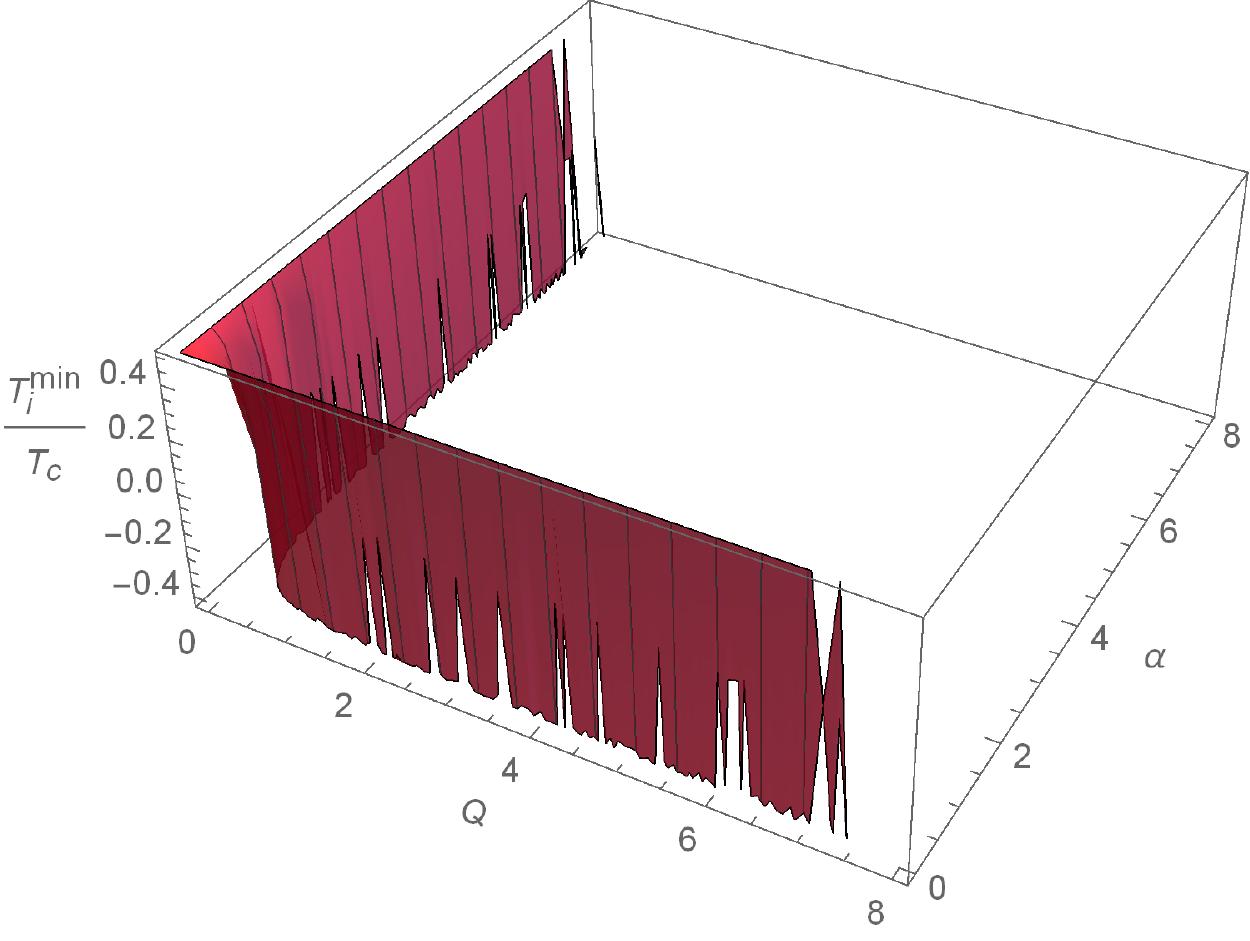}\label{TMC12}}
\end{center}
\caption{The ratio between $T_{i}^{min}$ and $T_{c}$ of the black hole for $a=0,\alpha\neq0$.}%
\label{fig:TMIN1}
\end{figure}
As shown in Fig. \ref{fig:TMIN1}, the ratio between $T_{i}^{min}$ and $T_{c}$ of the black hole for $a=0,\alpha\neq0$ is asymptotically 1/2. This means that when only the contribution of quintessence is considered, the ratio $\frac{T_{i}^{min}}{T_{c}}$ of RN-AdS black holes is asymptotically 1/2.

If we set $\alpha=0$,$a\neq0$, one can obtain
\begin{equation}\label{eqn:RAIO3}
 -32\pi^{2}(a-1)^{3}(T_{i}^{min})^{2}-6912\pi^{4}Q^{2}(T_{i}^{min})^{4}=0,T_{c}=\frac{(1-a)^{3/2}}{3\sqrt{6}\pi Q},
\end{equation}
\begin{equation}\label{eqn:tttt1}
 t=\frac{T_{i}^{min}}{T_{c}},\frac{16(a-1)^{6}t^{2}(2t-1)(2t+1)}{27Q^{2}}=0,
\end{equation}
where $\omega=-1$.
\begin{equation}\label{eqn:RAIO4}
 4\pi(T_{i}^{min})\left(8\pi(a-1)^{3}(T_{i}^{min})+1728\pi^{3}Q^{2}(T_{i}^{min})^{3}\right)=0,T_{c}=\frac{2a^{2}-4a+2}{6\pi\sqrt{6-6a}Q},
\end{equation}
\begin{equation}\label{eqn:tttt2}
 t=\frac{T_{i}^{min}}{T_{c}},\frac{16(a-1)^{6}t^{2}(2t-1)(2t+1)}{27Q^{2}}=0,
\end{equation}
where $\omega=-2/3$. It is obvious that the ratio between $T_{i}^{min}$ and $T_{c}$ of the black hole for $\alpha=0$,$a\neq0$ is 1/2. This means that when only the contribution of the cloud of strings is considered, the ratio $\frac{T_{i}^{min}}{T_{c}}$ of RN-AdS black holes is 1/2.
In the case of $\alpha\neq0$, $a\neq0$, the ratio between $T_{i}^{min}$ and $T_{c}$ of the black hole is expressed as follows by setting $t=\frac{T_{i}^{min}}{T_{c}}$
\begin{equation}\label{eqn:tttt3}
  \frac{-81\alpha Q^{2}\left((a-1)^{2}-12\alpha Q^{2}\right)^{2}-64(a-1)^{6}t^{4}+16(a-1)^{6}t^{2}}{27Q^{2}}=0,
\end{equation}
where $\omega=-1$.
\begin{equation}\label{eqn:tttt4}
\begin{aligned}
&\frac{2t\left(\sqrt{6}a^{2}-9\sqrt{1-a}\alpha Q-2\sqrt{6}a+\sqrt{6}\right)-9\sqrt{1-a}\alpha Q}{81(a-1)Q^{2}}\\
&\times \{-9\alpha\sqrt{1-a}Q(a^{3}-3a^{2}+3a+27\alpha^{2}Q^{2}-1)\\
&+108\alpha Qt^{2}(2\sqrt{1-a}a^{3}-6a^{2}(\sqrt{1-a}-\sqrt{6}\alpha Q)-27\sqrt{1-a}\alpha^{2}Q^{2}\\
&+6a(\sqrt{1-a}-2\sqrt{6}\alpha Q)-2\sqrt{1-a}+6\sqrt{6}\alpha Q)\\
&-2t\left(\sqrt{6}a^{2}-9\sqrt{1-a}\alpha Q-2\sqrt{6}a+\sqrt{6}\right)\left(2a^{3}-6a^{2}+6a+81\alpha^{2}Q^{2}-2\right)\\
&+8t^{3}(2\sqrt{6}a^{5}-10\sqrt{6}a^{4}+a^{3}(20\sqrt{6}-54\sqrt{1-a}\alpha Q)\\
&+a^{2}(162\sqrt{1-a}\alpha Q-81\sqrt{6}\alpha^{2}Q^{2}-20\sqrt{6}))\\
&+8t^{3}(243\alpha^{3}\sqrt{1-a}Q^{3}+2a\left(-81\sqrt{1-a}\alpha Q+81\sqrt{6}\alpha^{2}Q^{2}+5\sqrt{6}\right)\\
&+54\alpha\sqrt{1-a}Q-81\sqrt{6}\alpha^{2}Q^{2}-2\sqrt{6})\}.\\
\end{aligned}
\end{equation}
Where $\omega=-2/3$. Based on Eq. $\left(\ref{eqn:tttt3}\right)$ and Eq. $\left(\ref{eqn:tttt4}\right)$, the ratios for the different parameters in the case of $Q=1$ are listed in tables. By observing Table \ref{tab:TMINa1}, Table \ref{tab:TMINa2}, Table \ref{tab:TMINA1} and Table \ref{tab:TMINA2}, it is found that the ratios are all less than but close to 1/2 in the case of $\alpha\neq0$, $a\neq0$. Then it can be concluded that the ratio between $T_{i}^{min}$ and $T_{c}$ of RN-AdS black holes with cloud of strings and quintessence is always less than 1/2.
\begin{table}[]
\caption{The ratio between $T_{i}^{min}$ and $T_{c}$ of RN-AdS black holes with cloud of
strings and quintessence for $\alpha=0.01$, $Q=1$, $\omega=-1$.}
\begin{tabular}{p{0.7in}|p{0.75in}|p{0.75in}|p{0.75in}|p{0.75in}|p{0.75in}|p{0.75in}|p{0.75in}}
\hline
\hline
$a$    & 0.0001   & 0.0005   & 0.001    & 0.005    & 0.01     & 0.05    & 0.06      \\ \hline
$\frac{T_{i}^{min}}{T_{c}}$ & 0.448671 & 0.448629 & 0.448577 & 0.448157 & 0.447625 & 0.44305 & 0.437845 \\ \hline
\hline
\end{tabular}
\label{tab:TMINa1}
\end{table}

\begin{table}[]
\caption{The ratio between $T_{i}^{min}$ and $T_{c}$ of RN-AdS black holes with cloud of
strings and quintessence for $\alpha=0.01$, $Q=1$, $\omega=-2/3$.}
\begin{tabular}{p{0.7in}|p{0.75in}|p{0.75in}|p{0.75in}|p{0.75in}|p{0.75in}|p{0.75in}|p{0.75in}}
\hline
\hline
$a$    & 0.0001   & 0.0005   & 0.001    & 0.005    & 0.01     & 0.05    & 0.06      \\ \hline
$\frac{T_{i}^{min}}{T_{c}}$ & 0.495164 & 0.495161 & 0.495157 & 0.495126 & 0.495087 & 0.494756 & 0.494385 \\ \hline
\hline
\end{tabular}
\label{tab:TMINa2}
\end{table}

\begin{table}[]
\caption{The ratio between $T_{i}^{min}$ and $T_{c}$ of RN-AdS black holes with cloud of
strings and quintessence for $a=0.01$, $Q=1$, $\omega=-1$.}
\begin{tabular}{p{0.7in}|p{0.75in}|p{0.75in}|p{0.75in}|p{0.75in}|p{0.75in}|p{0.75in}|p{0.75in}}
\hline
\hline
$\alpha$    & 0.0001   & 0.0005   & 0.001    & 0.005    & 0.01     & 0.05    & 0.06      \\ \hline
$\frac{T_{i}^{min}}{T_{c}}$ & 0.499483 & 0.497416 & 0.494828 & 0.473997 & 0.447625 & 0.449336 & 0.475224 \\ \hline
\hline
\end{tabular}
\label{tab:TMINA1}
\end{table}

\begin{table}[]
\caption{The ratio between $T_{i}^{min}$ and $T_{c}$ of RN-AdS black holes with cloud of
strings and quintessence for $a=0.01$, $Q=1$, $\omega=-2/3$.}
\begin{tabular}{p{0.7in}|p{0.75in}|p{0.75in}|p{0.75in}|p{0.75in}|p{0.75in}|p{0.75in}|p{0.75in}}
\hline
\hline
$\alpha$    & 0.0001   & 0.0005   & 0.001    & 0.005    & 0.01     & 0.05    & 0.06      \\ \hline
$\frac{T_{i}^{min}}{T_{c}}$ & 0.499953 & 0.499766 & 0.499531 & 0.497608 & 0.495087 & 0.469043 & 0.426382 \\ \hline
\hline
\end{tabular}
\label{tab:TMINA2}
\end{table}
\begin{table}[]
\caption{The ratio between $T_{i}^{min}$ and $T_{c}$ for various black holes.}
\begin{tabular}{p{3.1in}|p{1.5in}|p{0.9in}}
\hline
\hline
type  & $\frac{T_{i}^{min}}{T_{c}}$   & literature \\ \hline
van-der-Waals-fluids  & 0.75   & \cite{Okcu:2016tgt} \\ \hline
holographic-super-fluids &0.4864 & \cite{DAlmeida:2018ldi}        \\ \hline
Kerr-AdS BH &0.5 &\cite{Okcu:2017qgo} \\ \hline
RN-AdS BH &0.5 &\cite{Okcu:2016tgt} \\ \hline
quintessence-RN-AdS BH &0.5 &\cite{Ghaffarnejad:2018exz} \\ \hline
d-dimensional-AdS BH &less than 0.5 & \cite{Mo:2018rgq} \\ \hline
f(r)-gravity-AdS BH &0.5 &\cite{Belhaj:2019idh} \\ \hline
Lovelock-gravity & less than 0.4389 & \cite{Mo:2018qkt} \\ \hline
global-monopole-AdS BH &0.5 &\cite{Rizwan:2018mpy}  \\ \hline
Gauss-bonnet-AdS BH &0.4765 & \cite{Lan:2018nnp}\\ \hline
massive-gravity &0.4626 &\cite{Haldar:2018cks} \\ \hline
Einstien-Maxwell-axions-theory&0.5&\cite{Cisterna:2018jqg} \\ \hline
Bardeen-AdS BH &0.536622 &\cite{Pu:2019bxf} \\ \hline
torus-like BH & not exist &\cite{Liang:2021elg} \\ \hline
Born-Infeld-AdS BH&  asymptotically 0.5 &\cite{Bi:2020vcg} \\ \hline
cloud of strings-RN-AdS BH&  0.5 & \\ \hline
quintessence-RN-AdS BH &  asymptotically 0.5 & \\ \hline
cloud of strings and quintessence-RN-AdS BH& less than 0.5 & \\ \hline
\hline
\end{tabular}
\label{tab:zong}
\end{table}
\section{ Conclusion and Discussion}
\label{sec:D}
In this paper, the well-known Joule-Thomson expansion for RN-AdS black holes with cloud of strings and quintessence was studied. By considering the cosmological constant as a thermodynamic pressure, we studied the thermodynamics of such a black hole. The Joule-Thomson expansion describes the expansion of the gas through the porous plug from the high pressure section to the low pressure section. During this process the black hole mass remains unchanged and the mass is interpreted as the enthalpy of the black hole. Next we studied the Joule-Thomson expansion of van der Waals fluids and the Joule-Thomson expansion of RN-AdS black holes with cloud of strings and quintessence, respectively. The existence of inversion temperature results from the competition of attractive and repulsive interactions between real molecules. Then, we plotted their inversion curves, as well as determined the cooling and heating regions. A comparison shows that the inversion curve of van der Waals gas is closed, while the inversion curve of this black hole is the lower one and no closure. It means that the RN-AdS black holes with cloud of strings and quintessence always cool above the inversion curve during the Joule-Thomson expansion.

The sign of the Joule-Thomson coefficient plays an essential role in the research process, which can be used to determine whether heating or cooling will occur. The Joule-Thomson coefficient of this black hole was then obtained, and the scattering point of the Joule-Thomson coefficient is found to be consistent with the zero point of Hawking temperature. In addition, we plotted the inversion curves and the isoenthalpic curves of the black hole in the $T-P$ plane. In the case of $M>Q$, numerical analysis was performed by setting different values of the parameters, and it was found that the Joule-Thomson expansion occurs and there is the cooling-heating critical point, i.e., the intersection of the isoenthalpic curve and the inversion curve. This critical point is more sensitive to the black hole charge $Q$, followed by the normalization constant of the quintessence $\alpha$.

Furthermore, we calculated the ratio between minimum inversion temperature $T_{i}^{min}$ and the critical temperature $T_{c}$ in different cases.
 The study found that in the case of $a=0$, $\alpha=0$, the ratio $\frac{T_{i}^{min}}{T_{c}}$ equals 1/2. When $a=0$, $\alpha\neq0$, the ratio $\frac{T_{i}^{min}}{T_{c}}$ is asymptotically 1/2. When $\alpha=0$, $a\neq0$, the ratio $\frac{T_{i}^{min}}{T_{c}}$ is equal to 1/2. When both quintessence and cloud of strings are considered, i.e., $\alpha\neq0$, $a\neq0$, the ratio $\frac{T_{i}^{min}}{T_{c}}$ is always less than 1/2.
We also compared the ratio with various other black holes and the results are shown in the following Table \ref{tab:zong}.

\begin{acknowledgments}
We are grateful to Wei Hong, Peng Wang, Haitang Yang, Jun Tao, Deyou Chen and
Xiaobo Guo for useful discussions. This work is supported in part by NSFC
(Grant No. 11747171), Natural Science Foundation of Chengdu University of TCM
(Grants nos. ZRYY1729 and ZRYY1921), Discipline Talent Promotion Program of
/Xinglin Scholars(Grant no.QNXZ2018050) and the key fund project for Education
Department of Sichuan (Grantno. 18ZA0173).
\end{acknowledgments}


\begin{thebibliography}{999}
\bibitem{Boulware:1985wk}
D.~G.~Boulware and S.~Deser,
``String Generated Gravity Models,''
Phys. Rev. Lett. \textbf{55}, 2656 (1985)
doi:10.1103/PhysRevLett.55.2656

\bibitem{Wheeler:1985nh}
J.~T.~Wheeler,
``Symmetric Solutions to the Gauss-Bonnet Extended Einstein Equations,''
Nucl. Phys. B \textbf{268}, 737-746 (1986)
doi:10.1016/0550-3213(86)90268-3

\bibitem{Cai:2003kt}
R.~G.~Cai,
``A Note on thermodynamics of black holes in Lovelock gravity,''
Phys. Lett. B \textbf{582}, 237-242 (2004)
doi:10.1016/j.physletb.2004.01.015
[arXiv:hep-th/0311240 [hep-th]].

\bibitem{Hennigar:2016xwd}
R.~A.~Hennigar, R.~B.~Mann and E.~Tjoa,
``Superfluid Black Holes,''
Phys. Rev. Lett. \textbf{118}, no.2, 021301 (2017)
doi:10.1103/PhysRevLett.118.021301
[arXiv:1609.02564 [hep-th]].

\bibitem{Myers:1988ze}
R.~C.~Myers and J.~Z.~Simon,
``Black Hole Thermodynamics in Lovelock Gravity,''
Phys. Rev. D \textbf{38}, 2434-2444 (1988)
doi:10.1103/PhysRevD.38.2434

\bibitem{Letelier:1979ej}
P.~S.~Letelier,
``CLOUDS OF STRINGS IN GENERAL RELATIVITY,''
Phys. Rev. D \textbf{20}, 1294-1302 (1979)
doi:10.1103/PhysRevD.20.1294

\bibitem{Ghosh:2014pga}
S.~G.~Ghosh, U.~Papnoi and S.~D.~Maharaj,
``Cloud of strings in third order Lovelock gravity,''
Phys. Rev. D \textbf{90}, no.4, 044068 (2014)
doi:10.1103/PhysRevD.90.044068
[arXiv:1408.4611 [gr-qc]].

\bibitem{Herscovich:2010vr}
E.~Herscovich and M.~G.~Richarte,
``Black holes in Einstein-Gauss-Bonnet gravity with a string cloud background,''
Phys. Lett. B \textbf{689}, 192-200 (2010)
doi:10.1016/j.physletb.2010.04.065
[arXiv:1004.3754 [hep-th]].

\bibitem{Lee:2015xlp}
T.~H.~Lee, S.~G.~Ghosh, S.~D.~Maharaj and D.~Baboolal,
``Lovelock black hole thermodynamics in a string cloud model,''
[arXiv:1511.03976 [gr-qc]].

\bibitem{Graca:2016cbd}
J.~P.~Morais Gra\c{c}a, G.~I.~Salako and V.~B.~Bezerra,
``Quasinormal modes of a black hole with a cloud of strings in Einstein\textendash{}Gauss\textendash{}Bonnet gravity,''
Int. J. Mod. Phys. D \textbf{26}, no.10, 1750113 (2017)
doi:10.1142/S0218271817501139
[arXiv:1604.04734 [gr-qc]].

\bibitem{Letelier:1984dm}
P.~S.~Letelier,
``STRING COSMOLOGIES,''
Phys. Rev. D \textbf{28}, 2414-2419 (1983)
doi:10.1103/PhysRevD.28.2414

\bibitem{Li:2020zxi}
Z.~Li and T.~Zhou,
``Kerr Black Hole Surrounded by a Cloud of Strings and Its Weak Gravitational Lensing in Rastall Gravity,''
[arXiv:2001.01642 [gr-qc]].

\bibitem{Cai:2019nlo}
X.~C.~Cai and Y.~G.~Miao,
``Quasinormal modes and spectroscopy of a Schwarzschild black hole surrounded by a cloud of strings in Rastall gravity,''
Phys. Rev. D \textbf{101}, no.10, 104023 (2020)
doi:10.1103/PhysRevD.101.104023
[arXiv:1911.09832 [hep-th]].



\bibitem {intro-Toledo:2019szg}J.~Toledo and V.~Bezerra, ``Black holes with a
cloud of strings in pure Lovelock gravity,'' Eur. Phys. J. C \textbf{79},
no.2, 117 (2019) doi:10.1140/epjc/s10052-019-6628-4




\bibitem {Ghaffarnejad:2018tpr}H.~Ghaffarnejad and E.~Yaraie, ``Effects
of a cloud of strings on the extended phase space of Einstein-Gauss-Bonnet AdS
black holes,'' Phys. Lett. B \textbf{785}, 105-111 (2018)
doi:10.1016/j.physletb.2018.08.017 [arXiv:1806.06687 [gr-qc]].








\bibitem{Ghaffarnejad:2018gbf}
H.~Ghaffarnejad and M.~Farsam,
``The Last Lost Charge And Phase Transition In Schwarzschild AdS Minimally Coupled to a Cloud of Strings,''
doi:10.1140/epjp/i2019-12514-5
[arXiv:1806.06688 [hep-th]].




\bibitem {intro-Ghosh:2014dqa}S.~G.~Ghosh and S.~D.~Maharaj, ``Cloud of
strings for radiating black holes in Lovelock gravity,'' Phys. Rev. D
\textbf{89}, no.8, 084027 (2014) doi:10.1103/PhysRevD.89.084027
[arXiv:1409.7874 [gr-qc]].



\bibitem{Perlmutter:1998np}
S.~Perlmutter \textit{et al.} [Supernova Cosmology Project],
``Measurements of $\Omega$ and $\Lambda$ from 42 high redshift supernovae,''
Astrophys. J. \textbf{517}, 565-586 (1999)
doi:10.1086/307221
[arXiv:astro-ph/9812133 [astro-ph]].

\bibitem{Kiselev:2002dx}
V.~V.~Kiselev,
``Quintessence and black holes,''
Class. Quant. Grav. \textbf{20}, 1187-1198 (2003)
doi:10.1088/0264-9381/20/6/310
[arXiv:gr-qc/0210040 [gr-qc]].










\bibitem{Stern:1999ta}
B.~Stern, Y.~Tikhomirova, M.~Stepanov, D.~Kompaneets, A.~Berezhnoy and R.~Svensson,
``Search for non-triggered gamma-ray bursts in the batse continuous records: preliminary results,''
Astrophys. J. Lett. \textbf{540}, L21 (2000)
doi:10.1086/312873
[arXiv:astro-ph/9903094 [astro-ph]].


\bibitem{Bahcall:1999xn}
N.~A.~Bahcall, J.~P.~Ostriker, S.~Perlmutter and P.~J.~Steinhardt,
``The Cosmic triangle: Assessing the state of the universe,''
Science \textbf{284}, 1481-1488 (1999)
doi:10.1126/science.284.5419.1481
[arXiv:astro-ph/9906463 [astro-ph]].

\bibitem{Steinhardt:1999nw}
P.~J.~Steinhardt, L.~M.~Wang and I.~Zlatev,
``Cosmological tracking solutions,''
Phys. Rev. D \textbf{59}, 123504 (1999)
doi:10.1103/PhysRevD.59.123504
[arXiv:astro-ph/9812313 [astro-ph]].

\bibitem{Wang:1999fa}
L.~M.~Wang, R.~R.~Caldwell, J.~P.~Ostriker and P.~J.~Steinhardt,
``Cosmic concordance and quintessence,''
Astrophys. J. \textbf{530}, 17-35 (2000)
doi:10.1086/308331
[arXiv:astro-ph/9901388 [astro-ph]].

\bibitem{Godunov:2015nea}
S.~I.~Godunov, A.~N.~Rozanov, M.~I.~Vysotsky and E.~V.~Zhemchugov,
``Extending the Higgs sector: an extra singlet,''
Eur. Phys. J. C \textbf{76}, 1 (2016)
doi:10.1140/epjc/s10052-015-3826-6
[arXiv:1503.01618 [hep-ph]].

\bibitem{Toshmatov:2017btq}
B.~Toshmatov, Z.~Stuchl\'\i{}k and B.~Ahmedov,
``Comments on \textquotedblleft{}Casimir effect in the Kerr spacetime with quintessence\textquotedblright{},''
Mod. Phys. Lett. A \textbf{32}, no.21, 1775001 (2017)
doi:10.1142/S0217732317750013
[arXiv:1707.00403 [gr-qc]].


\bibitem{Chen:2008ra}
S.~Chen, B.~Wang and R.~Su,
``Hawking radiation in a $d$-dimensional static spherically-symmetric black Hole surrounded by quintessence,''
Phys. Rev. D \textbf{77}, 124011 (2008)
doi:10.1103/PhysRevD.77.124011
[arXiv:0801.2053 [gr-qc]].

\bibitem{Chen:2005qh}
S.~b.~Chen and J.~l.~Jing,
``Quasinormal modes of a black hole surrounded by quintessence,''
Class. Quant. Grav. \textbf{22}, 4651-4657 (2005)
doi:10.1088/0264-9381/22/21/011
[arXiv:gr-qc/0511085 [gr-qc]].



\bibitem {intro-Saleh:2011zz}M.~Saleh, B.~T.~Bouetou and T.~C.~Kofane,
``Quasinormal modes and Hawking radiation of a Reissner-Nordstroem black hole
surrounded by quintessence,'' Astrophys.\ Space Sci.\ \textbf{333}, 449 (2011)
doi:10.1007/s10509-011-0643-8 {[}arXiv:1604.02140 {[}gr-qc{]}{]}.




\bibitem {intro-Li:2014ixn}G.~Q.~Li, ``Effects of dark energy on P-V
criticality of charged AdS black holes,'' Phys. Lett. B \textbf{735}, 256-260
(2014) doi:10.1016/j.physletb.2014.06.047 [arXiv:1407.0011 [gr-qc]].




\bibitem {intro-Singh:2020tkf}A.~Singh, A.~Ghosh and C.~Bhamidipati, ``Effect
of dark energy on the microstructures of black holes in AdS spacetimes,''
[arXiv:2002.08787 [gr-qc]].




\bibitem {intro-Haldar:2020jmt}A.~Haldar and R.~Biswas, ``Thermodynamic
studies with modifications of entropy: different black holes embedded in
quintessence,'' Gen. Rel. Grav. \textbf{52}, no.2, 19 (2020)
doi:10.1007/s10714-020-02669-z




\bibitem {intro-Chen:2020rov}Q.~Chen, W.~Hong and J.~Tao, ``Universal
thermodynamic extremality relations for charged AdS black hole surrounded by
quintessence,'' [arXiv:2005.00747 [gr-qc]].







\bibitem {intro-Moinuddin:2019mzf}A.~Moinuddin and M.~Hossain Ali,
``Thermodynamics and quantum tunneling of Reissner-Nordstrom black holes with
deficit solid angle and quintessence,'' Int. J. Mod. Phys. A \textbf{34},
no.32, 1950211 (2019) doi:10.1142/S0217751X19502117




\bibitem {intro-Toledo:2019mlz}J.~Toledo and V.~Bezerra, ``Black holes with
quintessence in pure Lovelock gravity,'' Gen. Rel. Grav. \textbf{51}, no.3, 41
(2019) doi:10.1007/s10714-019-2528-z




\bibitem {intro-Chabab:2017xdw} M.~Chabab, H.~El Moumni, S.~Iraoui, K.~Masmar
and S.~Zhizeh,  ``More Insight into Microscopic Properties of RN-AdS Black
Hole Surrounded by Quintessence via an Alternative Extended Phase Space,''
Int.\ J.\ Geom.\ Meth.\ Mod.\ Phys.\ \textbf{15}, no. 10, 1850171 (2018)
doi:10.1142/S0219887818501712  [arXiv:1704.07720 [gr-qc]].




\bibitem {intro-Liu:2017baz}H.~Liu and X.~H.~Meng, ``Effects of dark energy on
the efficiency of charged AdS black holes as heat engines,'' Eur. Phys. J. C
\textbf{77}, no.8, 556 (2017) doi:10.1140/epjc/s10052-017-5134-9
[arXiv:1704.04363 [hep-th]].




\bibitem {intro-Ghaderi:2016dpi}K.~Ghaderi and B.~Malakolkalami,
``Thermodynamics of the Schwarzschild and the Reissner-Nordstrom black holes
with quintessence,'' Nucl. Phys. B \textbf{903}, 10-18 (2016)
doi:10.1016/j.nuclphysb.2015.11.019




\bibitem {intro-Fernando:2014rsa}S.~Fernando, S.~Meadows and K.~Reis, ``Null
trajectories and bending of light in charged black holes with quintessence,''
Int. J. Theor. Phys. \textbf{54}, no.10, 3634-3653 (2015)
doi:10.1007/s10773-015-2601-7 [arXiv:1411.3192 [gr-qc]].




\bibitem {intro-Fernando:2014wma}S.~Fernando, ``Cold, ultracold and Nariai
black holes with quintessence,'' Gen. Rel. Grav. \textbf{45}, 2053-2073 (2013)
doi:10.1007/s10714-013-1578-x [arXiv:1401.0714 [gr-qc]].




\bibitem {intro-Guo:2019hxa}X.~Y.~Guo, H.~F.~Li, L.~C.~Zhang and R.~Zhao,
``Continuous Phase Transition and Microstructure of Charged AdS Black Hole
with Quintessence,'' Eur. Phys. J. C \textbf{80}, no.2, 168 (2020)
doi:10.1140/epjc/s10052-019-7601-y [arXiv:1911.09902 [gr-qc]].




\bibitem {intro-Nandan:2016ksb}H.~Nandan and R.~Uniyal, ``Geodesic Flows in a
Charged Black Hole Spacetime with Quintessence,'' Eur. Phys. J. C \textbf{77},
no.8, 552 (2017) doi:10.1140/epjc/s10052-017-5122-0 [arXiv:1612.07455
[gr-qc]].




\bibitem {intro-Malakolkalami:2015cza}B.~Malakolkalami and K.~Ghaderi,
``Schwarzschild-anti de Sitter black hole with quintessence,'' Astrophys.
Space Sci. \textbf{357}, no.2, 112 (2015) doi:10.1007/s10509-015-2340-5




\bibitem {intro-WangChun-Yan:2012tcg}C.~Y.~Wang, Y.~F.~Zhao and Y.~J.~Gao,
``Quasinormal Modes of a Black Hole with Quintessence-Like Matter and a
Deficit Solid Angle for Electromagnetic Perturbation,'' Commun. Theor. Phys.
\textbf{57}, no.6, 1101-1104 (2012)




\bibitem {intro-Xi:2008ce}P.~Xi, ``Quasinormal modes of a black hole with
quintessence-like matter and a deficit solid angle,'' Astrophys. Space Sci.
\textbf{321}, 47-51 (2009) doi:10.1007/s10509-009-9994-9 [arXiv:0810.4022
[gr-qc]].




\bibitem {intro-Harada:2006dv}T.~Harada, H.~Maeda and B.~J.~Carr,
``Non-existence of self-similar solutions containing a black hole in a
Universe with a stiff fluid or scalar field or quintessence,'' Phys. Rev. D
\textbf{74}, 024024 (2006) doi:10.1103/PhysRevD.74.024024
[arXiv:astro-ph/0604225 [astro-ph]].







\bibitem{Ghaffarnejad:2018zlz}
H.~Ghaffarnejad, E.~Yaraie and M.~Farsam,
``Thermodynamic phase transition for Quintessence Dyonic Anti de Sitter Black Holes,''
Eur. Phys. J. Plus \textbf{135}, no.2, 179 (2020)
doi:10.1140/epjp/s13360-020-00211-3
[arXiv:1808.09789 [physics.gen-ph]].













\bibitem{Yin:2021fsg}
R.~Yin, J.~Liang and B.~Mu,
``Stability of horizon with pressure and volume of d-dimensional charged AdS black holes with cloud of strings and quintessence,''
Phys. Dark Univ. \textbf{32}, 100831 (2021)
doi:10.1016/j.dark.2021.100831
[arXiv:2103.08162 [gr-qc]].



\bibitem {Toledo:2019amt}J.~Toledo and V.~Bezerra, ``Some remarks on the
thermodynamics of charged AdS black holes with cloud of strings and
quintessence,'' Eur. Phys. J. C \textbf{79}, no.2, 110 (2019)
doi:10.1140/epjc/s10052-019-6616-8




\bibitem {Chabab:2020ejk}M.~Chabab and S.~Iraoui, ``Thermodynamic criticality
of d-dimensional charged AdS black holes surrounded by quintessence with a
cloud of strings background,'' Gen. Rel. Grav. \textbf{52}, no.8, 75 (2020)
doi:10.1007/s10714-020-02729-4 [arXiv:2001.06063 [hep-th]].







\bibitem {intro-Sakti:2019iku}M.~F.~A.~R.~Sakti, H.~L.~Prihadi, A.~Suroso and
F.~P.~Zen, ``Rotating and Twisting Charged Black Holes with Cloud of Strings
and Quintessence,'' [arXiv:1911.07569 [gr-qc]].




\bibitem {Ma:2019pya}Y.~Ma, Y.~Zhang, R.~Zhao, S.~Cao, T.~Liu, S.~Geng,
Y.~Liu and Y.~Huang, ``Phase transitions and entropy force of charged de
Sitter black holes with cloud of string and quintessence,'' [arXiv:1907.11870
[hep-th]].


\bibitem {Toledo:2018pfy}J.~de M.Toledo and V.~B.~Bezerra, ``Black holes with
cloud of strings and quintessence in Lovelock gravity,'' Eur. Phys. J. C
\textbf{78}, no.7, 534 (2018) doi:10.1140/epjc/s10052-018-6001-z














\bibitem{Bekenstein:1973ur}
J.~D.~Bekenstein,
``Black holes and entropy,''
Phys. Rev. D \textbf{7}, 2333-2346 (1973)
doi:10.1103/PhysRevD.7.2333

\bibitem{Bardeen:1973gs}
J.~M.~Bardeen, B.~Carter and S.~W.~Hawking,
``The Four laws of black hole mechanics,''
Commun. Math. Phys. \textbf{31}, 161-170 (1973)
doi:10.1007/BF01645742

\bibitem{Bekenstein:1974ax}
J.~D.~Bekenstein,
``Generalized second law of thermodynamics in black hole physics,''
Phys. Rev. D \textbf{9}, 3292-3300 (1974)
doi:10.1103/PhysRevD.9.3292

\bibitem{Hawking:1974rv}
S.~W.~Hawking,
``Black hole explosions,''
Nature \textbf{248}, 30-31 (1974)
doi:10.1038/248030a0
\bibitem{Hawking:1974sw}
S.~W.~Hawking,
``Particle Creation by Black Holes,''
Commun. Math. Phys. \textbf{43}, 199-220 (1975)
[erratum: Commun. Math. Phys. \textbf{46}, 206 (1976)]
doi:10.1007/BF02345020


\bibitem{Hawking:1982dh}
S.~W.~Hawking and D.~N.~Page,
``Thermodynamics of Black Holes in anti-De Sitter Space,''
Commun. Math. Phys. \textbf{87}, 577 (1983)
doi:10.1007/BF01208266

\bibitem{Chamblin:1999tk}
A.~Chamblin, R.~Emparan, C.~V.~Johnson and R.~C.~Myers,
``Charged AdS black holes and catastrophic holography,''
Phys. Rev. D \textbf{60}, 064018 (1999)
doi:10.1103/PhysRevD.60.064018
[arXiv:hep-th/9902170 [hep-th]].

\bibitem{Chamblin:1999hg}
A.~Chamblin, R.~Emparan, C.~V.~Johnson and R.~C.~Myers,
``Holography, thermodynamics and fluctuations of charged AdS black holes,''
Phys. Rev. D \textbf{60}, 104026 (1999)
doi:10.1103/PhysRevD.60.104026
[arXiv:hep-th/9904197 [hep-th]].

\bibitem{Gunasekaran:2012dq}
S.~Gunasekaran, R.~B.~Mann and D.~Kubiznak,
``Extended phase space thermodynamics for charged and rotating black holes and Born-Infeld vacuum polarization,''
JHEP \textbf{11}, 110 (2012)
doi:10.1007/JHEP11(2012)110
[arXiv:1208.6251 [hep-th]].

\bibitem{Altamirano:2013ane}
N.~Altamirano, D.~Kubiznak and R.~B.~Mann,
``Reentrant phase transitions in rotating anti\textendash{}de Sitter black holes,''
Phys. Rev. D \textbf{88}, no.10, 101502 (2013)
doi:10.1103/PhysRevD.88.101502
[arXiv:1306.5756 [hep-th]].

\bibitem{Altamirano:2013uqa}
N.~Altamirano, D.~Kubiz\v{n}\'ak, R.~B.~Mann and Z.~Sherkatghanad,
``Kerr-AdS analogue of triple point and solid/liquid/gas phase transition,''
Class. Quant. Grav. \textbf{31}, 042001 (2014)
doi:10.1088/0264-9381/31/4/042001
[arXiv:1308.2672 [hep-th]].

\bibitem{Belhaj:2015hha}
A.~Belhaj, M.~Chabab, H.~El Moumni, K.~Masmar, M.~B.~Sedra and A.~Segui,
``On Heat Properties of AdS Black Holes in Higher Dimensions,''
JHEP \textbf{05}, 149 (2015)
doi:10.1007/JHEP05(2015)149
[arXiv:1503.07308 [hep-th]].

\bibitem{Dolan:2011jm}
B.~P.~Dolan,
``Compressibility of rotating black holes,''
Phys. Rev. D \textbf{84}, 127503 (2011)
doi:10.1103/PhysRevD.84.127503
[arXiv:1109.0198 [gr-qc]].

\bibitem{Dolan:2013dga}
B.~P.~Dolan,
``The compressibility of rotating black holes in D-dimensions,''
Class. Quant. Grav. \textbf{31}, 035022 (2014)
doi:10.1088/0264-9381/31/3/035022
[arXiv:1308.5403 [gr-qc]].

\bibitem{Wei:2012ui}
S.~W.~Wei and Y.~X.~Liu,
``Critical phenomena and thermodynamic geometry of charged Gauss-Bonnet AdS black holes,''
Phys. Rev. D \textbf{87}, no.4, 044014 (2013)
doi:10.1103/PhysRevD.87.044014
[arXiv:1209.1707 [gr-qc]].

\bibitem{Banerjee:2011cz}
R.~Banerjee and D.~Roychowdhury,
``Critical phenomena in Born-Infeld AdS black holes,''
Phys. Rev. D \textbf{85}, 044040 (2012)
doi:10.1103/PhysRevD.85.044040
[arXiv:1111.0147 [gr-qc]].

\bibitem{Niu:2011tb}
C.~Niu, Y.~Tian and X.~N.~Wu,
``Critical Phenomena and Thermodynamic Geometry of RN-AdS Black Holes,''
Phys. Rev. D \textbf{85}, 024017 (2012)
doi:10.1103/PhysRevD.85.024017
[arXiv:1104.3066 [hep-th]].

\bibitem{Chen:2019pdj}
D.~Chen,
``Thermodynamics and weak cosmic censorship conjecture in extended phase spaces of anti-de Sitter black holes with particles\textquoteright{} absorption,''
Eur. Phys. J. C \textbf{79}, no.4, 353 (2019)
doi:10.1140/epjc/s10052-019-6874-5
[arXiv:1902.06489 [hep-th]].

\bibitem{Liang:2021voh}
J.~Liang, X.~Guo and B.~Mu,
``Thermodynamics with pressure and volume of black holes based on two assumptions under scalar field scattering,''
[arXiv:2101.11414 [gr-qc]].
\bibitem{Mu:2020szg}
B.~Mu, J.~Liang and X.~Guo,
``Thermodynamics with pressure and volume of 4D Gauss-Bonnet AdS Black Holes under the scalar field,''
[arXiv:2011.00273 [gr-qc]].
\bibitem{Liang:2020uul}
J.~Liang, B.~Mu and J.~Tao,
``Thermodynamics and overcharging problem in the extended phase spaces of charged AdS black holes with cloud of strings and quintessence under charged particle absorption,''
Chin. Phys. C \textbf{45}, no.2, 023121 (2021)
doi:10.1088/1674-1137/abd085
[arXiv:2008.09512 [gr-qc]].

\bibitem{Liang:2020hjz}
J.~Liang, X.~Guo, D.~Chen and B.~Mu,
``Remarks on the weak cosmic censorship conjecture of RN-AdS black holes with cloud of strings and quintessence under the scalar field,''
Nucl. Phys. B \textbf{965}, 115335 (2021)
doi:10.1016/j.nuclphysb.2021.115335
[arXiv:2008.08327 [gr-qc]].
\bibitem{Bai:2020ieh}
T.~Bai, W.~Hong, B.~Mu and J.~Tao,
``Weak cosmic censorship conjecture in the nonlinear electrodynamics black hole under the charged scalar field,''
Commun. Theor. Phys. \textbf{72}, no.1, 015401 (2020)
doi:10.1088/1572-9494/ab544b
\bibitem{Mu:2019bim}
B.~Mu, J.~Tao and P.~Wang,
``Minimal Length Effect on Thermodynamics and Weak Cosmic Censorship Conjecture in anti-de Sitter Black Holes via Charged Particle Absorption,''
doi:10.1155/2020/2612946
[arXiv:1906.10544 [gr-qc]].
\bibitem{Hong:2019yiz}
W.~Hong, B.~Mu and J.~Tao,
``Thermodynamics and weak cosmic censorship conjecture in the charged RN-AdS black hole surrounded by quintessence under the scalar field,''
Nucl. Phys. B \textbf{949}, 114826 (2019)
doi:10.1016/j.nuclphysb.2019.114826
[arXiv:1905.07747 [gr-qc]].



\bibitem{Chabab:2017knz}
M.~Chabab, H.~El Moumni, S.~Iraoui and K.~Masmar,
``Phase Transition of Charged-AdS Black Holes and Quasinormal Modes : a Time Domain Analysis,''
Astrophys. Space Sci. \textbf{362}, no.10, 192 (2017)
doi:10.1007/s10509-017-3175-z
[arXiv:1701.00872 [hep-th]].

\bibitem{Okcu:2016tgt}
\"O.~\"Okc\"u and E.~Ayd\i{}ner,
``Joule\textendash{}Thomson expansion of the charged AdS black holes,''
Eur. Phys. J. C \textbf{77}, no.1, 24 (2017)
doi:10.1140/epjc/s10052-017-4598-y
[arXiv:1611.06327 [gr-qc]].


\bibitem{Okcu:2017qgo}
\"O.~\"Okc\"u and E.~Ayd\i{}ner,
``Joule\textendash{}Thomson expansion of Kerr\textendash{}AdS black holes,''
Eur. Phys. J. C \textbf{78}, no.2, 123 (2018)
doi:10.1140/epjc/s10052-018-5602-x
[arXiv:1709.06426 [gr-qc]].









\bibitem{Ghaffarnejad:2018exz}
H.~Ghaffarnejad, E.~Yaraie and M.~Farsam,
``Quintessence Reissner Nordstr\"om Anti de Sitter Black Holes and Joule Thomson effect,''
Int. J. Theor. Phys. \textbf{57}, no.6, 1671-1682 (2018)
doi:10.1007/s10773-018-3693-7
[arXiv:1802.08749 [gr-qc]].

\bibitem{DAlmeida:2018ldi}
R.~D'Almeida and K.~P.~Yogendran,
``Thermodynamic Properties of Holographic superfluids,''
[arXiv:1802.05116 [hep-th]].

\bibitem{Chabab:2018zix}
M.~Chabab, H.~El Moumni, S.~Iraoui, K.~Masmar and S.~Zhizeh,
``Joule-Thomson Expansion of RN-AdS Black Holes in $f(R)$ gravity,''
LHEP \textbf{02}, 05 (2018)
doi:10.31526/LHEP.2.2018.02
[arXiv:1804.10042 [gr-qc]].

\bibitem{Rizwan:2018mpy}
A.~Rizwan C.L., N.~Kumara A., D.~Vaid and K.~M.~Ajith,
``Joule-Thomson expansion in AdS black hole with a global monopole,''
Int. J. Mod. Phys. A \textbf{33}, no.35, 1850210 (2019)
doi:10.1142/S0217751X1850210X
[arXiv:1805.11053 [gr-qc]].

\bibitem{Mo:2018qkt}
J.~X.~Mo and G.~Q.~Li,
``Effects of Lovelock gravity on the Joule\textendash{}Thomson expansion,''
Class. Quant. Grav. \textbf{37}, no.4, 045009 (2020)
doi:10.1088/1361-6382/ab60b9
[arXiv:1805.04327 [gr-qc]].

\bibitem{Liang:2021elg}
J.~Liang, W.~Lin and B.~Mu,
``Joule-Thomson expansion of the torus-like black hole,''
[arXiv:2103.03119 [gr-qc]].

\bibitem{Hegde:2020xlv}
K.~Hegde, A.~Naveena Kumara, C.~L.~A.~Rizwan, A.~K.~M. and M.~S.~Ali,
``Thermodynamics, Phase Transition and Joule Thomson Expansion of novel 4-D Gauss Bonnet AdS Black Hole,''
[arXiv:2003.08778 [gr-qc]].

\bibitem{Mo:2018rgq}
J.~X.~Mo, G.~Q.~Li, S.~Q.~Lan and X.~B.~Xu,
``Joule-Thomson expansion of $d$-dimensional charged AdS black holes,''
Phys. Rev. D \textbf{98}, no.12, 124032 (2018)
doi:10.1103/PhysRevD.98.124032
[arXiv:1804.02650 [gr-qc]].
\bibitem{Lan:2018nnp}
S.~Q.~Lan,
``Joule-Thomson expansion of charged Gauss-Bonnet black holes in AdS space,''
Phys. Rev. D \textbf{98}, no.8, 084014 (2018)
doi:10.1103/PhysRevD.98.084014
[arXiv:1805.05817 [gr-qc]].

\bibitem{Wei:2017vqs}
S.~W.~Wei and Y.~X.~Liu,
``Charged AdS black hole heat engines,''
Nucl. Phys. \textbf{B}, 114700 (2019)
doi:10.1016/j.nuclphysb.2019.114700
[arXiv:1708.08176 [gr-qc]].

\bibitem{Kuang:2018goo}
X.~M.~Kuang, B.~Liu and A.~\"Ovg\"un,
``Nonlinear electrodynamics AdS black hole and related phenomena in the extended thermodynamics,''
Eur. Phys. J. C \textbf{78}, no.10, 840 (2018)
doi:10.1140/epjc/s10052-018-6320-0
[arXiv:1807.10447 [gr-qc]].


\bibitem{Yekta:2019wmt}
D.~Mahdavian Yekta, A.~Hadikhani and \"O.~\"Okc\"u,
``Joule-Thomson expansion of charged AdS black holes in Rainbow gravity,''
Phys. Lett. B \textbf{795}, 521-527 (2019)
doi:10.1016/j.physletb.2019.06.049
[arXiv:1905.03057 [hep-th]].

\bibitem{Pu:2019bxf}
J.~Pu, S.~Guo, Q.~Q.~Jiang and X.~T.~Zu,
``Joule-Thomson expansion of the regular(Bardeen)-AdS black hole,''
Chin. Phys. C \textbf{44}, no.3, 035102 (2020)
doi:10.1088/1674-1137/44/3/035102
[arXiv:1905.02318 [gr-qc]].

\bibitem{Nam:2018sii}
C.~H.~Nam,
``Thermodynamics and phase transitions of non-linear charged black hole in AdS spacetime,''
Eur. Phys. J. C \textbf{78}, no.7, 581 (2018)
doi:10.1140/epjc/s10052-018-6056-x
\bibitem{Zhao:2018kpz}
Z.~W.~Zhao, Y.~H.~Xiu and N.~Li,
``Throttling process of the Kerr\textendash{}Newman\textendash{}anti-de Sitter black holes in the extended phase space,''
Phys. Rev. D \textbf{98}, no.12, 124003 (2018)
doi:10.1103/PhysRevD.98.124003
[arXiv:1805.04861 [gr-qc]].

\bibitem{Li:2019jcd}
C.~Li, P.~He, P.~Li and J.~B.~Deng,
``Joule-Thomson expansion of the Bardeen-AdS black holes,''
Gen. Rel. Grav. \textbf{52}, no.5, 50 (2020)
doi:10.1007/s10714-020-02704-z
[arXiv:1904.09548 [gr-qc]].

\bibitem{Hyun:2019gfz}
S.~Hyun and C.~H.~Nam,
``Charged AdS black holes in Gauss\textendash{}Bonnet gravity and nonlinear electrodynamics,''
Eur. Phys. J. C \textbf{79}, no.9, 737 (2019)
doi:10.1140/epjc/s10052-019-7248-8
[arXiv:1908.09294 [gr-qc]].



\bibitem{Nam:2019zyk}
C.~H.~Nam,
``Heat engine efficiency and Joule\textendash{}Thomson expansion of nonlinear charged AdS black hole in massive gravity,''
Gen. Rel. Grav. \textbf{53}, no.3, 30 (2021)
doi:10.1007/s10714-021-02787-2
[arXiv:1906.05557 [gr-qc]].

\bibitem{Nam:2018ltb}
C.~H.~Nam,
``Non-linear charged AdS black hole in massive gravity,''
Eur. Phys. J. C \textbf{78}, no.12, 1016 (2018)
doi:10.1140/epjc/s10052-018-6498-1

\bibitem{Rostami:2019ivr}
M.~Rostami, J.~Sadeghi, S.~Miraboutalebi, A.~A.~Masoudi and B.~Pourhassan,
``Charged accelerating AdS black hole of $f(R)$ gravity and the Joule\textendash{}Thomson expansion,''
Int. J. Geom. Meth. Mod. Phys. \textbf{17}, no.09, 2050136 (2020)
doi:10.1142/S0219887820501364
[arXiv:1908.08410 [gr-qc]].

\bibitem{Haldar:2018cks}
A.~Haldar and R.~Biswas,
``Joule-Thomson expansion of five-dimensional Einstein-Maxwell-Gauss-Bonnet-AdS black holes,''
EPL \textbf{123}, no.4, 40005 (2018)
doi:10.1209/0295-5075/123/40005

\bibitem{Guo:2019gkr}
S.~Guo, J.~Pu and Q.~Q.~Jiang,
``Joule-Thomson Expansion of Hayward-AdS black hole,''
[arXiv:1905.03604 [gr-qc]].

\bibitem{Lan:2019kak}
S.~Q.~Lan,
``Joule-Thomson expansion of neutral AdS black holes in massive gravity,''
Nucl. Phys. B \textbf{948}, 114787 (2019)
doi:10.1016/j.nuclphysb.2019.114787

\bibitem{Sadeghi:2020bon}
J.~Sadeghi and R.~Toorandaz,
``Joule-Thomson expansion of hyperscaling violating black holes with spherical and hyperbolic horizons,''
Nucl. Phys. B \textbf{951}, 114902 (2020)
doi:10.1016/j.nuclphysb.2019.114902

\bibitem{Bi:2020vcg}
S.~Bi, M.~Du, J.~Tao and F.~Yao,
``Joule-Thomson expansion of Born-Infeld AdS black holes,''
Chin. Phys. C \textbf{45}, no.2, 025109 (2021)
doi:10.1088/1674-1137/abcf23
[arXiv:2006.08920 [gr-qc]].


\bibitem{Ranjbari:2019ktp}
H.~Ranjbari, M.~Sadeghi, M.~Ghanaatian and G.~Forozani,
``Critical behavior of AdS Gauss\textendash{}Bonnet massive black holes in the presence of external string cloud,''
Eur. Phys. J. C \textbf{80}, no.1, 17 (2020)
doi:10.1140/epjc/s10052-019-7592-8
[arXiv:1911.10803 [hep-th]].

\bibitem{Guo:2019pzq}
S.~Guo, Y.~Han and G.~P.~Li,
``Joule-Thomson expansion of a specific black hole in different dimensions,''
[arXiv:1912.09590 [hep-th]].

\bibitem{K.:2020rzl}
R.~K., C.~L.~A.~Rizwan, A.~Naveena Kumara, D.~Vaid and M.~S.~Ali,
``Joule-Thomson Expansion of Regular Bardeen AdS Black Hole Surrounded by Static Anisotropic Quintessence Field,''
Phys. Dark Univ. \textbf{32}, 100825 (2021)
doi:10.1016/j.dark.2021.100825
[arXiv:2002.03634 [gr-qc]].

\bibitem{Nam:2020gud}
C.~H.~Nam,
``Effect of massive gravity on Joule\textendash{}Thomson expansion of the charged AdS black hole,''
Eur. Phys. J. Plus \textbf{135}, no.2, 259 (2020)
doi:10.1140/epjp/s13360-020-00274-2

\bibitem{Meng:2020csd}
Y.~Meng, J.~Pu and Q.~Q.~Jiang,
``P-V criticality and Joule-Thomson expansion of charged AdS black holes in the Rastall gravity,''
Chin. Phys. C \textbf{44}, no.6, 065105 (2020)
doi:10.1088/1674-1137/44/6/065105

\bibitem{Guo:2020qxy}
S.~Guo, Y.~Han and G.~P.~Li,
``Joule\textendash{}Thomson expansion of a specific black hole in f(R) gravity coupled with Yang\textendash{}Mills field,''
Class. Quant. Grav. \textbf{37}, no.8, 085016 (2020)
doi:10.1088/1361-6382/ab77ec

\bibitem{Ghanaatian:2019xhi}
M.~Ghanaatian, M.~Sadeghi, H.~Ranjbari and G.~Forozani,
``Effects of the external string cloud on the Van der Waals like behavior and efficiency of AdS-Schwarzschild black holes in massive gravity,''
Mod. Phys. Lett. A \textbf{35}, no.24, 2050203 (2020)
doi:10.1142/S021773232050203X
[arXiv:1906.00369 [hep-th]].

\bibitem{Guo:2020zcr}
S.~Guo, Y.~Han and G.~P.~Li,
``Thermodynamic of the charged AdS black holes in Rastall gravity: P \ensuremath{-} V critical and Joule\textendash{}Thomson expansion,''
Mod. Phys. Lett. A \textbf{35}, no.14, 2050113 (2020)
doi:10.1142/S0217732320501138

\bibitem{Feng:2020swq}
Z.~W.~Feng, X.~Zhou, G.~He, S.~Q.~Zhou and S.~Z.~Yang,
``Joule\textendash{}Thomson expansion of higher dimensional nonlinearly AdS black hole with power Maxwell invariant source,''
Commun. Theor. Phys. \textbf{73}, no.6, 065401 (2021)
doi:10.1088/1572-9494/abecd9
[arXiv:2009.02172 [gr-qc]].

\bibitem{Debnath:2020zdv}
U.~Debnath,
``The General Class of Accelerating, Rotating and Charged Plebanski-Demianski Black Holes as Heat Engine,''
[arXiv:2006.02920 [gr-qc]].

\bibitem{Cao:2021dcq}
Y.~Cao, H.~Feng, W.~Hong and J.~Tao,
``Joule-Thomson Expansion of RN-AdS Black Hole Immersed in Perfect Fluid Dark Matter,''
[arXiv:2101.08199 [gr-qc]].

\bibitem{Huang:2020xcs}
Y.~l.~Huang and S.~Guo,
``Thermodynamic of the charged accelerating AdS black hole: P-V critical and Joule-Thomson expansion,''
[arXiv:2009.09401 [hep-th]].

\bibitem{Zhang:2021raw}
M.~Zhang, C.~M.~Zhang, D.~C.~Zou and R.~H.~Yue,
``$P-V$ criticality and Joule-Thomson Expansion of Hayward-AdS black holes in 4D Einstein-Gauss-Bonnet gravity,''
[arXiv:2102.04308 [hep-th]].

\bibitem{Chen:2020igz}
N.~Chen,
``Throttling Process of Rotating Bardeen AdS Black Holes,''
[arXiv:2003.00247 [gr-qc]].

\bibitem{Jawad:2020mdc}
A.~Jawad and S.~Chaudhary,
``Implications of new phase transitions approach onto specific black holes,''
Mod. Phys. Lett. A \textbf{35}, no.39, 2050326 (2020)
doi:10.1142/S0217732320503265

\bibitem{Liang:2021xny}
J.~Liang, B.~Mu and P.~Wang,
``Joule-Thomson expansion of Lower-dimensional black hole,''
[arXiv:2104.08841 [gr-qc]].

\bibitem{Debnath:2020inx}
U.~Debnath,
``Thermodynamics of FRW Universe: Heat Engine,''
Phys. Lett. B \textbf{810}, 135807 (2020)
doi:10.1016/j.physletb.2020.135807
[arXiv:2010.02102 [gr-qc]].

\bibitem{Mirza:2021kvi}
B.~Mirza, F.~Naeimipour and M.~Tavakoli,
``Joule-Thomson Expansion of the Quasitopological Black Holes,''
Front. in Phys. \textbf{9}, 33 (2021)
doi:10.3389/fphy.2021.628727
[arXiv:2105.05047 [gr-qc]].

\bibitem{Graca:2021izb}
J.~P.~M.~Gra\c{c}a, E.~F.~Capossoli and H.~Boschi-Filho,
``Joule-Thomson expansion for quantum corrected AdS-Reissner-Nordstrom black holes in Kiselev spacetime,''
[arXiv:2105.04689 [gr-qc]].



























\bibitem{Dolan:2011xt}
B.~P.~Dolan,
``Pressure and volume in the first law of black hole thermodynamics,''
Class. Quant. Grav. \textbf{28}, 235017 (2011)
doi:10.1088/0264-9381/28/23/235017
[arXiv:1106.6260 [gr-qc]].

\bibitem{Kubiznak:2012wp}
D.~Kubiznak and R.~B.~Mann,
``P-V criticality of charged AdS black holes,''
JHEP \textbf{07}, 033 (2012)
doi:10.1007/JHEP07(2012)033
[arXiv:1205.0559 [hep-th]].

\bibitem{Cvetic:2010jb}
M.~Cvetic, G.~W.~Gibbons, D.~Kubiznak and C.~N.~Pope,
``Black Hole Enthalpy and an Entropy Inequality for the Thermodynamic Volume,''
Phys. Rev. D \textbf{84}, 024037 (2011)
doi:10.1103/PhysRevD.84.024037
[arXiv:1012.2888 [hep-th]].

\bibitem{Caceres:2015vsa}
E.~Caceres, P.~H.~Nguyen and J.~F.~Pedraza,
``Holographic entanglement entropy and the extended phase structure of STU black holes,''
JHEP \textbf{09}, 184 (2015)
doi:10.1007/JHEP09(2015)184
[arXiv:1507.06069 [hep-th]].

\bibitem{Hendi:2012um}
S.~H.~Hendi and M.~H.~Vahidinia,
``Extended phase space thermodynamics and P-V criticality of black holes with a nonlinear source,''
Phys. Rev. D \textbf{88}, no.8, 084045 (2013)
doi:10.1103/PhysRevD.88.084045
[arXiv:1212.6128 [hep-th]].

\bibitem{Pedraza:2018eey}
J.~F.~Pedraza, W.~Sybesma and M.~R.~Visser,
``Hyperscaling violating black holes with spherical and hyperbolic horizons,''
Class. Quant. Grav. \textbf{36}, no.5, 054002 (2019)
doi:10.1088/1361-6382/ab0094
[arXiv:1807.09770 [hep-th]].





















\bibitem{Johnston:2014dc}
D.~C.~Johnston
``Thermodynamic Properties of the Van derWaals Fluid,''
[arXiv:1402.1205] (2014)

\bibitem{Landau:1980dc}
L.~D.~Landau, E.~M.~Lifshitz and L.~P.~Pitaevski
``Statistical physics,'' (1980)

















\bibitem{Ghaffarnejad:2013cma}
H.~Ghaffarnejad, H.~Neyad and M.~A.~Mojahedi,
``Evaporating Quantum Lukewarm Black Holes Final State From Back-Reaction Corrections of Quantum Scalar Fields,''
Astrophys. Space Sci. \textbf{346}, 497-506 (2013)
doi:10.1007/s10509-013-1462-x
[arXiv:1305.6914 [physics.gen-ph]].






\bibitem{Belhaj:2019idh}
A.~Belhaj and H.~El Moumni,
``Entanglement entropy and phase portrait of f(R)-AdS black holes in the grand canonical ensemble,''
Nucl. Phys. B \textbf{938}, 200-211 (2019)
doi:10.1016/j.nuclphysb.2018.11.010
[arXiv:1812.07962 [hep-th]].

















\bibitem{Cisterna:2018jqg}
A.~Cisterna, S.~Q.~Hu and X.~M.~Kuang,
``Joule-Thomson expansion in AdS black holes with momentum relaxation,''
Phys. Lett. B \textbf{797}, 134883 (2019)
doi:10.1016/j.physletb.2019.134883
[arXiv:1808.07392 [gr-qc]].

































\end{thebibliography}
\end{document}